\documentclass[12pt]{article}
\usepackage[utf8]{inputenc}

\usepackage{CJKutf8}
\usepackage{empheq}

\usepackage[square, comma, compress,numbers]{natbib}

\usepackage{footmisc}
\usepackage{float}
\usepackage{comment}

\usepackage[dvipsnames]{xcolor}
\definecolor{amber(sae/ece)}{rgb}{1.0, 0.49, 0.0}
\definecolor{ballblue}{rgb}{0.13, 0.67, 0.8 }

\usepackage{amsmath, amssymb, amscd, amsthm, amsfonts}
\usepackage[colorlinks = true, linkcolor = ballblue, citecolor = red, final]{hyperref}
\usepackage{physics}
\usepackage{color}
\usepackage{bm}
\usepackage{graphicx}
\usepackage{multicol}
\usepackage{ marvosym }
\usepackage{tensor}
\usepackage[export]{adjustbox}

\usepackage{caption}
\usepackage{subcaption}

\newcommand{\cH}{\mathcal{H}}
\newcommand{\bea}{\begin{align}}
\newcommand{\mH}{\mathcal{H}}
\newcommand{\eea}{\end{align}}
\newcommand{\nn}{\nonumber}
\newcommand{\mHb}{\mathcal{H}_{\text{bulk}}(K)}
\newcommand{\Zb}{Z_{\text{bulk}}}
\numberwithin{equation}{section}

\newcommand{\be}{\begin{equation}}
\newcommand{\ee}{\end{equation}}

\newcommand{\ii}{\mathrm{i}}

\newcommand{\tvarphi}{\tilde{\varphi}}
\allowdisplaybreaks

\begin{document}

\thispagestyle{plain}
\begin{center}
\vspace*{1cm}
\vspace*{1cm}
\LARGE
\textbf{How the Hilbert space of two-sided black holes factorises }
\vspace*{1cm}
\normalsize
        
\vspace{0.4cm}

\textbf{Jan Boruch${}^1$, Luca V.~Iliesiu${}^1$, Guanda Lin${}^1$, Cynthia Yan${}^{2,3}$}
\,\vspace{0.5cm}\\
{${}^1$  Department of Physics, University of California, Berkeley, CA 94720, USA}
\,\vspace{0.5cm}\\
{${}^2$ Stanford Institute for Theoretical Physics, Stanford University, Stanford, CA 94305, USA}
\,\vspace{0.5cm}\\
{${}^3$ Kavli Institute for Theoretical Physics, University of California, Santa Barbara, CA 93106, USA}
       
\vspace{0.9cm}
\end{center} 
\noindent\textbf{Abstract}
In AdS/CFT, two-sided black holes 
are described by states in the tensor product of two Hilbert spaces associated with the two asymptotic boundaries of the spacetime. Understanding how such a tensor product arises from the bulk perspective is an important open problem in holography, known as the factorisation puzzle. In this paper, we show how the Hilbert space of bulk states factorises due to non-perturbative contributions of spacetime wormholes: the trace over two-sided states with different particle excitations behind the horizon factorises into a product of traces of the left and right sides. This precisely occurs when such states form a complete basis for the bulk Hilbert space. We prove that the factorisation of the trace persists to all non-perturbative orders in $1/G_N$,
consequently providing a possible resolution to the factorisation puzzle from the gravitational path integral. In the language of von Neumann algebras, our results provide strong evidence that the algebra of one-sided observables transitions from a Type II or Type III algebra, depending on whether or not perturbative gravity effects are included, to a Type I factor when including non-perturbative corrections in the bulk.

\vspace{0.4cm}

\newpage
\tableofcontents

\newpage 
\section{Introduction and summary of results}

According to AdS/CFT \cite{Maldacena:1997re,Gubser:1998bc,Witten:1998qj}, two-sided black hole states should belong to the tensor product of two Hilbert spaces, each associated with one of the two asymptotic boundaries of the spacetime \cite{Maldacena:2001kr, Maldacena:2013xja}. This property, however, is not at all apparent from the bulk Hilbert space of matter and gravitational excitations in the two-sided geometry. Perturbatively, from the bulk perspective, the Hilbert space is seemingly given by a tensor product between states describing the geometry of timeslices between the two sides and states describing the excitations of the matter quantum field theory on such timeslices, resulting in a Hilbert space that is drastically different from a tensor product between the left and right sides. Consequently, in such a setup, one is seemingly unable to define pure states for one of the sides by simply tracing out the other. Understanding how such a tensor product arises from the bulk perspective is an important open problem in holography, known as the ``factorisation puzzle'' \cite{Guica:2015zpf, Harlow:2015lma, Harlow:2018tqv, Penington:2023dql}.\footnote{Here, we are following the naming convention discussed in \cite{Penington:2023dql} and suggested by Henry Maxfield. This puzzle is different from the ``factori\textbf{z}ation puzzle'' which concerns the non-trivial statistics of multi-boundary observables in gravity. } 
While it has commonly been believed that addressing this puzzle would require understanding the UV completion of the gravitational theory, the tremendous progress made in the past years in understanding the non-perturbative gravitational path integral suggests that a resolution to this puzzle might come from wormhole corrections. In this paper, we indeed explain how the Hilbert space of bulk states factorises due to the non-perturbative contributions of spacetime wormholes to the statistics of the inner product between bulk states. More precisely, by solely using the gravitational path integral, we show that the trace in the bulk Hilbert space for products between left and right operators factorises into a product of traces, 
\be 
\label{eq:kL-kR-factorisation}
\Tr_{\mathcal{H}_\text{bulk}}(k_L k_R)=\Tr_{\mathcal{H}_L}(k_L)\Tr_{\mathcal{H}_R}(k_R)\,, \qquad k_{L,R} \in \text{Left or right algebra of ops,}
\ee
to all non-perturbative orders in $1/G_N$. In the process of proving \eqref{eq:kL-kR-factorisation}, we will discuss the interplay between two different factorization puzzles -- the factorisation of the two-sided Hilbert space discussed above and the factorization of multi-boundary observables -- and will explain that resolving the former does not require resolving the latter puzzle.

Determining that the Hilbert space factorises is critical for understanding the algebraic classification of operators in gravity. Suppose we are interested in operators localized to the left or right exteriors of the two-sided black hole. If the spacetime is rigid, these are simply the quantum field theory operators localized to each region of the spacetime. The algebra of such operators is not one we typically encounter in quantum mechanics: it is a Type III algebra, for which one cannot define density matrices or entropies associated with this subregion. When spacetime is no longer rigid, and the quantum field theory is coupled to gravity, there is a surprising change in the algebra of observables: once semi-classical effects are taken into account, in numerous known examples \cite{Penington:2023dql, Leutheusser:2021qhd, Witten:2021unn, Chandrasekaran:2022eqq, Bahiru:2022mwh, Jensen:2023yxy, Witten:2023qsv, Witten:2023xze,Kolchmeyer:2023gwa, Kudler-Flam:2023qfl, Engelhardt:2023xer, Soni:2023fke, AliAhmad:2023etg, Akers:2024bel, Cirafici:2024jdw}, the algebra transitions from Type III to Type II. In contrast to QFTs, density matrices and entropies are now well defined in subregions, but such algebras still lack a notion of pure states associated with each subregion. The algebra seen in semi-classical gravity is still not what we expect of two-sided black holes in AdS/CFT: precisely because of the factorisation of the two-sided Hilbert space on the CFT side, there are pure states associated with each one of the two sides, and the algebra should become Type I. Thus, by proving that the Hilbert space factorises, we provide strong evidence that the algebra of one-sided operators becomes a Type I factor. Deriving that the algebra of operators becomes Type I solely by using the gravitational path integral has broader implications. In principle, the procedure proposed in this paper should not be limited to AdS and should be applicable to cosmological spacetimes closer to our universe, such as de Sitter spacetimes. 

The steps towards proving that the bulk Hilbert space factorises are as follows. 

In \textbf{section \ref{sec:conventions-and-review}} \textbf{and} \textbf{\ref{sec:fact-leading-order}}, we set the foundations for our computations by defining a basis for the bulk Hilbert space $\mathcal{H}_\text{bulk}$, which we shall later trace over. In order to stand a chance of finding a factorising Hilbert space, we need to break all possible gauge constraints between the left and right boundaries \cite{Guica:2015zpf, Harlow:2015lma, Harlow:2018tqv}.
One such gauge constraint, which fixes the left and right ADM Hamiltonians to be equal, is present for two-sided black hole states prepared without any operator insertions.\footnote{On the boundary side, this would be the case in the thermofield double state with any temperature or in any linear combination of such states.} To break this gauge constraint, we must thus consider states where matter excitations are inserted between the two sides. Thus, we will define our bulk Hilbert space $\mathcal{H}_\text{bulk}$ as
\be 
\label{eq:Hilbert-space-construction}
\mHb \equiv \text{Span } \, \{\ket{q_i} = O_i \ket{HH}, \, i=1,\,\dots,\,K\}\,,
\ee
where the state $\ket{HH}$ is the two-sided state prepared through Euclidean preparation without any matter excitation, on which we create the matter excitation $O_i$, where $i$ labels the different matter excitation that we can use to create a basis.  The set of matter operators (with label $i$) can be chosen such that the states in \eqref{eq:Hilbert-space-construction} are orthogonal to each other when computing inner products in perturbation theory.\footnote{A concrete example for this comes from JT gravity coupled to matter for which, perturbatively, the Hilbert space takes the form $L^2(\mathbb R) \otimes \mathcal H_\text{matter}$ \cite{Penington:2023dql,Kolchmeyer:2023gwa, Iliesiu:2024cnh}. There, states can be labeled by the length between the two sides, by the irreducible $SL(2, \mathbb R)$ representation of the matter excitation, and by the state within that irreducible representation. Choosing any of these three quantum numbers to be different, for instance, by choosing the $O_i$'s to have different scaling dimensions, yields a set of orthogonal states at the perturbative level.  } Perturbatively, when taking $G_N \to 0$, such states would result in an infinite dimensional Hilbert space, even if we focused on a finite energy window, that would never factorise into a product of left and right Hilbert spaces.  Wormhole corrections, however, make the states $\ket{q_i}$ slightly non-orthogonal. Instead of an infinite-dimensional Hilbert space, the non-perturbative overlaps make it finite-dimensional when focusing on states within any finite-size energy window. Even with a finite Hilbert space, the puzzle of Hilbert space factorisation remains.\footnote{For example, in pure gravity, where the gauge constraint for two-sided black holes $H_L = H_R$ is valid, wormhole corrections result in a finite-dimensional Hilbert space; nevertheless, because of the gauge constraint, the bulk Hilbert space still does not factorise.   }  To address this puzzle, we compute the trace in \eqref{eq:kL-kR-factorisation} using the statistics of inner-product between the states in the  $\ket{q_i}$ basis. When $k_L$ and $k_R$ are arbitrary functions of the left and right Hamiltonians, such as the Euclidean time evolution operators  $k_L = e^{-\beta_L H_L}$ and $k_R = e^{-\beta_R H_R}$ with arbitrary temperatures, we show that 
\be 
{\Tr_{{\mHb}}(e^{-\beta_L H_L}\, e^{-\beta_R H_R})}= \begin{cases}\Tr_{\mathcal{H}_L}(e^{-\beta_L H_L})\Tr_{\mathcal{H}_R}(e^{-\beta_R H_R})\,, \quad \forall\, \beta_L,\,\beta_R\,, \quad \text{ when } K \geq d^2,\\ 
\text{Complicated non-factorising trace,} \hspace{1.80cm} \text{ when } K < d^2\,,
\end{cases}
\label{eq:Page-curve-factorisation}
\ee
where $d=\int_{\mathcal E} \dd E \rho(E)$ can be identified as the size of the one-sided Hilbert space in a chosen finite size energy window $\mathcal E$ given in terms of the coarse-grained density of states $\rho(E)$.\footnote{In the limit $G_N\to 0$, because $d\sim O\left(e^{\#/G_N}\right)$ we need to consider the double-scaling limit $d\to \infty$, $K \to \infty$ and $K/d^2\sim O(1)$.} Thus, precisely when the states $\ket{q_i}$ span the entire Hilbert space in a finite energy window  ($K\geq d^2$ and $\mathcal H_\text{bulk} = \mHb$), the trace factorises and consequently $\mathcal H_\text{bulk}$ also factorises; in the absence of additional boundary global symmetries that would lead to a degeneracy of states with the same energy, which, in turn, implies that the non-perturbative Hilbert space factorises into a tensor product once accounting for non-perturbative effects. Similarly, when the states $\ket{q_i}$ do not span the Hilbert space ($K<d^2$ and $\mathcal H_\text{bulk} \neq \mHb$), our  computation of ${\Tr_{{\mHb}}(e^{-\beta_L H_L}\, e^{-\beta_R H_R})}$ reveals an explicit non-factorising answer. To further clarify how Hilbert space factorisation arises from the gravitational path integral, in \textbf{section \ref{sec:resolvent_as_operator}}, we provide a further geometric derivation for the factorisation of the trace.

The result~\eqref{eq:Page-curve-factorisation} has precisely the expected behavior for the trace of two conventional quantum systems: when considering the span of only a few highly entangled states,\footnote{The states $\ket{q_i}$ are almost equally entangled as the thermofield double.} one should not obtain a factorising trace, but, once sufficiently many states are put together to span the entire Hilbert space factorisation is necessary. Thus,~\eqref{eq:Page-curve-factorisation} can be viewed as an analog of the Page curve \cite{Almheiri:2019psf, Penington:2019npb, Penington:2019kki,Almheiri:2020cfm} for Hilbert space factorisation,  both necessary for gravitational theories to be described by conventional quantum systems. However, both the Page curve and the factorisation result that we shall derive in section \ref{sec:fact-leading-order} suffer from an additional subtlety: in both cases the path integral computations only capture a coarse-grained answer and the result \eqref{eq:Page-curve-factorisation}, as well as the entanglement entropy of the Page curve, are only valid to leading order in $1/G_N$.

To complete our proof of Hilbert space factorisation, we thus wish to derive that the trace exactly factorises for sufficiently large $K$. We accomplish this in \textbf{section \ref{sec:fact-to-all-orders}} where we show that factorisation holds to all non-perturbative orders in $1/G_N$. We proceed with this proof in three steps, focusing on the large $K$ limit ($K \gg d^2$). First, we show that when including $1/G_N$ corrections the average bulk trace precisely matches the average product of traces,
\be 
\label{eq:average-factorisation}
\textbf{Step 1}: \quad \overline{\Zb(\beta_L, \beta_R)} = \overline{\Tr_{{\mHb}}(e^{-\beta_L H_L}\, e^{-\beta_R H_R})}= \overline{\Tr_{\mathcal{H}_L}(e^{-\beta_L H_L})\Tr_{\mathcal{H}_R}(e^{-\beta_R H_R})}\,, \quad \forall\, \beta_L,\,\beta_R\,,
\ee 
where the right-hand side is obtained from the spectral statistics of energy levels. Because of the non-trivial statistics of energy levels, $\overline{\Tr_{\mathcal{H}_L}(e^{-\beta_L H_L})\Tr_{\mathcal{H}_R}(e^{-\beta_R H_R})}$ is no longer a factorised product of a function of $\beta_L$ and one of $\beta_R$.  While this is to be expected, this instructs us to find a tool that probes the factorisation of the trace more directly. For this we define the differential, 
\be 
d(\beta_L, \beta_R) = \Zb(\beta_L, \beta_R)\partial_{\beta_L}\partial_{\beta_R}\Zb(\beta_L, \beta_R)-\partial_{\beta_L}\Zb(\beta_L, \beta_R)\partial_{\beta_R}\Zb(\beta_L, \beta_R)\,.
\label{eq:differential-equation-intro}
\ee
This differential vanishes if and only if the trace factorises into a product of a function of $\beta_L$ and one of $\beta_R$. Thus, we can show that the trace exactly factorises by analyzing whether $d(\beta_L, \beta_R)$ vanishes to all non-pertubative orders in $1/G_N$. We will first prove that the average of the differential exactly vanishes to all non-pertubative orders in $1/G_N$,
\begin{align}
\label{eq:diff-equations-vanish-1}
\hspace{-7cm} \textbf{Step 2}: \hspace{6cm} \quad \overline{d(\beta_L, \beta_R)} = 0 \,.
\end{align}
Even more surprisingly, we will show that the average of the square of the differential equation also vanishes to all non-pertubative orders in $1/G_N$,
\be
\label{eq:diff-equations-vanish}
\hspace{-6.6cm} \textbf{Step 3}: \hspace{5.8cm}\quad \overline{\left(d(\beta_L, \beta_R)\right)^2} = 0\,.
\ee
Because both the average differential and its square are vanishing, the trace exactly factorises, taking our results beyond the coarse-grained level. Consequently, this completes our proof of the factorisation of the Hilbert space. Even though the right-hand side of \eqref{eq:average-factorisation} contains connected contributions to the energy level statistics of the left and right sides due to the spacetime wormholes, \eqref{eq:diff-equations-vanish-1} and \eqref{eq:diff-equations-vanish} are independent of these statistics thus allowing us to resolve the puzzle of Hilbert space factorisation without addressing multi-boundary factorization.

Three properties reveal the robustness of our results. Firstly, both the fact that the trace eventually factorises as well as the moment when factorisation is achieved are independent of the construction of the states $\ket{q_i}$. For example, our proof is independent of the scaling dimensions of the matter operators used in the construction of the states and independent of the period of Euclidean evolution to the left and right of the operator insertion. We show that not only the factorisation of the Hilbert space is independent of these parameters, but, as one might expect, the dimension of the Hilbert space itself is also independent.\footnote{This allows one to drop the technical assumption of taking all scaling dimensions equal, sometimes made in past computations of this dimension \cite{Hsin:2020mfa, Boruch:2023trc}.  } Secondly, the wormholes that compute the trace in \eqref{eq:Page-curve-factorisation} to leading non-perturbative order in $1/G_N$ are all genuine saddles of the gravitational path integral when the matter operators are sufficiently heavy. Because such saddles can be generalized to any number of spacetime dimensions, the derivation of Hilbert space factorisation follows suit. Finally, the exact cancellation of the differentials  \eqref{eq:diff-equations-vanish-1} and \eqref{eq:diff-equations-vanish} non-perturbatively in $1/G_N$ relies on generic properties of spectral correlators that we expect should be valid regardless of the exact gravitational theory that we study.

To emphasize that factorisation can be probed from the gravitational integral more broadly, in \textbf{section \ref{sec:factorisation-BPS-BHs}} we show that the Hilbert space of two-sided BPS black holes also factorises due to non-perturbative wormhole corrections. In this case, the factorisation of the trace cannot be probed by taking $k_L = e^{-\beta_L H_L}$ and $k_R = e^{-\beta_R H_R}$ since all BPS states in a given charge sector have the same energy. Instead, we shall construct $k_L = e^{-\alpha_L K_L}$ and $k_R= e^{-\alpha_R K_R}$ by using two matter operators that are projected to the BPS sector; using these operators, we will prove \eqref{eq:kL-kR-factorisation} and the analogous steps 1--3 also apply to the BPS sector.

 Finally, in \textbf{section \ref{sec:discussion}}, we discuss several consequences and interpretations of our results.

\section{Conventions and review}
\label{sec:conventions-and-review}

In this section we introduce the conventions and objects used subsequently in the paper. The main theory in which we study factorisation in this paper is that of a two-dimensional dilaton gravity, called JT gravity \cite{Teitelboim, Jackiw,Henneaux:1985nw,Louis-Martinez:1993bge, AlmheiriPolchinski}, coupled to arbitrary probe matter, which is described by the action \cite{Penington:2023dql,Kolchmeyer:2023gwa,Iliesiu:2024cnh,Yang:2018gdb,Maldacena:2016upp,Kitaev:2017awl,Kitaev:2018wpr,Saad:2019pqd,Mertens:2022irh,Jafferis:2022wez}
\be 
I=-S_0 \chi(\mathcal{M})  -\frac{1}{2}\left(\int_\mathcal{M} \phi(R+2)+2\int_{\partial \mathcal{M}}\phi_b (K-1)\right) + I_{\text{matter}} 
.
\ee

We will first focus on the pure gravitational theory neglecting all matter interactions. The theory can be canonically quantized perturbatively on a disk topology and the resulting Hamiltonian describes a simple quantum mechanical system of a single degree of freedom $\ell$, governed by Hamiltonian 
\be 
H = - \frac{1}{2} \partial^2_\ell + 2 e^{-\ell} .
\ee
This Hamiltonian evolves the states of the left and right asymptotic boundary through $H = H_L$ and $H= H_R$, in pure JT gravity subject to the condition $H_L = H_R$.
From the bulk perspective, $\ell$ captures simply the geodesic length through the bulk between the two asymptotic spatial boundaries. 
The pure gravity perturbative Hilbert space is now simply given by normalizable wavefunctions $\psi(\ell)$ of the geodesic length $\ket{\ell}$, with the standard inner product
\be 
\mH_{\text{pure gravity}} = L^2(\mathbb R)\,, 
\qquad 
\braket{\psi} = \int \dd \ell |\psi(\ell)|^2 ,
\ee
with the geodesic length states normalized as $\braket{\ell}{\ell'} = \delta(\ell - \ell')$.
The energy eigenstates $\ket{E}$ of the bulk Hamiltonian $H$ in the length basis take the form 
\be 
\ket{E} = \int_{-\infty}^\infty \dd \ell \braket{\ell}{E}
,
\qquad
\braket{\ell}{E}=2^{3/2}K_{2is}(4e^{-\ell/2})  , \qquad \text{ where }\quad  E\equiv s^2 / 2\,,
\ee
and the states are normalized as 
\be 
\braket{E}{E'} = 
\int_{-\infty}^\infty d\ell\,\braket{E}{\ell}\braket{\ell}{E'}=\frac{\delta(s-s')}{\rho(s)} , \qquad \rho(s)=\frac{s}{2\pi^2}\sinh(2\pi s)\,,
\ee 
where $e^{S_0} \rho(s)$ is the density of states in the pure gravitational theory. To be able to quickly identify the order at which each geometry contributes, we will always separate factors of $e^{S_0}$ from $\rho(s)$. It will also oftentimes be convenient to work in terms of the variables $s$ and $s'$ instead of the actual energies, but we will oftentimes revert to the energy notation for our final results. We will also oftentimes use the shorthand notation $\rho(E) \equiv \rho(s(E))$ when writing the density of states in terms of the energy. 
We also introduce the rescaled wavefunction
\be 
\label{eq:phi-E-l}
\tvarphi_s (\ell)=e^{-\ell/2}\braket{\ell}{E} ,
\ee
which is known as the fixed energy Hartle-Hawking wavefunction. The fixed temperature Hartle-Hawking wavefunction of temperature $\beta$ can be now further defined through 
\be
\bra{\ell} \ket{HH}=\varphi_{\beta} (\ell) =  \int ds\, \rho(s) e^{-\beta s^2/2} \tvarphi_s (\ell) .
\ee
In terms of the path integral, this wavefunction can be prepared by performing a a gravitational path integral on a half-disk geometry with asymptotic geodesic boundary of $\beta$ and a bulk geodesic boundary of length $\ell$, where the fixed-temperature Hartle-Hawking state can be represented pictorially as 
\be
\ket{HH}=\includegraphics[valign=c,width=0.2\textwidth]{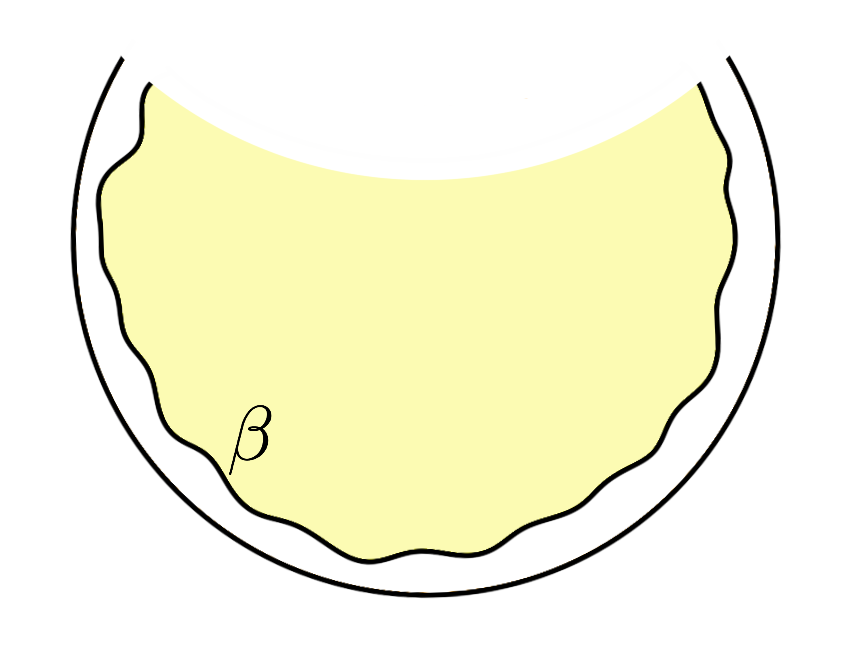}\, 
.
\ee

Moving on to the case with matter, the perturbative Hilbert space becomes simply a tensor product 
\be
\label{eq:perturbative-Hilbert-space-with-matter}
\mH_{\text{bulk}}=\mH_{\text{matter}}\otimes \mH_{\text{pure gravity}} .
\ee
Since left and right operators act on both $\mH_{\text{matter}}$ and $\mH_{\text{pure gravity}} $ the Hilbert space \eqref{eq:perturbative-Hilbert-space-with-matter} is nowhere close to factorising into a left and right tensor product. Nevertheless, it is first useful to understand this Hilbert space at the perturbative level. 
The states of the matter fields can be described by how they transform under AdS$_2$ isometries \cite{Kitaev:2017hnr,Lin:2019qwu}. We decompose $\mH_{\text{matter}}$ into representations of $\widetilde{SL}(2,\mathbb{R})$ symmetry group 
\be
\mH_{\text{matter}}=\mathbb{C}\oplus\bigoplus_\Delta\mathcal{D}^+_\Delta ,
\ee
where $\mathbb{C}$ is the trivial representation of $\mathfrak{sl}(2,\mathbb{R})$, and $\mathcal{D}^+_\Delta$ denotes the discrete series of $\mathfrak{sl}(2,\mathbb{R})$ of weight $\Delta$ and each of them carries an orthonormal basis labeled by $m$. The states of the total Hilbert space are now labeled by 
\be
\ket{\Delta;\ell,m} ,
\qquad
\bra{\Delta';\ell',m}\ket{\Delta;\ell,m}=\delta_{\Delta\Delta'}\delta_{mm'}\delta(\ell-\ell') .
\ee
It turns out that $H_L$ and $H_R$ form a complete set of commuting operators with positive energies, so we can make a change of basis to $\ket{\Delta;E_L,E_R}$:\footnote{This step is explicitly shown in Appendix B of \cite{Kolchmeyer:2023gwa}. Similar basis changes were also discussed in \cite{Jafferis:2022wez} and \cite{Iliesiu:2024cnh}.}
\be
\label{eq:Delta-EL-ER}
\ket{\Delta;E_L,E_R}=\sum_m\int \dd \ell\,\phi^{\Delta}_{E_L,E_R}(\ell,m)\ket{\Delta;\ell,m} ,
\ee
where $\phi^{\Delta}_{E_L,E_R}(\ell,m)$ is a change of basis matrix analogous to $\braket{\ell}{s}$ \cite{Iliesiu:2024cnh}. $\ket{\Delta;E_L,E_R}$ are normalized by
\begin{align}
\bra{\Delta';E'_L,E'_R}
\ket{\Delta;E_L,E_R}
&=
\sum_m\int \dd \ell \,
\phi^\Delta_{E_L,E_R}(\ell,m)\phi^\Delta_{E'_L,E'_R}(\ell,m)\\
&=\frac{\delta(E_L-E_L')}{\rho(E_L)} \frac{\delta(E_R-E_R')}{\rho(E_R)}\gamma_\Delta(E_L,E_R)
\end{align}
where $\gamma_\Delta (E_L,E_R)$ is a normalization factor. With this we can further define the states of fixed asymptotic boundary length $\tilde \beta_L + \tilde \beta_R$ 
\be 
\ket{q_\Delta} 
\equiv 
\ket{\Delta ; \tilde \beta_L , \tilde \beta_R} = 
\int \dd E_L \dd E_R \rho(E_L) \rho(E_R) 
e^{-\tilde \beta_L E_L - \tilde \beta_R E_R} 
\ket{\Delta ; E_L , E_R} , 
\ee
which can also be prepared through a gravitational path integral by inserting an operator $O_i$ on the asymptotic boundary with left Euclidean boundary segment $\tilde \beta_L$ and right Euclidean boundary segment $\tilde \beta_R$ 
\be 
\ket{q_\Delta}= O_\Delta \ket{HH}=\includegraphics[valign=c,width=0.2\textwidth]{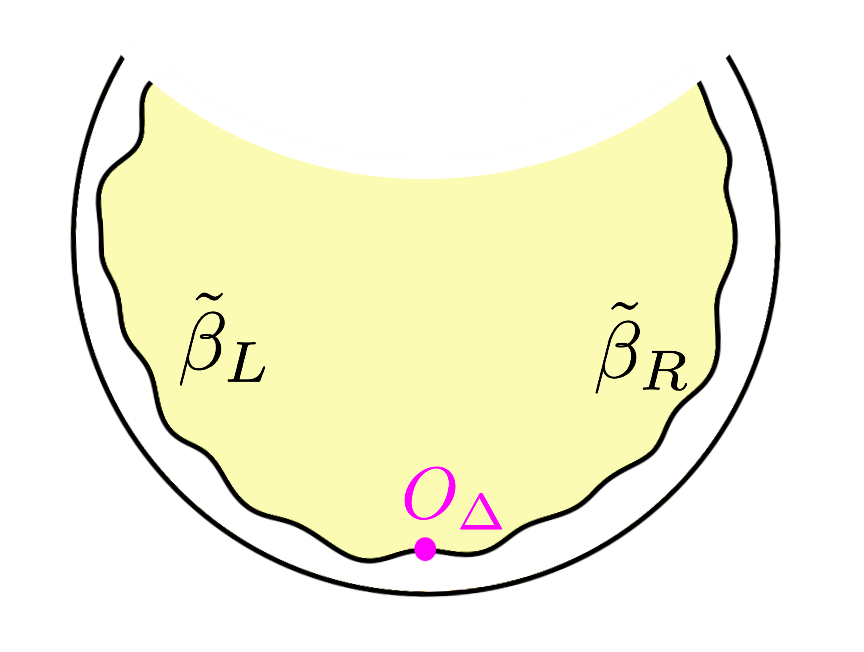}\,.
\label{eq:state-qi}
\ee
 Using these states we can calculate two-point functions. To agree with perturbative JT computation we need to set the normalization factor
\be
\gamma_\Delta(E_L,E_R)=\frac{\Gamma(\Delta\pm i\sqrt{2E_L}\pm i\sqrt{2E_R})}{2^{2\Delta-1}\Gamma(2\Delta)} .
\ee
This two-point function can be computed through a gravitational path integral at the perturbative level as
\be
\bra{q_\Delta}\ket{q_\Delta}_\text{pert} =\includegraphics[valign=c,width=0.12\textwidth]{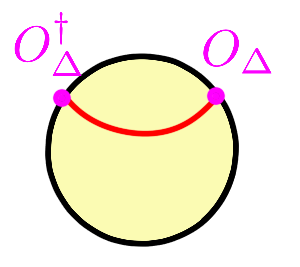} 
.
\ee
Consider a family of such states that are orthogonal at the perturbative level: for instance, they can all lie within different $SL(2, \mathbb R)$ representations, but for computational convenience, with close but slightly different values of $\Delta$. For simplicity, we also take these states to have $\tilde \beta_L = \tilde \beta_R = \beta/4$.\footnote{In section \ref{sec:checking-factorisation-is-basis-indep} and appendix \ref{sec:basis-prepared-with-different-dim}, we will relax both these assumptions and show that all derived results remain unchanged. } These will serve as a basis  for the construction of the bulk Hilbert space, 
\be 
\label{eq:Hilbert-space-construction-explicit}
\mHb \equiv \text{Span }\{ \ket{q_i} = O_i \ket{HH}, \, i=1,\,\dots,\,K\}\,.
\ee
At the perturbative level, two states in this family $\ket{q_i}$ and $\ket{q_j}$ of different weights $\Delta_{i/j}$ are orthogonal to each other: $\overline{\braket{q_i}{q_j}} \propto \delta_{ij}$. This, however, stops being true once we move on to the nonperturbative inner product of JT gravity, which allows for geometries of higher topology to contribute to the gravitational path integral. Denoting the nonperturbative inner product by $\overline{\langle \cdot \rangle }$ we find that \cite{Stanford:2020wkf,Hsin:2020mfa}
\be
\overline{|\braket{q_i}{q_j}|^2} = \includegraphics[valign=c,width=0.23\textwidth]{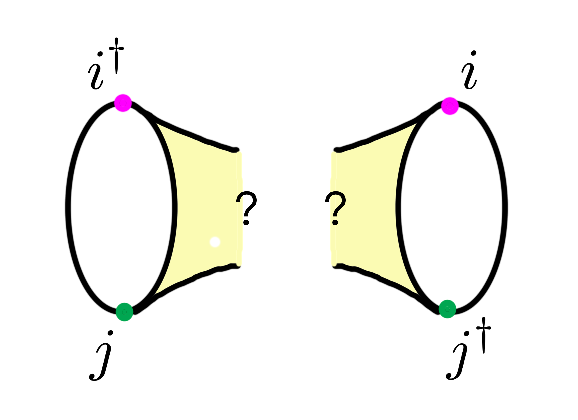}=\underbrace{\includegraphics[valign=c,width=0.25\textwidth]{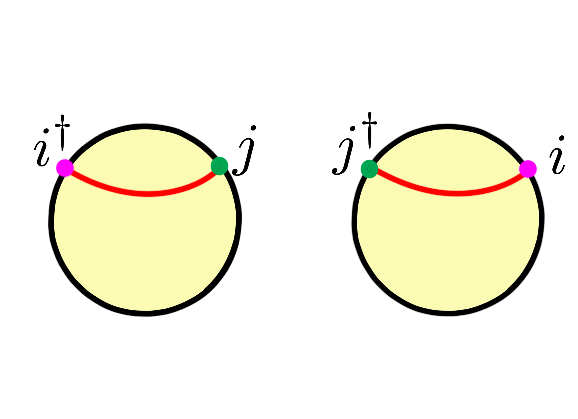}}_{0 \, \text{ for } i\neq j}+\includegraphics[valign=c,width=0.23\textwidth]{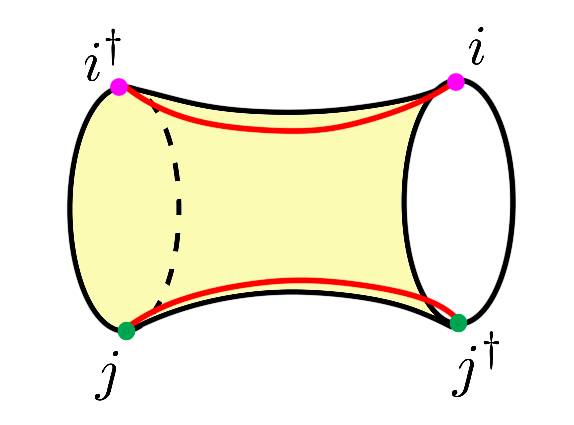}
+\cdots\,,
\label{eq:geometries-in-inner-product-squared}
\ee
and so we see that states orthogonal at the disk level can have nonperturbatively small overlaps once we take into account wormhole contributions. Note, however, that we still have $\overline{\braket{q_i}{q_j}} \propto \delta_{ij}$. The nonperturbative computation does not therefore define a good notion of an inner product, and should be thought of instead as capturing the coarse-grained statistics of the overlaps in some underlying quantum gravitational Hilbert space $\mH_{\text{bulk}}$. From the boundary side, even in a theory that is not ensemble-averaged, one can compute the statistics of the inner products between states constructed as in \eqref{eq:state-qi} and, a coarse-graining of these statistics is expected to reproduce \eqref{eq:geometries-in-inner-product-squared}.

Knowing the statistics of these overlaps instead of their exact value is still extremely powerful. The nonperturbative overlaps allow one to compute the dimension of the bulk Hilbert space from the gravitational path integral \cite{Hsin:2020mfa,Balasubramanian:2022gmo,Balasubramanian:2022lnw,Boruch:2023trc}. To do that, one considers a maximally mixed density matrix of the form 
\be 
\rho \propto \sum_{i=1}^K \ket{q_i} \bra{q_i}\,,
\ee
formed out of $K$ states $\ket{q_i}_{i=1\dots K}$ which are orthogonal at the disk level $\overline{\braket{q_i}{q_j}} \propto \delta_{ij}$. The rank can then be simply computed by analytically continuing the trace of the $n$-th power of the density matrix as
\be
\overline{\text{rank}(\rho)} = \overline{\dim\, \mHb} =   \lim_{n \rightarrow 0}
\sum_{i_1, \dots , i_n=1}^{K} 
\overline{
\braket{q_{i_1}}{q_{i_2}}\dots \braket{q_{i_n}}{q_{i_1}}
} .
\ee
To resum the possible geometries contributing to the rank one uses the resolvent method which we introduce in the next section and heavily use throughout this paper. An important ingredient in the computation of the resolvent will be the leading fully connected geometry contributing to $\overline{\braket{q_1}{q_2}\dots \braket{q_n}{q_1}}$, called the pinwheel geometry. This can be computed through the gravitational path integral with $n$-boundaries and two operator insertions on each boundary. These are given by
\be
Z_n^O =
\includegraphics[valign=c,width=0.25\textwidth]{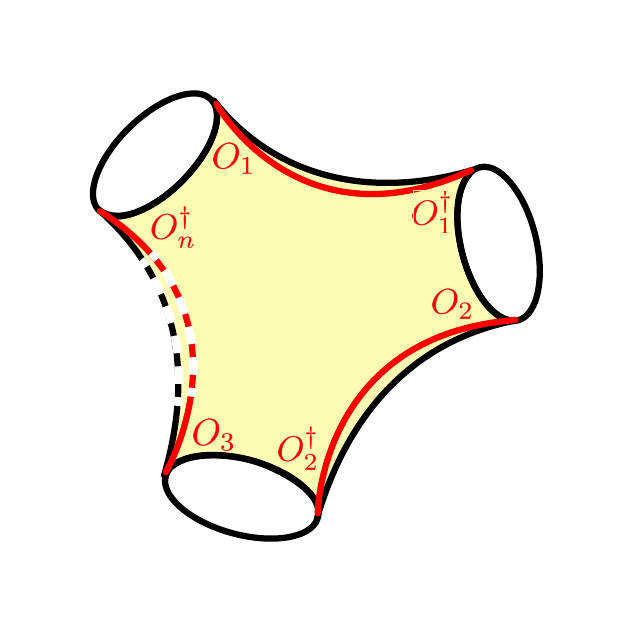}=e^{2S_0} \int ds_L ds_R\, \rho(s_L) \rho(s_R) 
y_q^n ,
\label{eq:n_boundary_pinwheel}
\ee
where 
\be 
y_q(s_L, s_R) = e^{-S_0} e^{-\beta s_L^2/4} 
e^{-\beta s_R^2/4} 
\gamma_\Delta (s_L, s_R) .
\label{eq:y_q_equation}
\ee
As with the density of states, we will interchangeably refer to these functions either in terms of the $s$ variable or directly in terms of the energies. By resuming all these geometries, the rank can be found to be
\be
\overline{\mathrm{rank}(\rho)}=\begin{cases}K&K<d^2\\d^2 &K\geq d^2\end{cases}\,.
\label{eq:rank-summary} 
\ee
This allows us to deduce the dimension of the Hilbert space as being $d^2$, where $d =\int_{\mathcal{E}} \dd E \rho(E) $.

Lastly, since we'll make use of similar techniques when computing the bulk trace, let us introduce a simple generalization of the pinwheel which we'll make use of below. 
This is a pinwheel with two additional operator insertions $k_L=e^{-\beta_L H_L}$ and $k_R=e^{-\beta_R H_R}$ at the ``top" boundary. Explicitly it gives the result
\be
Z_{n+1}^O(k_L,k_R) =
\includegraphics[valign=c,width=0.25\textwidth]{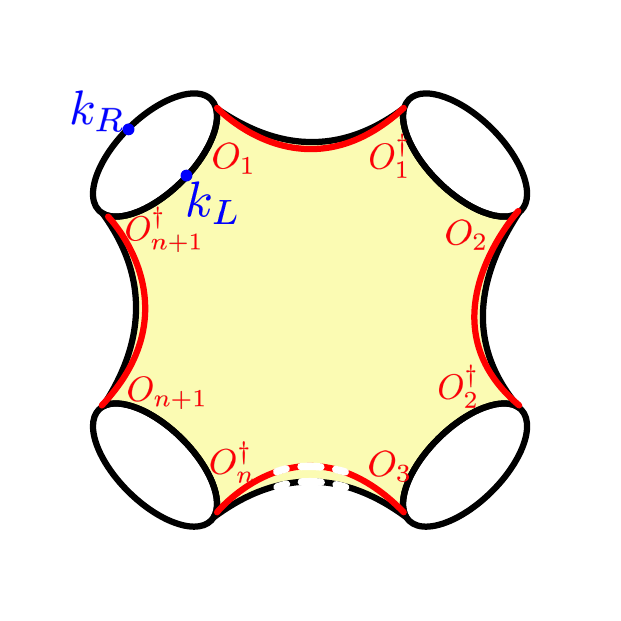}=e^{2S_0} \int ds_L ds_R\, \rho(s_L) \rho(s_R) 
y_q^n y_k
\label{eq:n+1_boundary_pinwheel}
\ee
with
\be
y_k(s_L, s_R) = e^{-S_0} e^{-(\beta/2 + \beta_L) s_L^2/2} 
e^{-(\beta/2+\beta_R) s_R^2/2} 
\gamma_\Delta (s_L, s_R) 
.
\label{eq:y_k_equation}
\ee
These will play an important role below in evaluating the bulk trace by using the resolvent method.

\section{Factorisation to leading order in $e^{1/G_N}$}
\label{sec:fact-leading-order}

In this section, we resolve the factorisation puzzle to leading order in $K$ and $e^{1/G_N}$(or equivalently, the leading order in $e^{2S_0}$). 
More precisely, we work in regime 
\be
K\rightarrow\infty, \qquad e^{2S_0}\to \infty, \quad\quad d^2=e^{2S_0}\int_{E_L, E_R \in \mathcal E} \rho(E_L)\rho(E_R)\dd E_L \dd E_R \rightarrow\infty
,
\quad\quad \frac{K}{d^2}\sim O(1)\,,
\label{eq:K-d2-limit-section3}
\ee
where $\mathcal E$ is an arbitrary energy window which we will simply use to truncate the overall dimension of the Hilbert space. To start, in section \ref{subsec:resolvent-and-bulk-trace} we will introduce the resolvent method and show that the bulk trace can be expressed in terms of the resolvent. In section \ref{sec:physical-mechanism-of-factorisation}, we will use this  expression to explicitly link the factorisation of the trace, the overcompleteness of the basis of states used in our construction of $\mHb$ and the analytic structure of the resolvent. In section \ref{sec:null-states-and-fact-for-K>d^2}, by providing a detailed analysis of the analytic structure of the resolvent we will prove that the trace factorises when $K>d^2$ and will explicitly find a non-factorising answer when $K<d^2$.

\subsection{The resolvent and bulk trace}
\label{subsec:resolvent-and-bulk-trace}

To study the factorisation of the Hilbert space, we want to study the statistics of the bulk trace 
\be 
\Tr_{\mHb}(k_L \, k_R)\,.
\ee
Up to section \ref{sec:fact-to-all-orders}, we will focus on operators 
\be 
k_L \equiv e^{-\beta_L H_L},\qquad k_R\equiv e^{-\beta_R H_R}\,,
\ee 
obtained from Euclidean time evolutions on the left and right sides of the black hole.
In terms of the basis of states used in the construction of the Hilbert space $\mHb$ in \eqref{eq:Hilbert-space-construction}, the bulk trace can be defined by 
\be 
\label{eq:Hbulk-explicit}
\Tr_{\mHb} \left(k_L \,k_R\right) = 
(M^{-1})_{ij} \bra{q_i} e^{-\beta_L H_L} e^{-\beta_R H_R} \ket{q_j} ,
\ee
where we introduced the overlap matrix of the basis vectors $\ket{q_i}$
\be 
\label{eq:matrix-M-defintion}
M_{ij} \equiv \braket{q_i}{q_j} \,.
\ee 
In  \eqref{eq:Hbulk-explicit}, $M^{-1}$ is a regular inverse if the matrix $M$ is invertible, i.e.~when the states $\ket{q_i}$ are not linearly dependent, and is a generalized inverse when the matrix $M$ is non-invertible,  i.e.~when the states $\ket{q_i}$ become linearly dependent.\footnote{The generalized inverse is taken by inverting the non-zero eigenvalues and leaving the zero eigenvalues unchanged.}

To study this trace from the bulk perspective, we will want to compute its statistics through the gravitational path integral. Denoting the statistics computed from the bulk with an overline, we can write the average of the trace as
\begin{align}
\label{eq:Hbulkdefinition}
\overline{\Tr_{\mHb}(k_L \,k_R)}
&= 
\lim_{n \rightarrow -1} 
\overline{(M^{n})_{ij} \bra{q_i} e^{-\beta_L H_L} e^{-\beta_R H_R} \ket{q_j}}\nn \\  &= \lim_{n \rightarrow -1}  \overline{\braket{q_i}{q_{i_1}} \dots  \braket{q_{i_{n-1}}}{q_{j}}  \bra{q_i} e^{-\beta_L H_L} e^{-\beta_R H_R} \ket{q_j}}\,, 
\end{align}
where all the indices $i$, $j$, $i_1$, \dots, $i_{n-1} =1 , \dots K$ are being summed over. 
From the boundary dual perspective, the overline can be thought of as taking an ensemble average or an appropriate coarse-graining of a single theory. 
In the bulk, we compute the above trace by summing all possible geometries consistent with the boundary conditions specified by the inner-products on the right-hand side. Namely, we sum over geometries containing $n+1$ asymptotic boundaries, with two operator insertions on each of the boundaries, and an additional $e^{-\beta_L H_L}$, $e^{-\beta_R H_R}$ Euclidean time evolution on a single boundary. For example, one such geometry that will appear will be $n+1$ disconnected disks that give the perturbative value of the trace. Using the fact that the states $\ket{q_i}$ are orthogonal at the perturbative level, the contributions of such geometries to the trace gives
\begin{equation}
\label{eq:perturbative-result-to-contrast}
    \Tr_{\rm pert}(k_L \,k_R) = \sum_{i=1}^{K} \frac{\bra{q_i} e^{-\beta_L H_L} e^{-\beta_R H_R} \ket{q_i}_\text{pert}}{\langle q_i|q_i\rangle_\text{pert} } \propto  K \,,
\end{equation}
which gives the perturbative expression for the bulk trace in gravity and never gives a factorising answer. However, as we have already hinted in section \ref{sec:conventions-and-review}, non-perturbatively there are many geometries contributing to \eqref{eq:Hbulkdefinition}. One example is the pinwheel shown in \eqref{eq:n+1_boundary_pinwheel}, but there will be many more disconnected geometries appearing. In the limit \eqref{eq:K-d2-limit-section3}, such geometries will consist of all possible pinwheels and disks for which if one cuts the geometry along the matter geodesic the loops obtained in the summation over matter indices form a planar graph;  all other geometries that could contribute to \eqref{eq:K-d2-limit-section3} are suppressed either in $e^{-S_0}$, due to the geometry having lower Euler characteristic, or in $K$, due to the matter index loops forming a non-planar graph. To perform the analytic continuation $n \to -1$ we need to resum all such pinwheels and disks.

\begin{figure}[t!]
     \centering
     \begin{subfigure}[b]{0.3\textwidth}
         \centering
         \includegraphics[width=\textwidth]{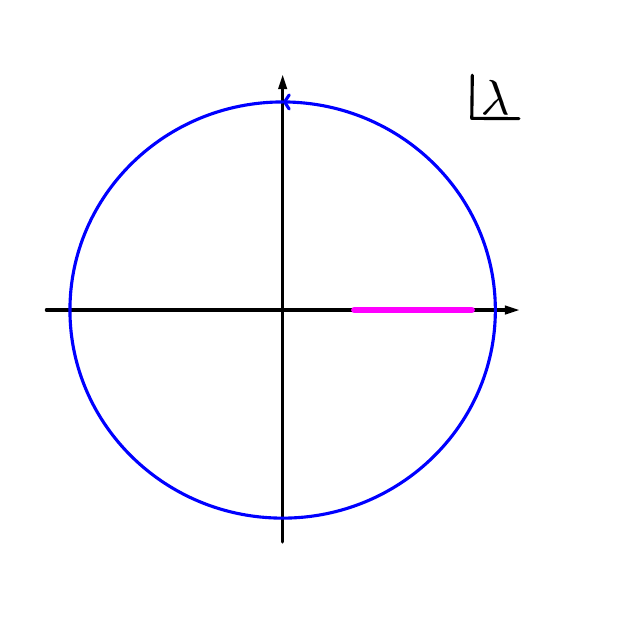}
         \caption{$C_0$}
         \label{fig:C0}
     \end{subfigure}
     \hfill
     \begin{subfigure}[b]{0.3\textwidth}
         \centering
         \includegraphics[width=\textwidth]{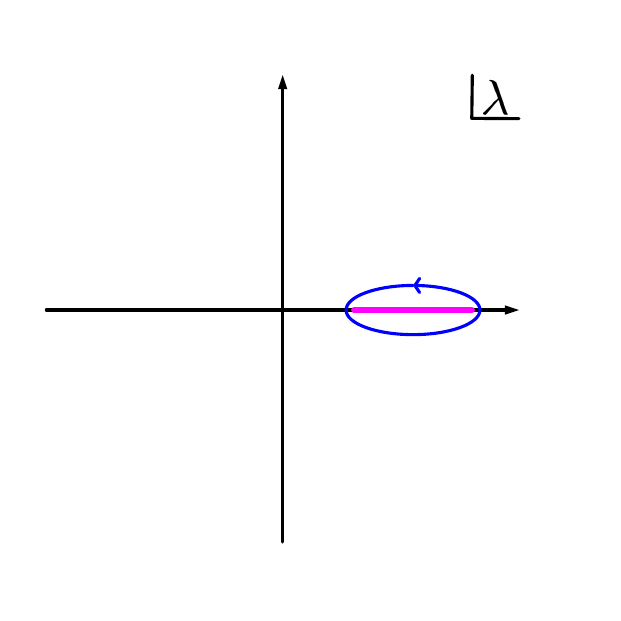}
         \caption{$C_1$}
         \label{fig:C1}
     \end{subfigure}
     \hfill
     \begin{subfigure}[b]{0.3\textwidth}
         \centering
         \includegraphics[width=\textwidth]{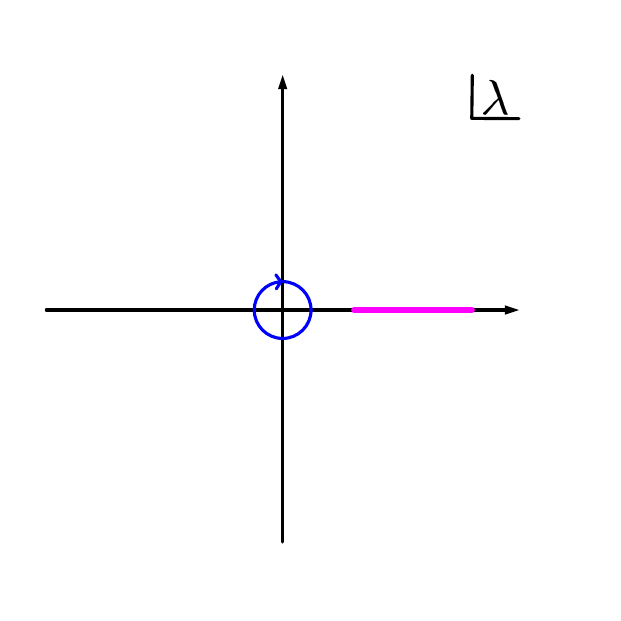}
         \caption{$C_2$}
         \label{fig:C2}
     \end{subfigure}
        \caption{The contour choice for $\text{Tr}_{\mHb}$. The resolvent is defined as a convergent sum in \eqref{eq:resolventdef} for $\lambda\rightarrow\infty$, so we start from the contour $C_0$ counterclockwise at very large $\lambda$. The resolvent has a branch cut which we show in pink. For $n$ a non-negative integer, we can deform the contour in \eqref{eq:pretraceHbulk} from $C_0$ to $C_1$ to just include the cut. This requires no pole at $\lambda=0$, which is easy to see for $n>0$. The $n=0$ case is a bit peculiar in the sense that $\mathbf{R}_{ij}(\lambda)$ itself has a pole at $\lambda=0$ but the average $\overline{\mathbf{R}_{ij} \langle q_i|\cdots |q_j\rangle}$ does not.   Then we can keep the $C_1$ contour and analytically continue to $n\rightarrow-1$ and deform the contour again from $C_1$ to $C_2$. We then finally arrive at a contour just encircling $\lambda=0$ but clockwise in \eqref{eq:leadingTrHbulk}. }
        \label{fig:contours}
\end{figure}

To practically resum all the possible geometries that contribute to the above quantity, we introduce the resolvent 
\be\label{eq:resolventdef}
\mathbf{R}_{ij}(\lambda)  = \left( 
\frac{1}{\lambda  - M}
\right)_{ij} 
= \frac{\delta_{ij}}{\lambda} + \frac{1}{\lambda} \sum_{n=1}^\infty \frac{(M^n)_{ij}}{\lambda^n}
,
\ee
in terms of which we can write \eqref{eq:Hbulkdefinition} as
\be\label{eq:pretraceHbulk}
\overline{(M^{n})_{ij} \bra{q_i} e^{-\beta_L H_L} e^{-\beta_R H_R} \ket{q_j}} = 
\frac{1}{2\pi i} \oint_{C_0} \dd \lambda\, \lambda^n \,
\overline{\mathbf{R}_{ij} (\lambda)
\bra{q_i} e^{-\beta_L H_L} e^{-\beta_R H_R} \ket{q_j}}
,
\ee
with $C_0$ denoting a defining contour shown in figure \ref{fig:C0}. 

The geometric expansion for the resolvent $\overline{\mathbf{R}_{ij}(\lambda)}$ computed from gravity can be schematically expressed as \cite{Penington:2019kki, Hsin:2020mfa, Balasubramanian:2022lnw, Balasubramanian:2022gmo}
\begin{align}
\includegraphics[valign=c,width=0.15\textwidth]{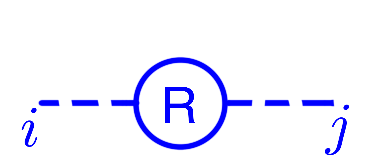}&=\includegraphics[valign=c,width=0.15\textwidth]{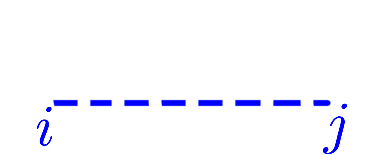} +\includegraphics[valign=c,width=0.13\textwidth]{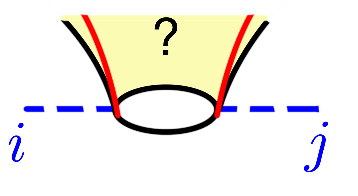} +\includegraphics[valign=c,width=0.2\textwidth]{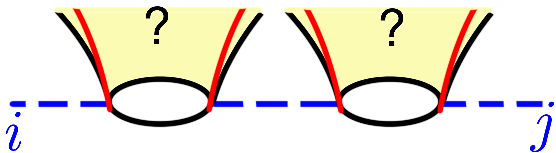}+\cdots\\
&=\includegraphics[valign=c,width=0.15\textwidth]{sd1m.png} +\includegraphics[valign=c,width=0.15\textwidth]{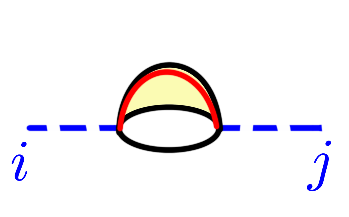}\nonumber\\
&\quad\quad+\includegraphics[valign=c,width=0.25\textwidth]{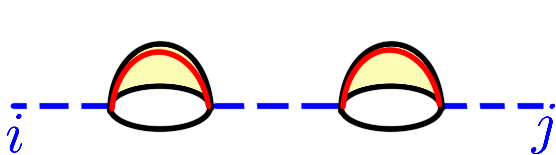}+\includegraphics[valign=c,width=0.25\textwidth]{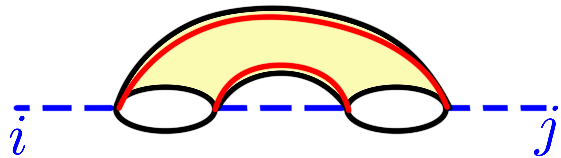}+\cdots
\label{eq:geometries-that-contribute}
\end{align}

We now compute the resolvent from the Schwinger-Dyson (SD) equations \cite{Penington:2019kki}. One organizes the infinite expansion in terms of how many boundaries the first boundary is connected to. Then, between each of the connected boundaries one introduces a single power of the resolvent. This will take into account all the geometries relevant in the limit \eqref{eq:K-d2-limit-section3} we're working with. The resulting consistency equation pictorially takes the form 
\begin{multline}\label{eq:DSpicture}
 \includegraphics[valign=c,width=0.15\textwidth]{sd0m.png}
=\includegraphics[valign=c,width=0.15\textwidth]{sd1m.png} +\includegraphics[valign=c,width=0.2\textwidth]{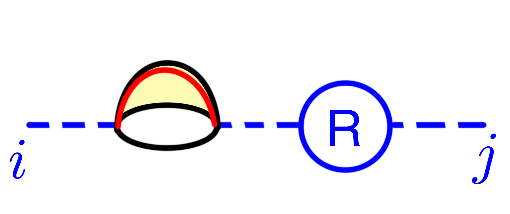}
+\includegraphics[valign=c,width=0.28\textwidth]{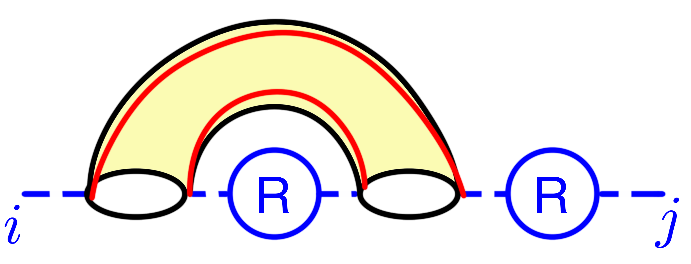}+\cdots   
\end{multline}
which can be written explicitly as 
\begin{align}
\overline{\mathbf{R}_{ij}(\lambda)} &= \frac{\delta_{ij}}{\lambda} + \frac{1}{\lambda} \sum_{n=1}^\infty Z_n^\mathcal{O} R^{n-1} \overline{\mathbf{R}_{ij} (\lambda)}\\
&= \frac{\delta_{ij}}{\lambda} + \frac{e^{2S_0}}{\lambda} \int ds_L ds_R\, \rho(s_L) \rho(s_R) 
\frac{y_q (s_L , s_R)}{1-R y_q (s_L , s_R)} \overline{\mathbf{R}_{ij} (\lambda)},
\label{eq:SD}
\end{align}
where we denoted the trace as $R (\lambda) \equiv \sum_i \overline{\mathbf{R}_{ii} (\lambda)}$, $Z_n^\mathcal{O}$ is the $n$-boundary pinwheel geometry introduced in \eqref{eq:n_boundary_pinwheel} and $y_q(s_L , s_R)$ is explicitly given through \eqref{eq:y_q_equation}.
From the Schwinger-Dyson equation \eqref{eq:SD} we can now extract two things: first, the trace of resolvent is given by
\be\label{eq:DSleadingorder}
R(\lambda)=\frac{K}{\lambda} + \frac{1}{\lambda} \sum_{n=1}^\infty Z_n^\mathcal{O} R^{n}=\frac{K}{\lambda} + 
\frac{e^{2S_0}}{\lambda} 
\int \dd s_L \dd s_R\, \rho(s_L) \rho(s_R) 
\frac{R(\lambda) y_q(s_L,s_R)}{1-R(\lambda) y_q(s_L,s_R)}\,.
\ee
The analytic structure of the resolvent turns out to be crucial for factorisation, and we discuss it in detail later in Sections~\ref{sec:nullstates} and \ref{sec:towardsfactorisation}.
Second, a slight modification of the Schwinger-Dyson equations allows us to also write down the expression for the bulk trace.
The contributing geometries are the same as ones in the resolvent except for the modification that one of the boundaries has now additional $(k_L,k_R)$ operator insertions. This simply modifies the pinwheel expression into $Z_{n+1}^O(k_L,k_R)$ defined in \eqref{eq:n+1_boundary_pinwheel} and we get
\begin{align}
\overline{\mathbf{R}_{ij} \bra{q_i} k_L k_R \ket{q_j}} 
&= \sum_{n=0}^\infty R^{n+1} Z_{n+1}^O(k_L,k_R)\label{pinwheelresum1}
\\ 
&= e^{2S_0} \int ds_L ds_R\, \rho(s_L) \rho(s_R) 
\frac{R(\lambda) y_k(s_L,s_R)}{1-R(\lambda) y_q(s_L,s_R)}\label{pinwheelresum2} ,
\end{align}
with $y_k(s_L,s_R)$ given by \eqref{eq:y_k_equation}.
From this we can extract 
\be\label{eq:replicatedbulktrace}
\overline{ (M^n)_{ij} \bra{q_i} k_L k_R \ket{q_j}} \, = 
e^{2S_0} 
\int ds_L ds_R \rho(s_L) \rho(s_R) 
 \oint_{C_1}
 \frac{\dd \lambda}{2\pi \ii}
 \lambda^n
\frac{R(\lambda) y_k(s_L,s_R)}{1-R(\lambda) y_q(s_L,s_R)} \,,
\ee
where we deformed the contour into $C_1$ which surrounds 
the branch cut of $R(\lambda)$ on the real $\lambda >0$ semi-axis as shown in figure \ref{fig:C1}. With this, finally, we can now analytically continue to $n=-1$ to obtain the final expression for the bulk trace
\begin{equation}\label{eq:leadingTrHbulk}
\begin{aligned}
\overline{\text{Tr}_{\mHb}(k_L k_R) }& =\, \overline{ \braket{q_i}{q_j}^{-1} \bra{q_i} k_L k_R \ket{q_j}} \, \\
& = 
e^{2S_0} 
\int ds_L ds_R \rho(s_L) \rho(s_R) 
 \oint_{C_2}
 \frac{\dd \lambda}{2\pi \ii}\,
 \frac{1}{\lambda}
\frac{R(\lambda) y_k(s_L,s_R)}{1-R(\lambda) y_q(s_L,s_R)} \,.
\end{aligned}
\end{equation}
Here in the last line, after analytically continuing to $n=-1$, we further deformed the contour to be around $\lambda=0$ as shown in figure \ref{fig:C2}. Having expressed $\text{Tr}_{\mHb}(k_L k_R) $ in terms of the resolvent, which in turn is determined by the Schwinger-Dyson equations, our goal now is to understand under what conditions \eqref{eq:leadingTrHbulk} factorises.

\subsection{Physical mechanism of factorisation}
\label{sec:physical-mechanism-of-factorisation}

In this section, we explain the physical mechanism of factorisation. 
As stated in the introduction, if the Hilbert space factorises $\mHb=\mathcal{H}_{L}\otimes \mathcal{H}_{R}$, the average trace should take the form
\be
    \overline{\Tr_{\mH_{\rm bulk} (K)}(k_Lk_R)}=\overline{\Tr_{\mH_L}(k_L)\Tr_{\mH_R}(k_R)}\,,
\ee
which, to the leading order of $e^{1/G_N}$, should imply 
\be
    \Tr_{\mHb}(k_Lk_R) = \Tr_{\mH_L}(k_L)  \Tr_{\mH_R}(k_R)\,,
\ee
and thus we will therefore drop the average above $\Tr_{\mHb}(k_Lk_R)$ for the remainder of this section. 
Upon inspecting the integral expression for $\Tr_{\mHb}(k_Lk_R)$  in \eqref{eq:leadingTrHbulk}, the integral above does not obviously factorise since the functions $y$ correlate $s_L$ and $s_R$.

We can see what is the condition for factorisation by explicitly evaluating \eqref{eq:leadingTrHbulk}. 
Because the contour $C_2$ is now around $\lambda = 0$, the behavior of the resolvent at small $\lambda$ will be important.
In particular, the results can be distinguished by their $R_0 \equiv R(\lambda=0)$ value:
\begin{itemize}
\item $R(\lambda)$ is a regular function at $\lambda=0$ and we can just take the residue, giving
\begin{equation}
\begin{aligned}
\label{eq:trace-when-it-does-not-factorise}
    \text{Tr}_{\mHb}(k_L k_R) & = e^{2S_0} 
\int ds_L ds_R \rho(s_L) \rho(s_R) 
\frac{R_0 y_k(s_L,s_R)}{R_0 y_q(s_L,s_R)-1}  \\ 
& \hspace{-1.9cm}= e^{2S_0} 
\int dE_L dE_R \rho(E_L) \rho(E_R) 
\frac{R_0  e^{- \beta_L  E_L} 
e^{-\beta_R E_R} 
\Gamma(\Delta\pm i\sqrt{E_L}\pm i\sqrt{E_R})}{R_0
\Gamma(\Delta\pm i\sqrt{E_L} \pm i\sqrt{E_R}) -e^{S_0}e^{\beta (E_L+E_R)/2}2^{2\Delta-1}\Gamma(2\Delta)}
\,.
\end{aligned}
\end{equation}
Because the integral over energies has non-trivial terms that depend on $E_L$ and $E_R$, the resulting integral does not factorise for any $R_0$ that is finite and non-zero. Since it is difficult to compute \eqref{eq:trace-when-it-does-not-factorise} analytically, we will emphasize the lack of factorisation by numerically computing $d(\beta_L, \beta_R)$ and showing that it takes a non-zero value for any $R_0$. As will be shown in Section~\ref{sec:towardsfactorisation}, the analyticity of $R(\lambda)$ at  $\lambda=0$ and the consequent lack of factorisation happens for $K<d^2$.

\item $R(\lambda)$ is divergent in the $\lambda\rightarrow 0$ limit, or $\lim_{\lambda\rightarrow 0}R(\lambda)^{-1}=0$.  
This means 
\begin{equation}\label{eq:TrHbulkfactorisedleading}
\begin{aligned}
    \text{Tr}_{\mHb}(k_L k_R) & = e^{2S_0} 
\int ds_L ds_R \rho(s_L) \rho(s_R) 
\oint
\frac{\dd \lambda}{2\pi \ii}\,
\frac{1}{\lambda}
\frac{y_k(s_L,s_R)}{R(\lambda)^{-1}- y_q(s_L,s_R)}  \\
& = e^{2S_0} 
\int ds_L ds_R \rho(s_L) \rho(s_R) 
\frac{y_k(s_L,s_R)}{y_q(s_L,s_R)}  \\
& = e^{2S_0} 
\int dE_L dE_R \rho(E_L) \rho(E_R) 
e^{-{\beta_L}E_L} e^{-{\beta_R} E_R} \,,
\end{aligned}
\end{equation}
which obviously factorises due to the cancellation of the correlations between in $E_L$ and $E_R$ after simplifying terms in the two $y$'s.  As will be shown in Section~\ref{sec:nullstates}, the above situation is valid only when $K\geq d^2$.

\end{itemize}

Physically, the distinction between the factorisation property of the trace boils down to whether the prepared states $\{|q_i\rangle\}_{i=1}^{K}$ span a complete basis of the bulk Hilbert space. 
From the definition of the resolvent, the existence of the $\lambda=0$ pole implies that the overlap matrix $M$ has zero eigenvalue and null states as corresponding eigenvectors, which is the case for $K\geq d^2$. Only once null states exist does the basis become overcomplete, and the trace can detect that the full bulk Hilbert space, rather than an arbitrary subspace, factorises. Thus, the statement about when the Hilbert space factorises can be nicely summarized as, 
\begin{align}
 \exists \,\, \text{Null states and }\mathcal{H}_\text{bulk} \text{ is spanned by }\ket{q_i}\quad &\Rightarrow \quad  \text{Tr}_{\mHb}(k_L k_R)= \text{Tr}_{L}(k_L) \text{Tr}_{L}(k_R), \nonumber\\ 
\not{\exists} \,\, \text{Null states and }\mathcal{H}_\text{bulk} \text{ is \textbf{not} spanned by }\ket{q_i} \quad &\Rightarrow \quad  \text{Tr}_{\mHb}(k_L k_R) \neq \text{Tr}_{L}(k_L) \text{Tr}_{L}(k_R)\,.    
\end{align}

$\mHb$, the two-sided Hilbert space for JT gravity coupled to matter, on its own, should be expected to be factorisable because the left and right energy can be distinct and the states $| E_L \rangle | E_R \rangle$ form factorisable basis of  $\mHb$. 

\subsection{Proving factorisation}\label{sec:nullstates}
\label{sec:null-states-and-fact-for-K>d^2}

In order to prove factorisation, the previous section instructs us to carefully study the analytic structure of the resolvent. After analyzing the analytic structure of the resolvent, we shall prove that the three probes of Hilbert space factorisation discussed in the introduction are satisfied to leading order in $1/G_N$ once $K$ is sufficiently large.

\subsubsection{The analytic structure of the resolvent and the trace }

As mentioned in the previous sections, the analytic structure, in the complex $\lambda$ plane, of the resolvent and $\text{Tr}_{\mHb}$ is crucial. 
To analyze these structures, we need to solve the self-consistency equation \eqref{eq:DSleadingorder} approximately. 
Concretely, we will first show that, with more details in the appendix~\ref{ap:analytic} when $K\geq d^2$, $R$ only has the following structures 
\begin{itemize}
\item As $\lambda\rightarrow 0$, we have the pole structure 
\be\label{eq:leadingrlambdato0}
 {R}(\lambda)\sim\frac{K-d^2}{\lambda}\Theta(K-d^2)+R_0\,.
\ee
This can be justified by taking an ansatz $R\sim \#/\lambda$ near $\lambda=0$ and plugging the ansatz back into the SD equation \eqref{eq:DSleadingorder}. 
Indeed, the $\lambda=0$ pole corresponds to null states and we expect to see $K-d^2$ of them for a $K\times K$ overlap matrix in a $d^2$ dimensional Hilbert space. Not surprisingly, $K-d^2 = \text{Res}_{\lambda=0} R(\lambda)$.\footnote{There is no residue at $\lambda=0$ if $K=d^2$. The single pole structure \eqref{eq:leadingrlambdato0} is not valid, and $R$ turns out to be $R\sim \pm i \sqrt{\lambda}^{-1}\left(e^{2 S_0} \int d s_L d s_R \frac{ \rho(s_L) \rho(s_R)}{y_q(s_L,s_R)}\right)^{\frac{1}{2}}$. This is because the branch cut hits the $\lambda=0$ pole.} We care about this because it controls the factorisation.

\item As $\lambda\rightarrow\infty$, we have asymptotic condition
\begin{equation}\label{eq:Rasymp}
    R(\lambda)\sim \frac{K}{\lambda}\,.
\end{equation}
Such an asymptotic behavior is expected from the definition of resolvent: if $\lambda$ is much larger than all the eigenvalues of $M$, then $(\lambda \mathbb{I} - M)^{-1}$ is approximately $\lambda^{-1}\mathbb{I}$. 
We care about this because it is involved in the analysis of the defining $C_0$ contour in \eqref{eq:pretraceHbulk}. 

\item As $\lambda\sim K$, we have a branch cut with 
\be
    R(\lambda+i\epsilon)-R(\lambda-i\epsilon)= 2\pi i D(\lambda) 
\ee
with $D(\lambda)$ the spectral density having support on $(\lambda_{-},\lambda_{+})$. Estimation of $\lambda_{-}$ and $\lambda_{+}$ are given by
\begin{equation}
\begin{aligned}
    \lambda_{-}\approx y_{q,min}(K-d^2)= y_q(E_{max},E_{max})(K-d^2), \\
      \lambda_{+}\approx y_{q,max}(K-d^2)= y_q(E_{min},E_{min})(K-d^2)\,.
\end{aligned}
\end{equation}
We care about the branch cut because it matters in the contour deformation.

\item In addition to the analytic structure of the resolvent listed above, when computing $\text{Tr}_{\mHb}(k_L k_R)$ we need to be worried whether $1-R(\lambda) y_q(s_L, s_R)$ ever becomes zero in the complex $\lambda$ plane in which case the contour deformation needed to compute $\text{Tr}_{\mHb}$ would pick up additional poles. The solution to the Schwinger-Dyson equation has the property that $1-R(\lambda) y_q(s_L, s_R)$ has no zero in the complex plane, which implies that no such additional poles are present. 

\end{itemize}

\begin{table}[H]
\centering
\begin{tabular}{c|c|c|c|c}
    & null states & Pole at $0$ & branch cut & Pole at infinity\\
    &&&&\\
    \hline
    &&&&\\
    $K<d^2$ & $\times$ & $0$ & $K$ & $-K$\\
    &&&&\\
    \hline
    &&&&\\
    $K\geq d^2$ & $\checkmark$ & $K-d^2$ & $d^2$ & $-K$\\
\end{tabular}
\caption{We show the singular structure of $R(\lambda)$ explicitly in this table. For $K<d^2$, there is null state so there is no pole at $\lambda=0$. But the branch cut on the real positive $\lambda$ axis gives residue $K$ and pole at infinity gives residue $-K$ and they sum together to zero. For $K\geq d^2$, there are $K-d^2$ null states and these give residue $K-d^2$ at $\lambda=0$. At the same time, the branch cut gives residue $d^2$ and the pole at infinity gives residue $-K$. Again these three contributions sum up to zero.}
\label{table:residue}
\end{table}

\subsubsection{Proving factorisation when $K\geq d^2$}
\label{sec:proving-factorisation}

With the analytic structure of the resolvent at hand, we can give a more rigorous argument on factorisation. The three different levels of factorisation discussed in the introduction will be shown as follows. 

\paragraph{Step 1 to leading order in $1/G_N$.}
The first calculation we can do is the factorisation of a single $\text{Tr}_{\mHb}$ shown in \eqref{eq:TrHbulkfactorisedleading}. Even though the factorisation has been explained in  section \ref{sec:physical-mechanism-of-factorisation}, here we will put all the results together and highlight the importance of analytic structure. 

We first inspect the $n\geq 0$ case. The replicated bulk trace is given by the contour integral on the defining contour $C_2$
\begin{align}
\overline{ (M^n)_{ij} \bra{q_i} k_L k_R \ket{q_j}} \, & = 
e^{2S_0} 
\int ds_L ds_R \rho(s_L) \rho(s_R) 
 \oint_{C_2}
 \frac{\dd \lambda}{2\pi \ii}
 \lambda^n
\frac{R(\lambda) y_k(s_L,s_R)}{1-R(\lambda) y_q(s_L,s_R)} \\
&\supset  e^{2S_0} K^{n+1} 
\int ds_L ds_R \rho_L \rho_R  y_k y_q^{n} 
 =  \begin{aligned}
    \includegraphics[valign=c,width=0.25\textwidth]{4pinwheel.pdf}
\end{aligned}\,,
\end{align}
in which \eqref{eq:Rasymp} is used and one can see again why $C_2$ is called the ``defining contour".
From the analysis on analytic structure, the function $ \frac{\lambda^n R(\lambda) y_k}{1-R(\lambda) y_q}$ has branch cut inherited from $R(\lambda)$, and is regular at everywhere else on the complex plane. 
We can thus deform the contour to $C_1$ and reach \eqref{eq:replicatedbulktrace}. And the contour integration around $C_1$ also means to consider only the non-zero eigenvalue of the overlap matrix $M_{ij}$. 

Next we analytically continue $n$ to $-1$.
From the overlap matrix point of view, this means to take the (formal) inverse of $M_{ij}$. 
One way to perform such an inverse is to take the inverse of non-zero eigenvalues, which means we do not change the contour $C_1$ so that the bulk trace $\text{Tr}_{\mHb}(k_L k_R)$ can be evaluated as 
\be 
\overline{ \braket{q_i}{q_j}^{-1} \bra{q_i} k_L k_R \ket{q_j}} = 
e^{2S_0} 
\int ds_L ds_R \rho(s_L) \rho(s_R) 
 \oint_{C_1} 
 \frac{\dd \lambda}{2\pi \ii}\,
 \frac{1}{\lambda}
\frac{R(\lambda) y_k(s_L,s_R)}{1 - R(\lambda) y_q(s_L,s_R)}.
\ee
From this, one can smoothly deform the contour from $C_1$ to $C_2$ which is around $\lambda=0$. 
Note that the possible pole of $R(\lambda)$ at $\lambda=\infty$ does not contribute, because near $\lambda=\infty$, the $\lambda$ integral takes the form $\int \frac{d\lambda}{2\pi i \lambda} \frac{1}{\lambda} \frac{K y_k}{\lambda}$ which decays fast enough. This is a special property for $n=-1$.
In the end, we obtain  
\be
\text{Tr}_{\mHb}(k_L k_R)=e^{2 S_0}\int d s_L d s_R \rho(s_L) \rho(s_R) \oint_{C_2} \frac{\dd \lambda}{2\pi i \lambda}
\frac{R(\lambda) y_k(s_L,s_R)}{1 - R(\lambda) y_q(s_L,s_R)}\,,
\ee
Using the $\lambda=0$ pole of $R(\lambda)$, together with the arguments in section \ref{sec:physical-mechanism-of-factorisation} it then follows that 
\be
\text{Tr}_{\mHb}(k_L k_R)= e^{2 S_0}\int_{\mathcal E} d E_L d E_R \rho(E_L) \rho(E_R) = \text{Tr}_{\mathcal H_L}(k_L) \text{Tr}_{\mathcal H_R}(k_R)\,.
\label{eq:trace-factorisation}
\ee
This is the main result of this section. Note that for any energy interval $\mathcal E$, we can choose $K$ to be sufficiently large for factorisation to occur; namely,  even if we take the size of the energy window to be arbitarily large, as long as we take $K$ to be even larger than the expected number of states in that window, factorisation will occur.  It is also useful to contrast \eqref{eq:trace-factorisation} with two alternative results. First, we can compare it to the perturbative result, which grows linearly with $K$ and thus never factorises for any value of $K$.  Second, we can compare it to the pure gravity result in the presence of wormholes but in the absence of matter. There, by considering the span of sufficiently many states, one can prove that in any finite energy interval, the Hilbert space is finite-dimensional. However, due to the gauge constraint that relates the left and right Hamiltonians, the bulk trace will always be a function of $\beta_L + \beta_R$ and will consequently never factories.

\begin{figure}[t!]
    \centering
\includegraphics[width=0.6\textwidth]{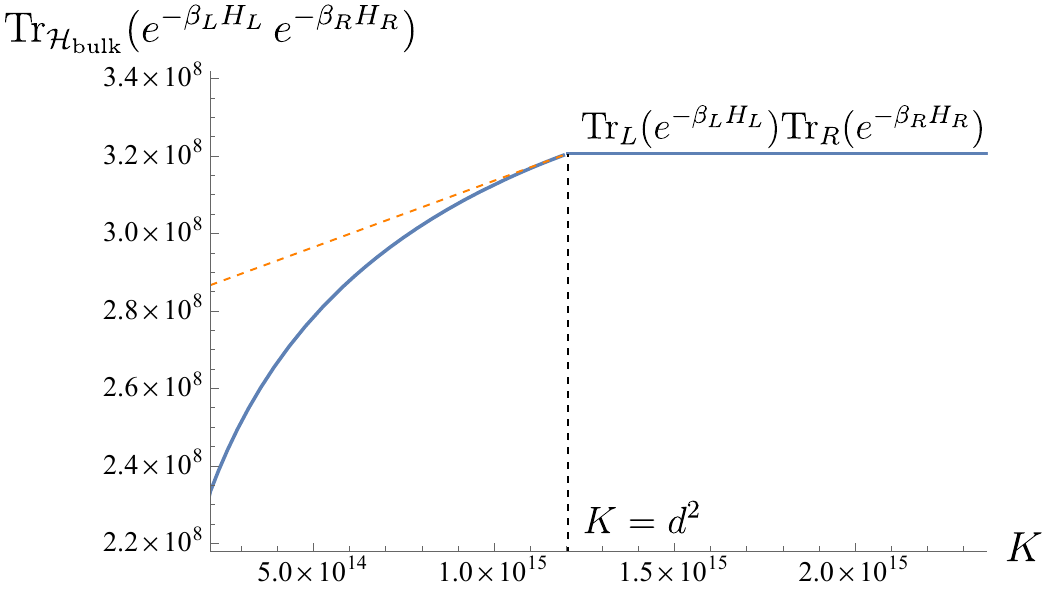}
    \caption{\textit{The ``Page curve'' of Hilbert space factorisation. } The two-sided partition function as a function of the number of states $K$ used to construct the Hilbert space $\mathcal H_\text{bulk}$. For small $K$, we solve the Schwinger-Dyson equation numerically and the resulting two-sided partition function is shown by the blue curve. As $K$ approach $d^2$ we compare the numerical results to the linear perturbative results in $K-d^2$ from \eqref{eq:tr-Hbulk-K-d^2-pert} shown by the dashed orange line. When $K>d^2$ we confirm that the bulk trace becomes equal to the product of two independent partition functions. }
    \label{fig:page}
\end{figure}

\paragraph{Step 2 \& 3 to leading order in $1/G_N$.}
Having proven that the trace factorises to leading order in $1/G_N$, we can now examine the differential equation, for which it is sufficient to just inspect the average of two $\text{Tr}_{\mHb}$. 
The calculation is similar to the one trace case above. Using the same contour deformation argument, one can again argue that the crucial point for factorisation is $R(\lambda_1,\lambda_2)\rightarrow \infty$ if either $\lambda_{1,2}\rightarrow 0$. 
Here $R(\lambda_1,\lambda_2)$ is the double resolvent. To the leading order in $e^{S_0}$, we have
\begin{equation}
    R(\lambda_1,\lambda_2)= R(\lambda_1)R(\lambda_2) \text{ at leading order in } e^{\frac{1}{G_N}} \text{ or } K. 
\end{equation}
The calculation of two traces goes as follows
\begin{align}
     &\overline{\text{Tr}_{\mHb}(k_L k_R) \text{Tr}_{\mHb}(k'_L k'_R)}\nonumber\\
     =&\int ds_Lds_Rd{s'_L}d{s'_R}\,\rho(s_L)\rho(s_R)\rho({s'_L})\rho({s'_R})\oint\frac{d\lambda_1d\lambda_2}{\lambda_1\lambda_2}\frac{R(\lambda_1)y_q(s_L,s_R)}{1-R(\lambda_1)y_k(s_L,s_R)}\frac{R(\lambda_2)y_q({s'_L},{s'_R})}{1-R(\lambda_2)y_k({s'_L},{s'_R})}\nonumber\\
     = & \text{Tr}_{\mathcal{H}_{L}}(e^{-\beta_L H_L}) \text{Tr}_{\mathcal{H}_{L}}(e^{-\beta_L^{\prime} H_L})
     \text{Tr}_{\mathcal{H}_{R}}(e^{-\beta_R H_R}) \text{Tr}_{\mathcal{H}_{R}}(e^{-\beta_R^{\prime} H_R})\,.
     \label{eq:factorised-answer-4-traces}
\end{align}
The expression obviously factorises. 
If we act the differential operator $\mathcal{D}_{\beta_L,\beta_R,{\beta'_L},{\beta'_R}}$, 
\be 
\label{eq:definition-differenatial-operator}
 \mathcal{D}_{\beta_L,\beta_R,{\beta'_L},{\beta'_R}} \equiv \partial_{\beta_L} \partial_{\beta_R} - \partial_{\beta_L} \partial_{\beta_R'}
\ee
on the above, and afterwards set $\beta_L ' = \beta_L$ and  $\beta_R ' = \beta_R$ we reproduce the differential \eqref{eq:differential-equation-intro} in the introduction. Acting with this operator on \eqref{eq:factorised-answer-4-traces}
we get zero to leading order in $1/G_N$, \be 
\overline{d(\beta_L, \beta_R)} = 0.
\ee
A similar argument can be made for $n$ traces 
\begin{equation}
\begin{aligned}
    &
    \overline{\Tr_{\mHb}(e^{-\beta_L^{(1)} H_L} e^{-\beta_R^{(1)} H_R})\dots \Tr_{\mHb}(e^{-\beta_L^{(n)} H_L} e^{-\beta_R^{(n)} H_R})}
    \\
    & \qquad \qquad = \text{Tr}_{\mathcal{H}_{L}}(e^{-\beta_L^{(1)} H_L}) 
     \text{Tr}_{\mathcal{H}_{R}}(e^{-\beta_R^{(1)} H_R}) \cdots 
     \text{Tr}_{\mathcal{H}_{L}}(e^{-\beta_L^{(n)} H_L})\text{Tr}_{\mathcal{H}_{R}}(e^{-\beta_R^{(n)} H_R}) \,.
\end{aligned}
\end{equation}
 By focusing on the case with $n=4$, we can show that 
 \be \overline{\left(d(\beta_L, \beta_R)\right)^2} = 0,
 \ee which is that the differential equation squared vanishes to leading order in $1/G_N$. 

Note that in the above argument, the differential operator gives zero is simply a fact that the average of multiple traces completely factorises. 
This is indeed a special property of the leading order. 
We will see later that it is possible to argue the differential equation vanishing without relying on such a special property.

\subsubsection{Towards factorisation as $K$ approaches $d^2$}\label{sec:towardsfactorisation}

\begin{figure}[t]
    \centering
\includegraphics[width=0.6\textwidth]{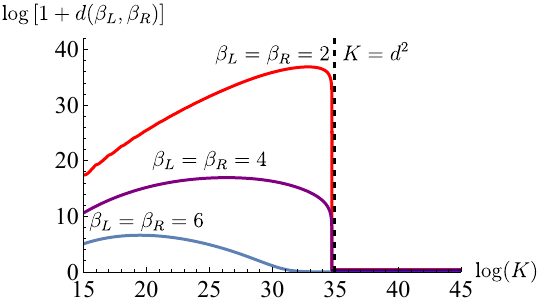}
    \caption{The value of the differential $d(\beta_L, \beta_R)$ as a function of the number of states $K$ used to construct the Hilbert space $\mathcal H_\text{bulk}$ for different values of $\beta_L$ and $\beta_R$. $d(\beta_L, \beta_R)$ increases for small $K$ since the number of states included in the trace and its corresponding value increases. Eventually, as $K$ increases, the bulk partition function is closer to factorising, and that overwhelms the increase in the number of states, making $d(\beta_L, \beta_R)$ decrease. For $K\geq d^2$, we confirm that the bulk trace function factorises by seeing that $d(\beta_L, \beta_R) = 0$ regardless of the value of $\beta_L$ and $\beta_R$.   }
    \label{fig:diffeq}
\end{figure}

We know from previous subsections that $R(\lambda)$ has no pole at $\lambda=0$ for $K<d^2$, which implies that the bulk trace does not factorise. 
In this section, we want to have a more detailed discussion of the non-factorisation of the trace, focusing on how factorisation would emerge when $K$ approaches $d^2$ from below.

Similar to the $K\geq d^2$ case, the singularity structure of the resolvent when $K<d^2$ can be briefly summarized as follows:  
\begin{enumerate}
    \item[(1)] The resolvent has a pole at $\infty$ with residue $-K$, i.e. $R(\lambda)=K/\lambda$ as $\lambda\rightarrow \infty$.
    \item[(2)] The resolvent has a branch cut supported on $(\lambda_{-},\lambda_{+})$ with  $\lambda_{\pm}>0$.  The contour integral around the branch cut gives $K$.    
\end{enumerate}

More importantly, at $\lambda=0$, $R(0) = R_0$ is finite. We justify this in appendix~\ref{ap:analytic}. 
There, by solving the Schwinger-Dyson equation, we can compute the value of $R_0$ explicitly in the limit $K\rightarrow d^2$. We find, 
\be
R_0=\frac{r_0}{K-d^2}+O\left((K-d^2)^0\right)\,,
\ee
with 
\be
r_0 =  e^{2 S_0} \int d s_L d s_R\, \frac{\rho(s_L) \rho(s_R)}{y_q(s_L,s_R)}\,.
\ee

With the approximation of $R_0$ in the $K\rightarrow d^2$ limit, we can see how the factorisation fails if  $K<d^2$ and, more importantly, how it is restored. 
First we examine $\text{Tr}_{\mHb}(k_L k_R)$
\begin{align}
    \text{Tr}_{\mHb}(k_L k_R)& = - e^{2S_0} \int d s_L d s_R \rho(s_L)\rho(s_R) \frac{R_0 y_k(s_L,s_R)}{1- R_0 y_q(s_L,s_R)}\nn\\
    & \approx e^{2S_0}  \int d E_L d E_R \rho(E_L)\rho(E_R) e^{-\beta_L E_L } e^{-\beta_R E_R} \nn \\  &\qquad+ \frac{K-d^2}{r_0} e^{2S_0} 
 \int d E_L d E_R \rho(E_L)\rho(E_R) \frac{ e^{-\beta_L E_L } e^{-\beta_R E_R}  }{ y_q(E_L,E_R)}, 
    \label{eq:tr-Hbulk-K-d^2-pert}
\end{align}
where in going from the first to the second line we expand at small $K-d^2$. From this equation, we can see explicitly the phase transition: when $K=d^2$, the second term vanishes, and this is when the Hilbert space factorises; when $K<d^2$ the second term explicitly correlates the left and right energies giving a non-factorising answer. 

Similarly, we find that the differential discussed in the introduction also gives a non-vanishing result as $K$ approaches $d^2$:
\begin{equation}
\begin{aligned}
    \overline{d(\beta_L,\beta_R)} &= e^{2S_0}  \int \prod_{i=L,R,L^{\prime},R^{\prime}} d E_i \rho(E_i) e^{-\beta_i E_i } (E_L-{E'_L}) E_R e^{-\beta_R E_R} \frac{R_0 y_q(E_L,E_R)}{1- R_0 y_q(E_L,E_R)}\frac{R_0 y_q(E'_L,E'_R)}{1-R_0 y_q({E'_L},{E'_R})}\\
    & \approx e^{2S_0}  \frac{(d^2-K)}{r_0} \int \prod_{i=L,R,L^{\prime},R^{\prime}} d E_i \rho(E_i) e^{-\beta_i E_i } (E_L-{E'_L}) E_R \left(\frac{1}{y_q(E_L,E_R)}+\frac{1}{y_q({E'_L},{E'_R})}\right)\,.
\end{aligned}
\end{equation}
Our analytic results can be compared to the numerical calculations\footnote{We thank Jinzhao Wang for very helpful discussions on the numerical method.} of $\text{Tr}_{\mHb}$ and the differential operator $d(\beta_L,\beta_R)$, both obtained by generating the resolvent numerically in figure~\ref{fig:page} and \ref{fig:diffeq}.

\subsubsection{Checking that factorisation is basis independent}
\label{sec:checking-factorisation-is-basis-indep}

Our proof about the factorisation of the Hilbert space and our calculation that shows when it occurs should be completely independent of the basis that we used to construct $\mathcal H_\text{bulk}$.\footnote{As long as we use states in which the gauge constraint $H_L = H_R$ is broken.} While this is obvious from a boundary perspective, this is not a priori clear from the bulk. Nevertheless, our path integral results exhibit this feature.  For example, the value of the bulk trace in $\mHb $ when $K>d^2$ is completely independent of the period of time evolution $\beta/4$ that can be inserted to the left and right of the operator $O_i$ in the construction of the state $\ket{q_i}$; moreover; the trace is independent of the exact values of the scaling dimension $\Delta$ that the operators $O_i$ have. 

In appendix \ref{sec:basis-prepared-with-different-dim} we generalize this observation by considering a more general construction of  $\mHb $. Instead of taking all states to be constructed by operators with close-by scaling dimensions which all have periods of Euclidean time evolution $\beta/2$ to the left and right of the operator insertion, we consider
\be 
\mHb = \text{span }\left\{ \ket{q_i} = \ket{\Delta_i,  \beta_L^{(i)}, \beta_R^{(i)}}, \quad i=1, \dots, K\right\}\,,
\label{eq:Hbulk-version1}
\ee
for arbitrary values $\{\Delta_i, \beta_L^{(i)}, \beta_R^{(i)}\}$. While the resolvent associated to the states $\ket{q_i}$ non-trivially depends on the values of  $\{\Delta_i, \beta_L^{(i)}, \beta_R^{(i)}\}$, the value of the trace once $K>d^2$ becomes completely independent of these parameters and matches the answer found in \eqref{eq:trace-factorisation}. The mechanism for this is similar to what we've seen in \eqref{eq:TrHbulkfactorisedleading}: once the trace of the resolvent has a pole, the analogous contour integral in $\lambda$ becomes independent of $\{\Delta_i, \beta_L^{(i)}, \beta_R^{(i)}\}$. Beyond factorisation, our computation in the appendix also shows that the dimension of the Hilbert space, computed from the rank of the Gram matrix from which the resolvent is constructed, is also independent of  $\{\Delta_i, \beta_L^{(i)}, \beta_R^{(i)}\}$. This thus generalizes all previous proofs that the rank saturates. 

To construct $\mHb$, we don't even need the matter excitation to be placed on the asymptotic boundary. Instead, we can consider matter excitation placed on geodesic slices of length $\ell_i$,  
\be 
\mHb = \text{span } \left\{ \ket{\tilde q_i} = \ket{\Delta_i, \ell_i, m_i}, \quad i=1, \dots, K\right\}\,.
\label{eq:Hbulk-version2}
\ee
As with \eqref{eq:Hbulk-version1}, we once again find that the bulk trace becomes independent of the scaling dimensions and geodesic lengths once $K>d^2$.

In the geodesic basis, the calculation of the bulk trace has the following interpretation. At a pertubative level, the states in \eqref{eq:Hbulk-version2} are all orthogonal. Each element in the trace $ \bra{\tilde q_i} e^{-\beta_L H_L - \beta_R H_R} \ket{q_i} $ gives the propagator between two states, prepared on geodesic slices of length $\ell_i$ whose matter state is determined by $\Delta_i$ and $m_i$. Performing the sum over $i$ in the trace then glues the two geodesics and naively, if enough lengths are included, results in a two-sided cylinder whose sides have renormalized lengths $\beta_L$ and $\beta_R$. However, at a pertubative level, summing over all such lengths and matter states yields a divergent result. For example, if we solely perform the sum over $\ell$, with all matter states being the same,  this divergence comes from the fact that when gluing along the two geodesics of $ \bra{\tilde q_i} e^{-\beta_L H_L - \beta_R H_R} \ket{\tilde q_i} $ we overcount cylindrical geometries that are related by the mapping class group.  Even if one, for some reason, instead accounts for the quotient by the mapping class group, the contribution of a cylindrical geometry to the gravitational path integral still gives an incorrect non-factorising answer -- this is because the cylinder does not correctly capture the bulk two-sided trace. Instead, at a non-perturbative level however we need to multiply  $ \bra{\tilde q_i} e^{-\beta_L H_L - \beta_R H_R} \ket{\tilde q_j} $ by $(M^{-1})_{ij}$ since the states   $ \ket{\tilde q_i}$ are no longer orthogonal. It is the insertion of $(M^{-1})_{ij}$  that results in a convergent answer for the trace. This thus "regularizes" the overcount of geometries that occurred in the perturbative computation. 

A similar issue arises in the absence of matter when considering states, 
\be 
\mathcal{H}_\text{pure grav} =\text{span } \left\{ \ket{\ell_i}, \quad i=1, \dots, K\right\}\,.
\label{eq:Hbulk-version3}
\ee  
 The naive perturbative trace once again gives a divergent answer that again comes from overcounting cylindrical geometries related by the mapping class group. As before, this can also be regularized by multiplying   $ \bra{\ell_i} e^{-\beta_L H_L - \beta_R H_R} \ket{\ell_j} $ by $(M^{-1})_{ij}$, this time with $M_{ij} = \braket{\ell_i}{\ell_j}$ \cite{Iliesiu:2024cnh}. Once we consider a large enough number of states, the trace yields $Tr_{\mathcal{H}_\text{pure grav}} = Z(\beta_L + \beta_R) = \int dE \rho(E) e^{-(\beta_L +\beta_R)E}$ where $Z(\beta)$ is the one-boundary partition function. As expected, in pure gravity, the result does not factorise; however, the trace gives the expected finite value in the Hilbert space spanned by the states $\ket{E,\, E}$.

\section{Rewriting the resolvent as an operator}
\label{sec:resolvent_as_operator}

In this section, we would like to reexamine the leading order result from a more geometrical point of view and analyze what is the essential feature of geometries that leads to factorisation.
Understanding such a feature better would be helpful to apply it to more general situations in later sections.
From the geometric analysis, we rewrite the resolvent as a projector operator, independent of $k_L$ and $k_R$, that captures the factorisation property.

\subsubsection*{Introducing the cutout pinwheel}

Let us start by inspecting the geometries contributing to the leading order $\text{Tr}_{\mHb}$ contribution. 
More specifically, it is helpful to compare the geometry contributing to the $Z_{n+1}^O(k_L,k_R)=\overline{M^n_{i j } \langle q_i|k_L k_R | q_j \rangle}$ and $Z_n^O=\overline{M^{n-1}_{i j } \langle q_i|q_j \rangle}$
\begin{equation*}
Z_{n+1}^O(k_L,k_R)=
\begin{aligned}
    \includegraphics[valign=c,width=0.25\textwidth]{4pinwheel.pdf}
\end{aligned} \,, \quad Z_n^O= \begin{aligned}
    \includegraphics[valign=c,width=0.25\textwidth]{3pinwheel.pdf}
\end{aligned}
\,.
\end{equation*}
One can see that the difference between the geometries is nothing but the operator insertion on one asymptotic boundary. 
Therefore, we can cutout a tiny operator dependent part and leave an universal part, which will be refered to as the cutout pinwheel
\begin{align}\label{eq:cutoutZn}
    Z_{n+1}^O(k_L,k_R)&=\includegraphics[valign=c,width=0.25\textwidth]{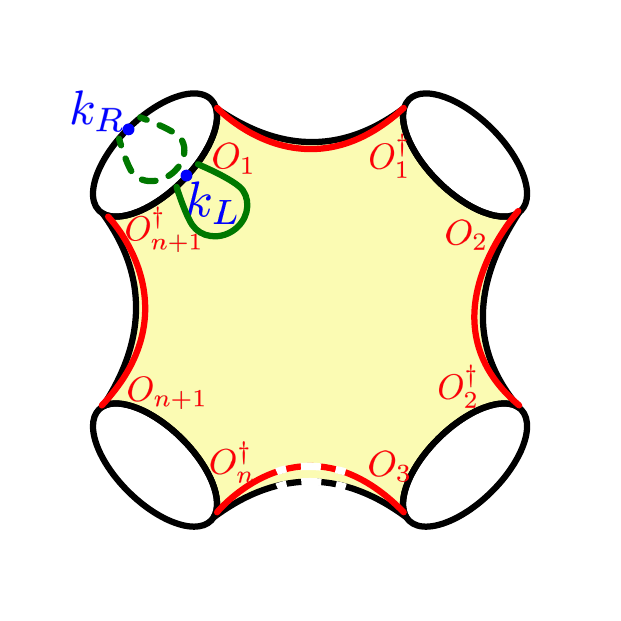}
    =\includegraphics[valign=c,width=0.25\textwidth]{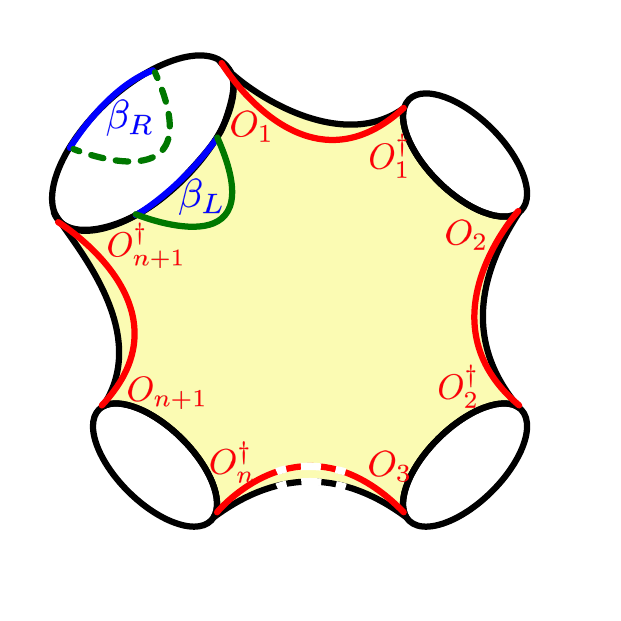}\\
    &=\int d\ell_L d\ell_R e^{\ell_L} e^{\ell_R}\, \underbrace{\includegraphics[valign=c,width=0.25\textwidth]{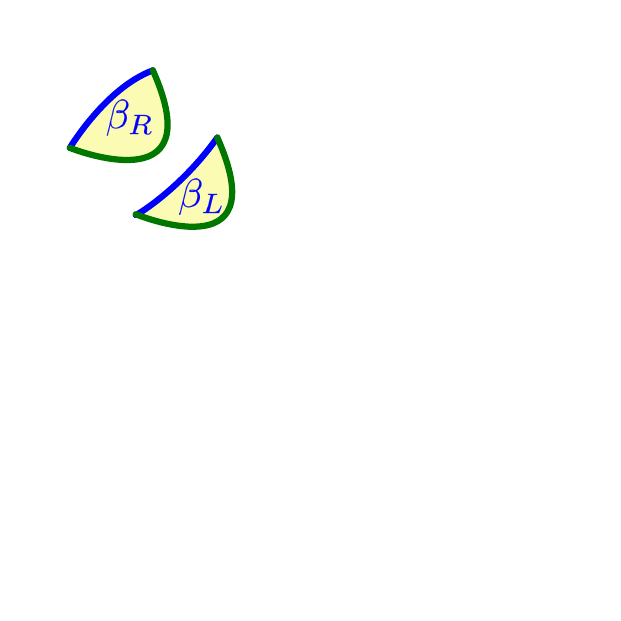}}_{{\text{operator dependent}}}\times \underbrace{\includegraphics[valign=c,width=0.25\textwidth]{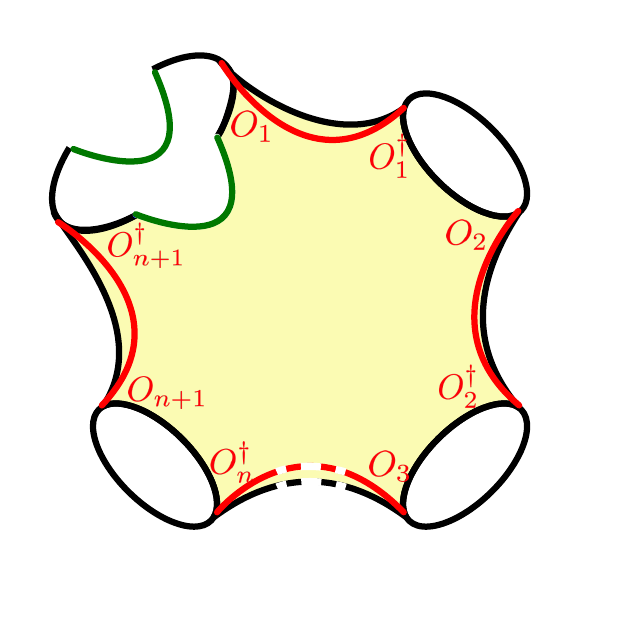}}_{\text{cutout pinwheel}}
\end{align}
In the second equate sign, we have specified the operator to be  $k_L=e^{-\beta_L H_L}$ ($k_R=e^{-\beta_R H_R}$) and reexpressed them as an asymptotic boundary with length $\beta_L$ ($\beta_R$). As for explicit expressions, we have 
\be\label{eq:operdependpart}
    \hspace{-2.0cm} \includegraphics[valign=c,width=0.24\textwidth]{4pinwheelremainder.pdf} \hspace{-2.0cm}  \equiv \varphi_{\beta_{L}}(\ell_L) \varphi_{\beta_{R}}(\ell_R)
     = \int \dd E_L \dd E_R \left( e^{-\beta_L E_L } e^{- \beta_R E_R} \right)\rho(E_L) \rho(E_R) \tilde{\varphi}_{E_L}(\ell_L) \tilde{\varphi}_{E_R}(\ell_R)\,,
\ee 
\vspace{-2cm}
    \be
      \hspace{-0.9cm}  \includegraphics[valign=c,width=0.24\textwidth]{4pinwheelcutout.pdf} \equiv z_{n+1}(\ell_L,\ell_R)   = e^{2S_0} \int \dd s_L \dd s_R \rho(s_L) \rho(s_R) 
y_q^n(s_L,s_R) \hat{y}_k(s_L,s_R; \ell_L ,\ell_R) \,,
\label{eq:cutout_pinwheel}
\ee
where 
\begin{equation}
    \hat{y}_k (\ell_L ,\ell_R) = 
e^{-S_0} e^{-\frac{\beta}{2}\frac{s_L^2}{2} } e^{-\frac{\beta}{2}\frac{s_R^2}{2} }  \tvarphi_{s_L}(\ell_L)
\tvarphi_{s_R}(\ell_R) 
\gamma_\Delta(s_L , s_R) = y_q \,  \tvarphi_{s_L}(\ell_L)
\tvarphi_{s_R}(\ell_R)  \,.
\end{equation}
This $\hat{y}_{k}(s_L,s_R;\ell_L,\ell_R)$ is related to the previously used $y_k$ and $y_q$ via  
\begin{align}
 \int \dd \ell_L \dd \ell_R e^{\ell_L} e^{\ell_R} 
 \hat{y}_k (\ell_L ,\ell_R)
 \varphi_{\beta_L}(\ell_L) \varphi_{\beta_R}(\ell_R) 
 &= y_k \,,
 \\
 \ \  \int \dd \ell_L \dd \ell_R e^{\ell_L} e^{\ell_R} 
 \hat{y}_k (\ell_L ,\ell_R)
 \varphi_{\beta=0}(\ell_L) \varphi_{\beta=0}(\ell_R) 
 &= y_q
,
\end{align}
in which we have suppressed the $s_{L,R}$ dependence of $y_k$ and $y_q$.

\subsubsection*{The $\mathcal{O}(\ell_L,\ell_R)$ operator and factorisation}
In the above discussion, we mostly focus on the $n$-th replicated geometry. 
To do the analytic continuation to $n\to -1$, we need to resum the pinwheel geometries in the same way of \eqref{pinwheelresum1}, which we rewrite here again as 
\be
\overline{\mathbf{R}_{ij}(\lambda) \bra{q_i} k_L k_R \ket{q_j}} 
= \sum_{n=0}^\infty R^{n+1}(\lambda) Z_{n+1}^O(k_L,k_R)\,.
\ee
Using the decomposition of the pinwheel $Z_{n+1}^O$ as in \eqref{eq:cutoutZn} and the fact that the operator dependent piece, such as in \eqref{eq:operdependpart}, is $n$ independent, we can resum only the cutout pinwheels $z_{n}(\ell_L,\ell_R)$ and define 
\begin{equation}
    \mathcal{O}_{\lambda}(\ell_L,\ell_R) \equiv \sum_{n=0}^{\infty} R^{n+1} (\lambda) z_{n+1}(\ell_L,\ell_R) \label{eq:O_lambda_definition}\,,
\end{equation}
such that 
\be 
\overline{ \mathbf{R}_{ij}(\lambda) \langle q_i| e^{-\beta_L H_L} e^{-\beta_R H_R} | q_j\rangle }
= \int d\ell_L d\ell_R\, e^{\ell_L} e^{\ell_R}\,
\varphi_{\beta_L}(\ell)\varphi_{\beta_R}(\ell')
\mathcal{O}_\lambda (\ell_L, \ell_R) \,.
\label{eq:resolventSD_cutout_resummation}
\ee
Knowing the explicit expression for the cutout pinwheel, it is easy to write down the explicit expression for $\mathcal{O}_\lambda (\ell_L,\ell_R)$ as
\be 
\mathcal{O}_\lambda (\ell_L,\ell_R) = 
e^{2S_0} 
\int \dd s_L \dd s_R 
\rho(s_L) \rho(s_R)  
\frac{R(\lambda) \hat{y}_k(\ell_L, \ell_R)}{1 - R(\lambda) y_q} \, ,
\ee
and extract the analytic continuation to $n=-1$ of the resulting expression that enters the bulk trace
\begin{align}
    \mathcal{O}(\ell_L , \ell_R) &\equiv \lim_{n \rightarrow-1} 
\oint \frac{\dd \lambda}{2\pi i} \lambda^n 
\mathcal{O}_\lambda (\ell_L,\ell_R)\\
&=e^{2S_0} 
\int \dd s_L \dd s_R 
\rho(s_L) \rho(s_R)  
\oint_{C_2} \frac{\dd \lambda}{2\pi i\lambda}\frac{R \hat{y}_k(\ell_L , \ell_R)}{1 - R y_q} \,.
\end{align}
Given the discussion of analytic structure of $R(\lambda)$, we can tell $\mathcal{O}(\ell_L,\ell_R)$ does not factorise when $K<d^2$ and factorises when $K\geq d^2$. For $K\geq d^2$ we have
\begin{align}
    \mathcal{O}(\ell_L,\ell_R) &= e^{2 S_0} \int \dd s_L \dd s_R 
\rho(s_L) \rho(s_R) \frac{\hat{y}_k(\ell_L,\ell_R)}{y_q(\ell_L,\ell_R)}\\
&= e^{2 S_0}  \varphi_{\beta=0}(\ell_L)\varphi_{\beta=0}(\ell_R) \label{eq:maximix}
\end{align}
Therefore, to leading order in $1/G_N$, we have
\be\label{eq:hbulkfromO}
\text{Tr}_{\mHb}(e^{-\beta_L H_L} e^{-\beta_R H_R})=\int d\ell_L d\ell_R e^{\ell_L} e^{\ell_R}\,  \varphi_{\beta_{L}}(\ell_L) \varphi_{\beta_{R}}(\ell_R) \mathcal{O}(\ell_L , \ell_R)
.
\ee
$\mathcal{O}(\ell_L , \ell_R)$ can now be seen to be the reason for factorisation: since its matrix elements are a product of two functions, one of $\ell_L$ and another of $\ell_R$, the trace for any functions of $H_L$ and $H_R$ consequently factorises. In the next section, it will prove especially convenient to use $\mathcal{O}(\ell_L,\ell_R)$ to test the factorisation of the bulk trace to all non-perturbative orders in $1/G_N$.

Now we turn to the algebraic meaning of $\mathcal{O}(\ell_L , \ell_R)$. 
For $K\geq d^2$, one can define an  operator whose matrix element in the $\ell$ basis is $\mathcal{O}(\ell_L,\ell_R)$,
\begin{align}
\hat{O}&=\int \dd\ell_L\dd\ell_R  e^{\ell_L} e^{\ell_R} \mathcal{O}(\ell_L,\ell_R) |\ell_L\rangle  \langle \ell_R| \nonumber \\
&= \int \dd E_L \dd E_R\,\rho(E_L)\rho(E_R)\ket{E_L} \bra{E_R}
\label{eq:operator-O}
\end{align}
where in the second line we also expressed the operator in the energy eigenstates basis. Using such an operator the trace can be expressed as $\text{Tr}_{\mHb}(e^{-\beta_L H_L} e^{-\beta_R H_R}) = \bra{\beta_L}\hat O \ket{\beta_R}$.
The above expression for $\hat{O}$ tells us that the operator is proportional to a projector onto the infinite temperature TFD state or, equivalently, onto a Hartle-Hawking state with $\beta\rightarrow 0$. This is why $ \bra{\beta_L}\hat O \ket{\beta_R}$ yields the expected factorising answer.

\section{Factorisation to all non-perturbative orders in $e^{1/G_N}$}
\label{sec:fact-to-all-orders}

So far we have considered the case when the number of states used in the construction of $\mathcal H_\text{bulk}(K)$ scales with $d^2$. This was sufficient to prove the factorisation of the trace to leading order in $K$. To proceed to prove factorisation to all non-perturbative orders in $e^{1/G_N}$ we can consider the limit,
\be
K\rightarrow\infty\quad\quad e^{2S_0}\rightarrow\infty\quad\quad
\frac{K}{d^2} \rightarrow \infty
.
\label{eq:all_orders_limit}
\ee
Since the Hilbert space $\cH_\text{bulk}(K)$ should remain unchanged once the states $\ket{q_i}$ fully span the Hilbert space, properties of the entire Hilbert space such as its potential factorisation into a tensor product can be probed regardless of the exact value of $K$. Therefore, we might as well study the question of factorisation in the limit in which the sum over geometries is most simplified. From this perspective, taking $K/d^2 \to \infty$ has multiple practical advantages. Firstly, the statistics of the trace is not contaminated by outliers where some subset of the states $\ket{q_i}$ used in the construction of $\cH_\text{bulk}(K)$ are approximately linearly dependent even without spanning the full Hilbert space $\cH_\text{bulk}(K)$. Secondly, the statistics of the trace 
is also not affected by the fact that the exact dimension of $\cH_\text{bulk}(K)$ is also slightly erratic since it is sensitive to the cut-off in the energy of the states; since we are taking $K\gg d^2$ the trace is not affected by the possible statistics of $d^2$.  Finally, this limit allows us to formally be able to exactly resum all higher genus geometries that contribute to the trace making the computation to all non-perturbative orders in $1/G_N$ (i.e., in $e^{-S_0}$) tractable.

We will proceed by verifying factorisation using the three steps \eqref{eq:average-factorisation}--\eqref{eq:diff-equations-vanish} presented in the introduction, all probed up to $1/K$ corrections that vanish in the large $K$ limit but to at all non-perturbative orders in $e^{-S_0}$.

If in \eqref{eq:average-factorisation}--\eqref{eq:diff-equations-vanish} we are solely interested in the large $K$ limit of the traces, we only have to include the geometries that contribute to leading order in $K$. In the expansion of the resolvent, such geometries are fully connected, have a maximal number of index loops, and have arbitrary topology. As we shall see, by gluing additional handles to the pinwheels considered in section \ref{sec:resolvent_as_operator}, such geometries can easily be incorporated in the computation and guarantee the factorisation of the Hilbert space to all non-perturbative orders in the topological expansion. We further confirm that there are no additional geometries that contribute to leading order in $K$ by classifying the possible surfaces that contribute to the resolvent, explaining to what order each geometry contributes, and explicitly writing a corrected expression for the resolvent and the bulk trace at subleading orders in $K$. While proving \eqref{eq:average-factorisation}--\eqref{eq:diff-equations-vanish} to leading order in $K$ are sufficient to show that the trace factorises, the subleading corrections that we compute suggest that the subleading perturbative corrections in $K$ are exactly vanishing as long as $K>d^2$.

\subsection{Factorisation through non-factorization}
\label{sec:fact-through-non-fact}

We start with the discussion of \textbf{Step 1}. As in previous section, our goal is to show that for large enough $K$ the following equality is satisfied 
\be 
\overline{\Tr_{\mH_{bulk}(K)}(e^{-\beta_L H_L} e^{-\beta_R H_R})}
=\overline{\Tr_{\mH_L}(e^{-\beta_L H_L})\Tr_{\mH_R}(e^{-\beta_R H_R})} 
,
\ee
with the left hand side again computed through the resolvent as 
\be 
\overline{\Tr_{\mH_{bulk}(K)}(e^{-\beta_L H_L} e^{-\beta_R H_R})} 
=\oint \frac{\dd \lambda}{2\pi i \lambda}  \overline{\mathbf{R}_{ij} \langle q_i| e^{-\beta_L H_L} e^{-\beta_R H_R} | q_j\rangle }
. 
\label{eq:Tr_bulk_in_resolvent}
\ee
In contrast to the previous sections, the right-hand side now contains contributions from pinwheels with empty handles. This will modify the Schwinger-Dyson equations for the resolvent. At leading order in $K$, the effect of the higher genus geometries we're including can be viewed as adding all possible independent configurations of empty handles on top of each pinwheel, contributing to the Schwinger-Dyson equations. The new resolvent, therefore, has to satisfy the equation
\be 
\overline{ \mathbf{R}_{ij} \langle q_i| e^{-\beta_L H_L} e^{-\beta_R H_R} | q_j\rangle }
= \sum_{n=0}^\infty R^{n+1} \overline{ Z_{n+1}^O(e^{-\beta_L H_L} , e^{-\beta_R H_R}) } ,
\ee
where the overline above $Z_{n+1}^O(e^{-\beta_L H_L} , e^{-\beta_R H_R})$ denotes the addition of higher genus contributions on a given pinwheel, e.g. 
\be 
\overline{ Z_{n+1}^O(e^{-\beta_L H_L} , e^{-\beta_R H_R}) }
= \includegraphics[valign=c,width=0.15\textwidth]{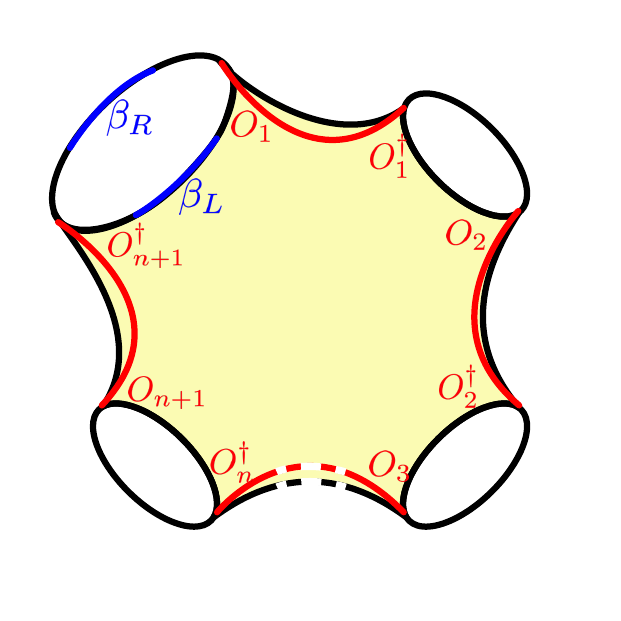}
+\includegraphics[valign=c,width=0.15\textwidth]{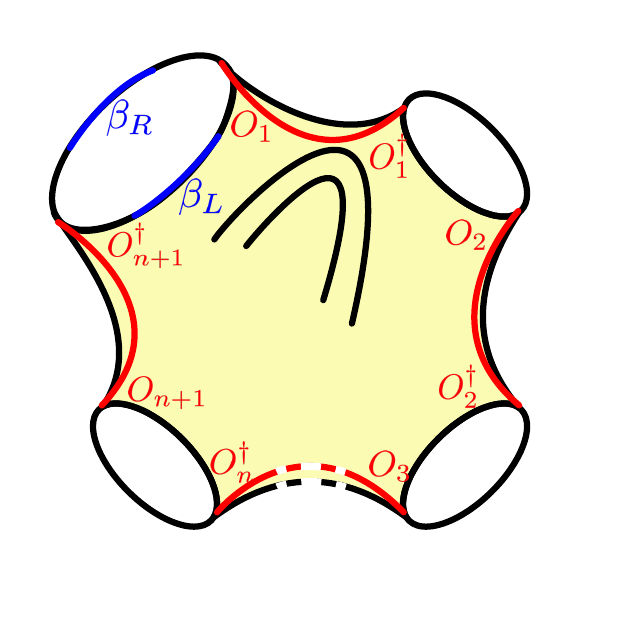}
+
\includegraphics[valign=c,width=0.15\textwidth]{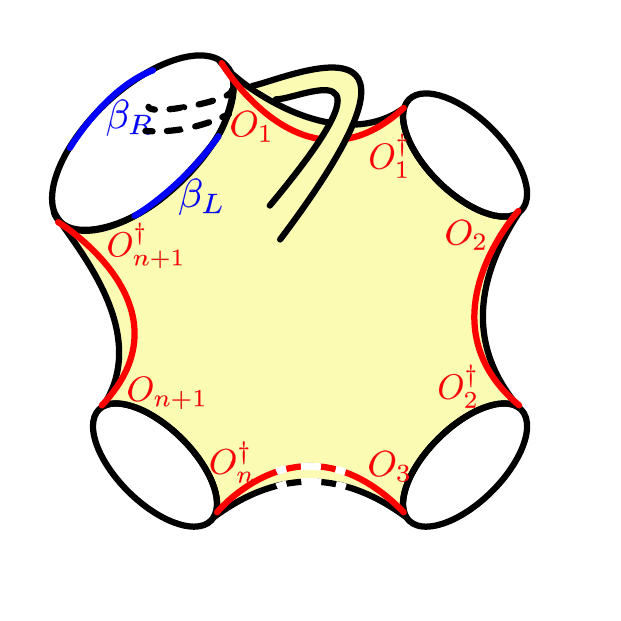}
+ \dots 
\ee
To resum the above geometries, we now generalize the cutout pinwheel construction of section \ref{sec:resolvent_as_operator}. Notice that for any higher genus geometry contributing to $\overline{Z_{n+1}^O(e^{-\beta_L H_L} , e^{-\beta_R H_R})}$ we can find two geodesics, one for each side of the pinwheel, anchored at the top boundary and homotopic to the sum of complementary matter geodesics (red curves) and asymptotic boundaries (black curves). Simply put, we can always find a geodesic such that all the handles start or end within the region enclosed by that geodesic. Cutting the geometry along the prescribed geodesic, we can decompose the contributing geometries as 
\begin{align}
\includegraphics[valign=c,width=0.15\textwidth]{4pinwheelplus.pdf}
&=
\int d\ell_L d\ell_R e^{\ell_L} e^{\ell_R}\, \includegraphics[valign=c,width=0.15\textwidth]{4pinwheelremainder.pdf}\times \includegraphics[valign=c,width=0.15\textwidth]{4pinwheelcutout.pdf}
,
\\
\includegraphics[valign=c,width=0.15\textwidth]{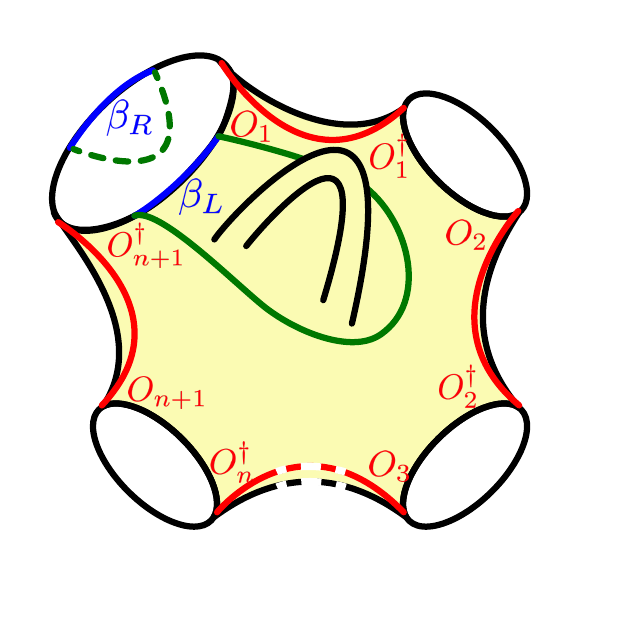} &=
\int d\ell_L d\ell_R e^{\ell_L} e^{\ell_R}\, \includegraphics[valign=c,width=0.15\textwidth]{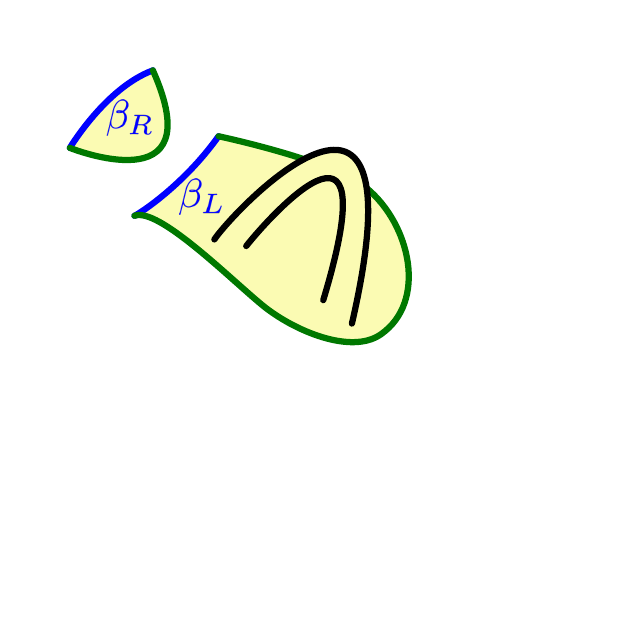}\times \includegraphics[valign=c,width=0.15\textwidth]{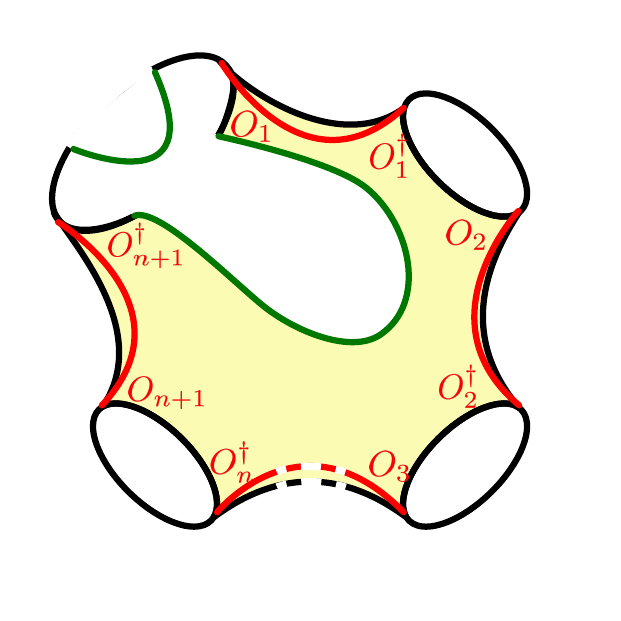} ,
\\ 
\includegraphics[valign=c,width=0.15\textwidth]{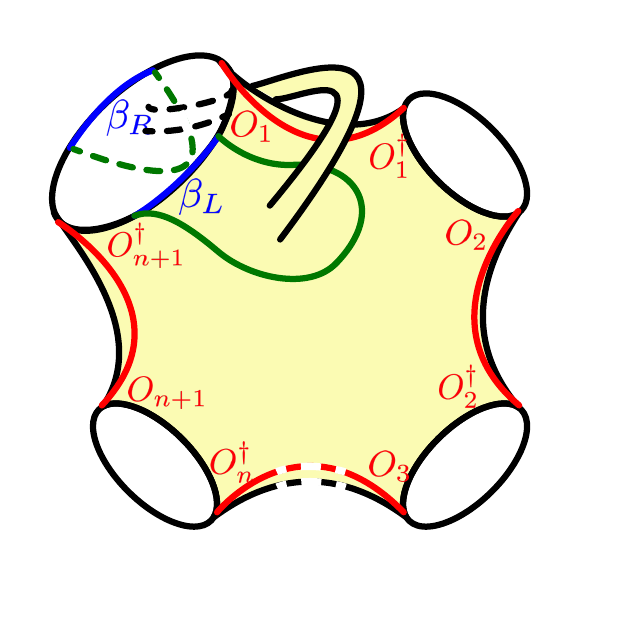}&= 
\int d\ell_L d\ell_R e^{\ell_L} e^{\ell_R}\, \includegraphics[valign=c,width=0.15\textwidth]{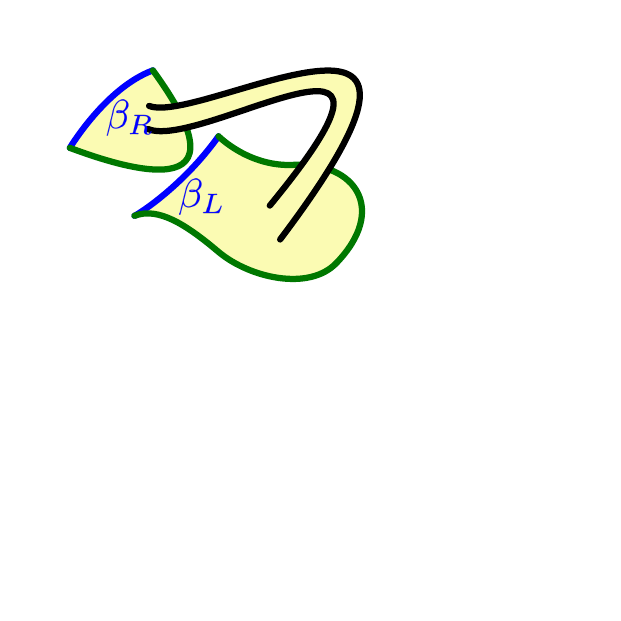}\times \includegraphics[valign=c,width=0.15\textwidth]{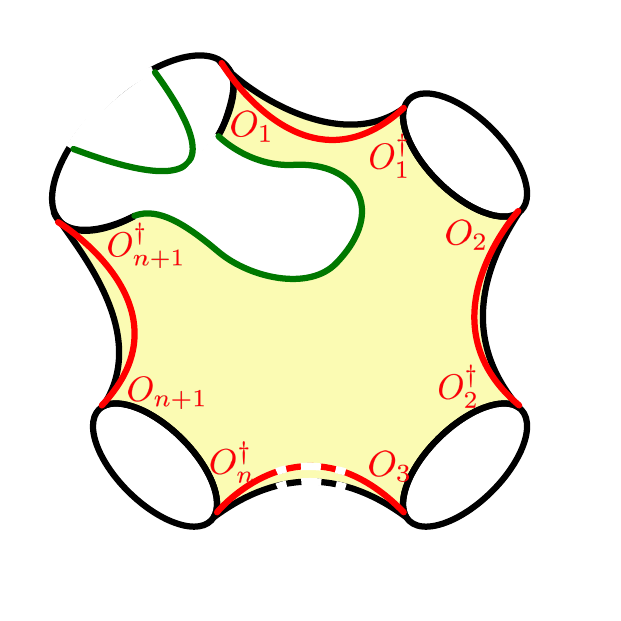} ,
\end{align} 
with an analogous decomposition holding for any geometry contributing to $\overline{Z_{n+1}^O(e^{-\beta_L H_L} , e^{-\beta_R H_R})}$. Note that now after cutting out the terms containing handles the remaining geometry is simply the cutout pinwheel $z_{n+1}(\ell,\ell')$ introduced in \eqref{eq:cutout_pinwheel}. This allows us to write the resummation of all higher genus contributions with a given number of boundaries as 
\begin{align}
\overline{ Z_{n+1}^O(e^{-\beta_L H_L} , e^{-\beta_R H_R}) }
= 
\int d\ell_L d\ell_R e^{\ell_L} e^{\ell_R}\, 
\overline{\varphi_{\beta_L}(\ell_L)\varphi_{\beta_R}(\ell_R)}
z_{n+1}(\ell_L,\ell_R) ,
\end{align}
where we introduced the notation 
\be 
\overline{\varphi_{\beta_L}(\ell_L)\varphi_{\beta_R}(\ell_R)}
=
\includegraphics[valign=c,width=0.17\textwidth]{4pinwheelremainder.pdf} 
+ 
\includegraphics[valign=c,width=0.17\textwidth]{4handle1cut.pdf} 
+ 
\includegraphics[valign=c,width=0.17\textwidth]{4handle2cut.pdf}
+\dots 
\ee
and in the above expression each geometry is weighted by a modified factor of $e^{-2g S_0}$.
The above sum over such geometries is now very similar to the exact two boundary partition function $\overline{ Z(\beta_1) Z (\beta_2) }$ of pure JT gravity. In fact, simply gluing 
$\overline{\varphi_{\beta_L}(\ell_L)\varphi_{\beta_R}(\ell_R)}$
to two copies of the Hartle-Hawking, we get back the product of the two boundary partition functions: 
\be 
\overline{ Z(\beta_L+\beta ) Z (\beta_R + \beta)} = e^{2S_0}
\int d\ell_L d\ell_R e^{\ell_L} e^{\ell_R} 
\overline{\varphi_{\beta_L}(\ell_L)\varphi_{\beta_R}(\ell_R)}
\varphi_\beta(\ell_L)
\varphi_\beta(\ell_R) .
\label{eq:two_boundary_partition_function}
\ee

With this decomposition, we can now go back to the resummation of the Schwinger-Dyson equations. Because all the higher genus terms are now simply repackaged into an $n$-independent term $\overline{\varphi_{\beta_L}(\ell_L)\varphi_{\beta_R}(\ell_R)}$ we can simply resum all the terms as in \eqref{eq:resolventSD_cutout_resummation}. This leads us to
\begin{align}
\overline{ \mathbf{R}_{ij} \langle q_i| e^{-\beta_L H_L} e^{-\beta_R H_R} | q_j\rangle }
= \int d\ell_L d\ell_R\, e^{\ell_L} e^{\ell_R}\,
\overline{\varphi_{\beta_L}(\ell_L)\varphi_{\beta_R}(\ell_R)}
O_\lambda (\ell_L, \ell_R) ,
\end{align}
with $O_\lambda (\ell_L, \ell_R)$ defined as in \eqref{eq:O_lambda_definition}
\be 
O_\lambda (\ell_L,\ell_R) = 
e^{2S_0} 
\int \dd s_L \dd s_R 
\rho(s_L) \rho(s_R)  
\frac{R \hat{y}_k(\ell_L , \ell_R)}{1 - R y_q} .
\ee
The final trace is now computed through 
\begin{align}
\overline{\Tr_{\mH_{bulk}(K)}(e^{-\beta_L H_L} e^{-\beta_R H_R})} 
&=\int d\ell_L d\ell_R e^{\ell_L} e^{\ell_R}\,
\overline{\varphi_{\beta_L}(\ell_L)\varphi_{\beta_R}(\ell_R)}
\oint_{C_2} \frac{\dd \lambda}{2\pi i \lambda} 
 O_\lambda (\ell_L, \ell_R) 
\\ 
&= \int d\ell_L d\ell_R e^{\ell_L} e^{\ell_R}\,
\overline{\varphi_{\beta_L}(\ell_L)\varphi_{\beta_R}(\ell_R)}
\mathcal{O}(\ell_L, \ell_R) . 
\end{align}
The final answer now simply follows from the discussion in section \ref{sec:resolvent_as_operator}. Whenever the resolvent develops zero eigenvalues, for $K\geq d^2$, the operator $\mathcal{O}(\ell_L,\ell_R)$ takes a simple factorised form
\be 
\mathcal{O}(\ell_L , \ell_R) \simeq e^{2S_0}  \int \dd s_L \dd s_R \rho(s_L) \rho(s_R) \tilde{\varphi}_{s_L}(\ell_L) \tilde{\varphi}_{s_R}(\ell_R) 
= e^{2S_0} \varphi_{\beta=0} (\ell_L) \varphi_{\beta=0} (\ell_R) , 
\ee
which after using \eqref{eq:two_boundary_partition_function} leads to the desired final answer 
\begin{align}
\overline{\Tr_{\mH_{bulk}(K>d^2)}(e^{-\beta_L H_L} e^{-\beta_R H_R})} 
= 
\overline{Z(\beta_L) Z(\beta_R)}
= \overline{\Tr_{\mH_L}(e^{-\beta_L H_L})\Tr_{\mH_R}(e^{-\beta_R H_R})} + O\left(\frac{1}K \right).
\end{align}
The trace could in principle receive $O(1/K)$ corrections could in principle come from subleading geomtries that we have so far not accounted for. A detailed analysis of such geometries will be presented in the next subsection \ref{sec:classif-geometries-subleading-in-K} where we shall see that the next subleading order in $1/K$ is in fact vanishing. 

We have therefore found that for $K\gg d^2$ the average of $\Tr_{\mH_{bulk}(K\gg d^2)}$, as computed from the gravitational path integral, behaves as if the bulk Hilbert space factorised into $\mH_{bulk}(K\gg d^2)=\mH_L \otimes \mH_R$ for ``each member of the ensemble"/before coarse-graining. The gravitational path integral ``knows" about the factorisation of the underlying Hilbert space even though the full answer contains wormhole contributions, i.e. the full answer is still only a coarse-grained quantity. Nevertheless, our approach constrains the Hilbert space factorisation of the non-perturbative quantum gravitational Hilbert space through the statistics computed by the GPI. Because we are using statistics of inner-products rather than their exact value, one can view the above result as an interesting interplay of the two factori(s/z)ation puzzles discussed in the literature: the factorisation of the Hilbert space can be seen due to the non-perturbative corrections that lead to the non-factorization of boundary observables, seen for example in the non-trivial correlation between two partition functions $\overline{Z(\beta_L) Z(\beta_R)}$.

\subsection{Proving factorisation for each member of the ensemble}
\label{sec:fact-for-each-member-of-the-ensemble}

As already discussed in previous sections, the proof of \textbf{Step 1} that we have described in section~\ref{sec:fact-through-non-fact}  only probes the Hilbert space factorisation on average. We can now further compute more fine-grained statistical quantities to constrain the information about $\mH_{bulk}$ even more. In particular, by showing that 
\be 
\overline{d(\beta_L, \beta_R)^2} =   O\left(\frac{1}K \right) , 
\ee
we can ensure that the underlying Hilbert space factorises for ``each member of the ensemble"/before coarse-graining.

\paragraph{Proving step 2:} 
As a warm-up, we start by showing a simpler differential equation 
\be 
\overline{d(\beta_L, \beta_R)} 
\equiv 
\mathcal{D}_{\beta_L,\beta_R,\beta_L^{\prime},\beta_R^{\prime}} 
\overline{Z_{\text{bulk}}(\beta_L,\beta_R)Z_{\text{bulk}}(\beta_L^{\prime},\beta_R^{\prime})} |_{\beta'_{L/R}=\beta_{L/R}}
= O(1/K),
\ee
where the differential operator $\mathcal{D}_{\beta_L,\beta_R,\beta_L^{\prime},\beta_R^{\prime}} $ is defined in \eqref{eq:definition-differenatial-operator}.

To show the above equation we simply need to compute $\overline{Z_{\text{bulk}}(\beta_L,\beta_R)Z_{\text{bulk}}(\beta_L^{\prime},\beta_R^{\prime})}$ at leading order in $K$ and all orders in $e^{-S_0}$ and then act on it with the differential operator. For the first step, we again refer to the resolvent method, which now requires us to use two copies of the resolvent. Denoting $Z_{\text{bulk}} \equiv Z_{\text{bulk}}(\beta_L,\beta_R)$ and  $Z'_{\text{bulk}} \equiv Z_{\text{bulk}}(\beta_L^{\prime},\beta_R^{\prime})$ we have 
\begin{align}
\overline{Z_{\text{bulk}}Z'_{\text{bulk}}} 
&= \oint \frac{\dd \lambda_1 \dd \lambda_2}{(2\pi i)^2 \lambda_1 \lambda_2}  
\overline{\mathbf{R}_{ij}(\lambda_1) \langle q_i| e^{-\beta_L H_L} e^{-\beta_R H_R} | q_j\rangle 
\mathbf{R}_{\mu \nu}(\lambda_2) \langle q_\mu| e^{-\beta'_L H_L} e^{-\beta'_R H_R} | q_\nu\rangle 
} ,
\end{align}
where, once again, the overline means that we need to evaluate the right-hand side through the gravitational path integral with boundary conditions now specified by both resolvents. This is, in general, a more complicated task compared to the case of a single resolvent, as the bulk geometries can connect together operators from both resolvents, making it difficult to organize the full expansion in terms of the standard Schwinger-Dyson equations. For progress in this direction in the context of PSSY model \cite{Penington:2019kki} see \cite{Bousso:2023efc}. In our case, however, we are interested in the strict limit of $K/d^2 \rightarrow \infty$, which greatly simplifies the geometries that contribute. Recall that for a geometry to contribute at leading order in $K$, it must contain a maximal number of matter loops for a given number of boundaries. 
This means that all geometries will take the form of two separate pinwheels (which can have a different number of boundaries), possibly connected by an empty wormhole through the bulk. E.g. 
\begin{align}
\includegraphics[valign=c,width=0.35\textwidth]{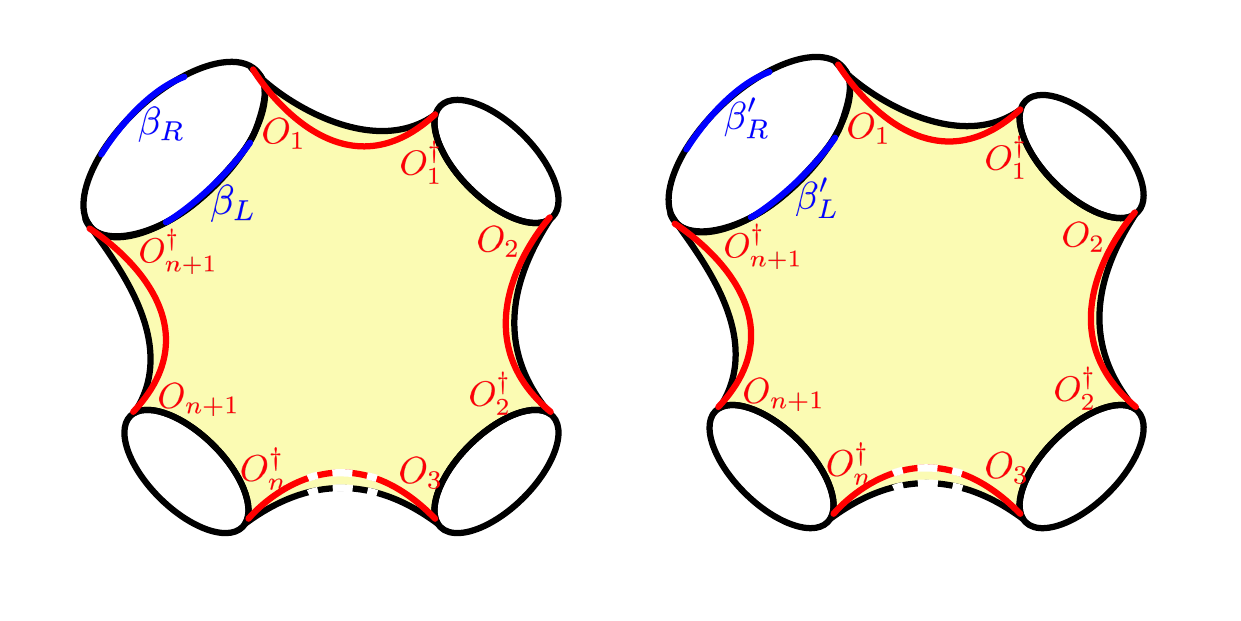}
,\includegraphics[valign=c,width=0.35\textwidth]{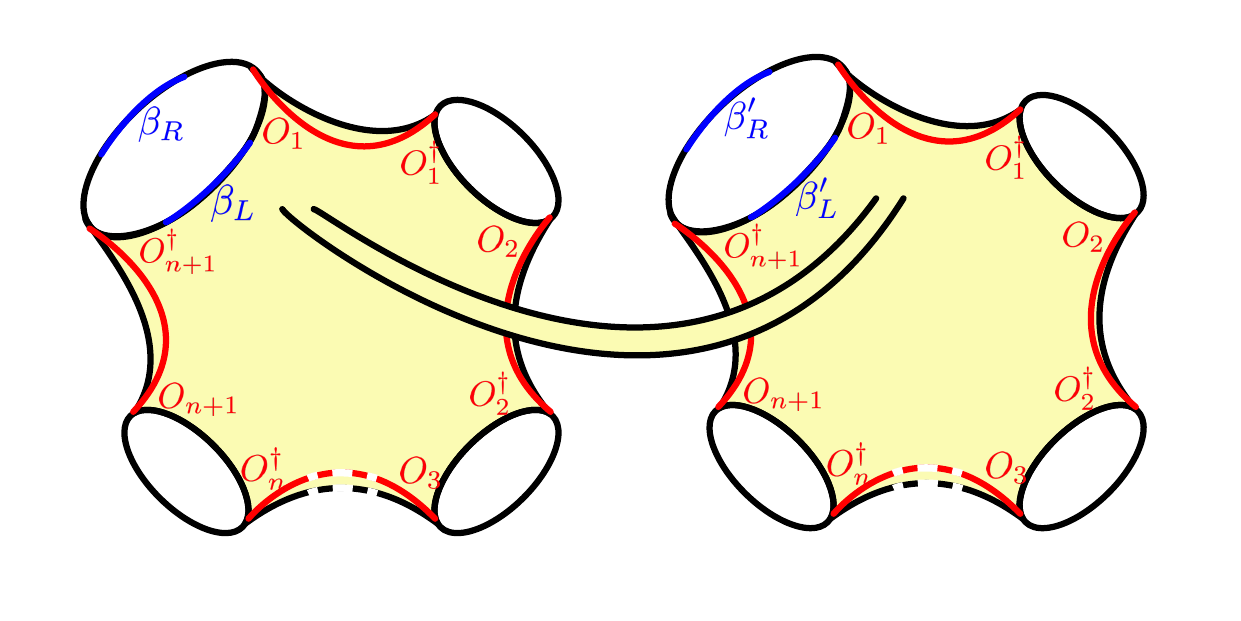}
, \dots 
\end{align}

For each such geometry, now we can repeat the construction of the cutout pinwheel as 
\begin{align}
\includegraphics[valign=c,width=0.24\textwidth]{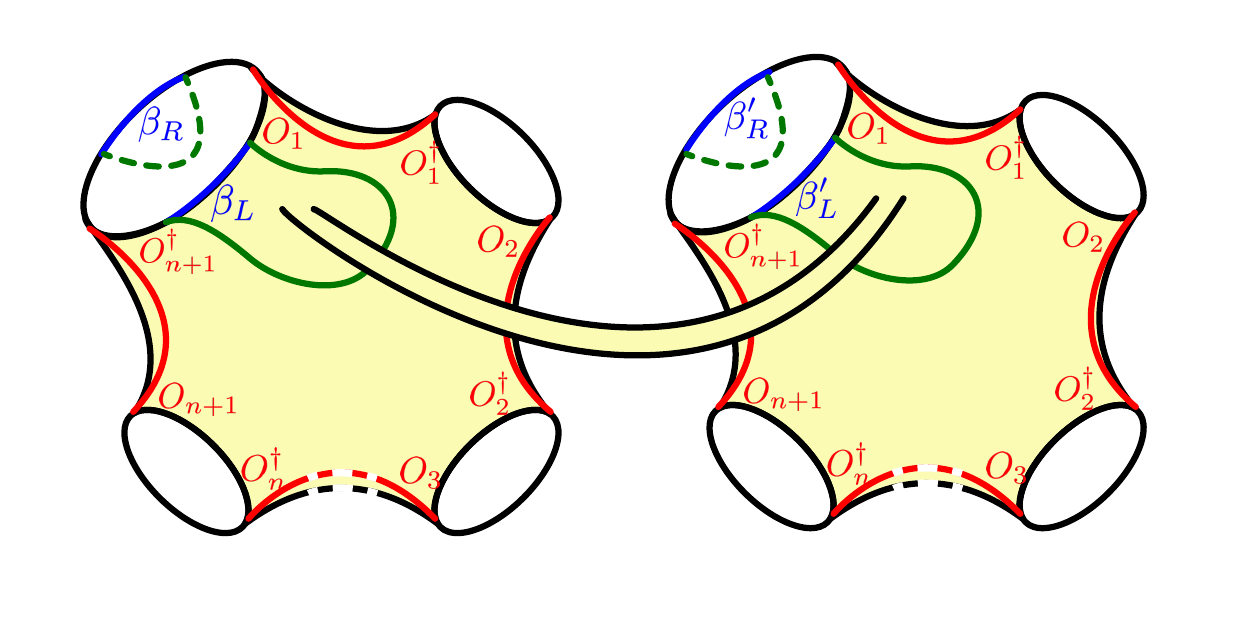}
&=\int \prod_{i=1}^4 d\ell_i e^{\ell_i}\, \includegraphics[valign=c,width=0.24\textwidth]{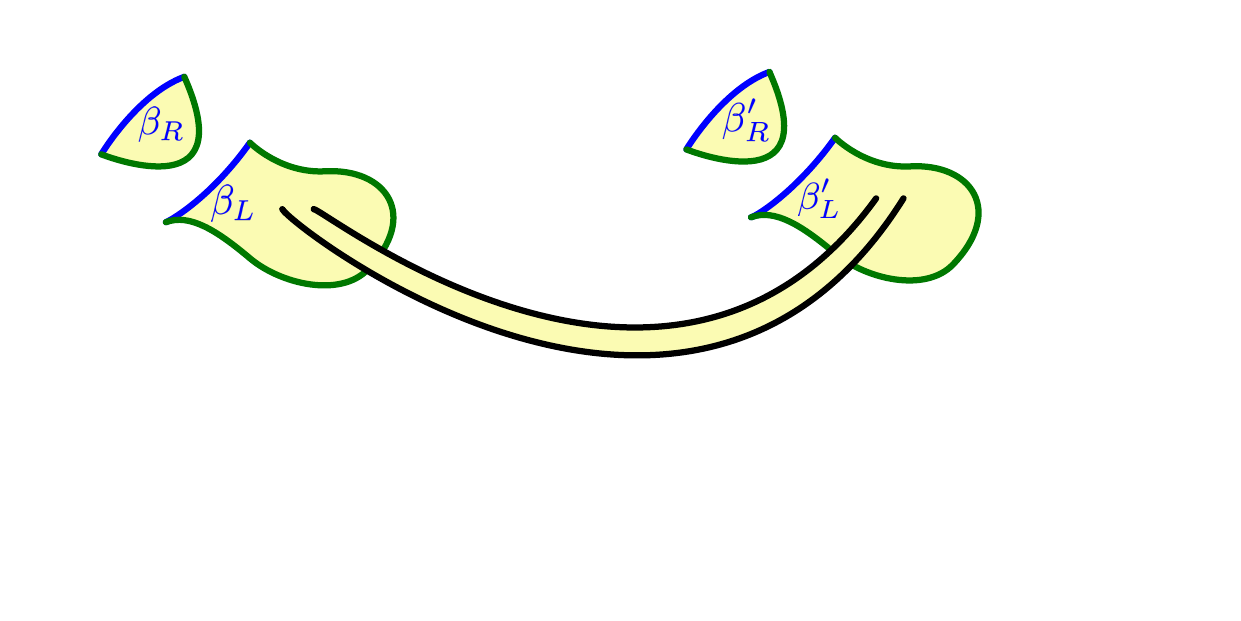}
\times \includegraphics[valign=c,width=0.24\textwidth]{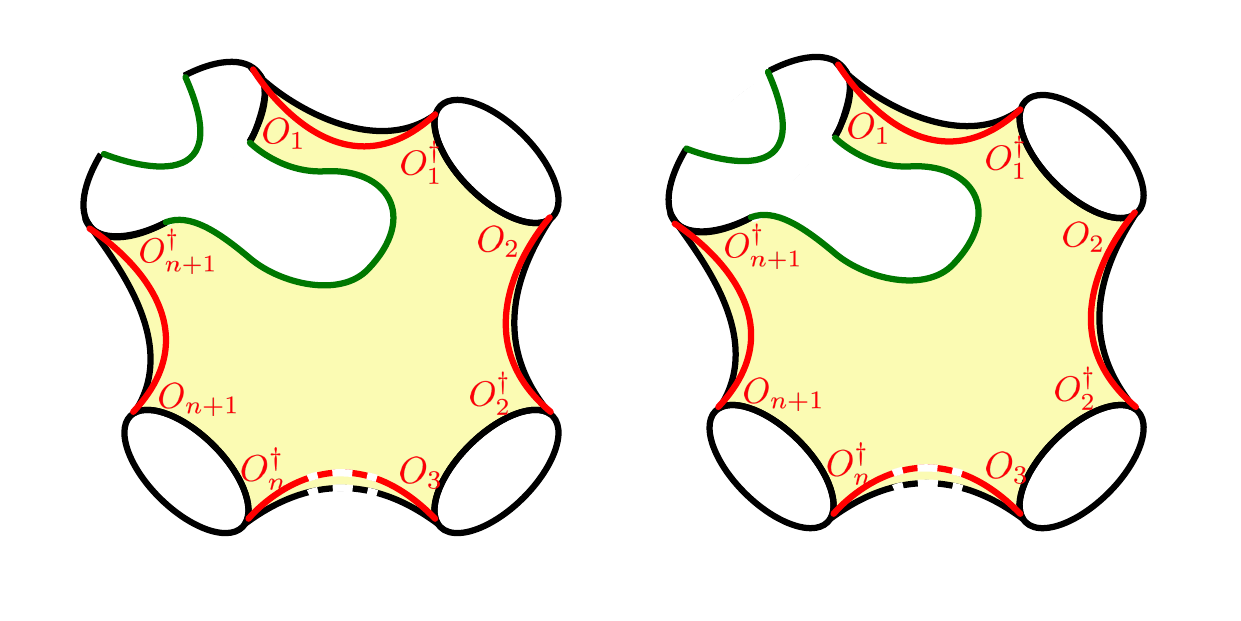}\,.
\label{eq:cutting-rules}
\end{align}
This again allows us to repackage all the higher genus corrections coming from empty wormholes into a single four-boundary partition function 
\begin{align}
\Phi_4(\ell_L,\ell_R,\ell'_L,\ell'_R)
&\equiv 
\overline{    
\varphi_{\beta_L}(\ell_L) \varphi_{\beta_R}(\ell_R) 
\varphi_{\beta'_L}(\ell'_L) \varphi_{\beta'_R}(\ell'_R)} \nonumber
\\
&= 
\includegraphics[valign=c,width=0.30\textwidth]{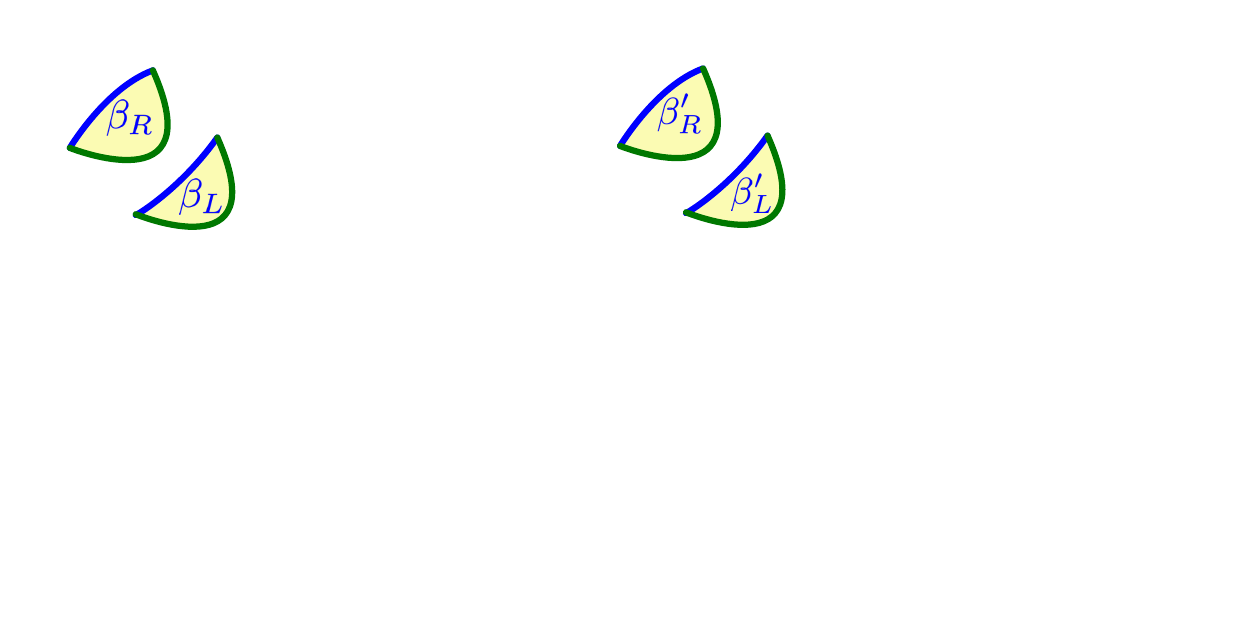}
+ \includegraphics[valign=c,width=0.30\textwidth]{2copieshandlecut.pdf} + \dots \nonumber \nonumber \\
& = \int dE_L\, dE_R \,dE_L'\, dE_R' \, \overline{\rho(E_L) \rho(E_R) \rho(E_L') \rho(E_R')} \times \nonumber \\ &\qquad \qquad \times  \varphi_{E_L}(\ell_L) \varphi_{E_R}(\ell_R) 
\varphi_{E'_L}(\ell'_L) \varphi_{E'_R}(\ell'_R) e^{-\left(\beta_L E_L+ \beta_L' E_L'+ \beta_R E_R+ \beta_R' E_R'\right)}\,,
\end{align}
where each of the boundaries consists of an asymptotic boundary and a geodesic boundary. Note that the four green geodesics are always chosen to be homotopic to the union between boundary segments and the matter geodesics traveling between these boundary segments. On a hyperbolic geometry, there is a unique geodesic connecting the two boundary points separated by the segment of renormalized length $\beta_L$, $\beta_R$, $\beta_L'$, or $\beta_R'$ that satisfies this criterion. For this reason, cutting the geometry as in \eqref{eq:cutting-rules} and individually computing the contributions of the now disconnected surfaces does not suffer from any mapping class group ambiguity.

Above, the entire topological expansion has been repackaged in the four-energy spectral correlator, 
$\overline{\rho(E_L) \rho(E_R) \rho(E_L') \rho(E_R')}$. With this, we can now simply resum the resolvents at the level of the cutout pinwheels, which leads us to a simple result, where for brevity, we denote $\Phi_4 (\{ \ell_i \}) \equiv \Phi_4(\ell_L,\ell_R,\ell'_L,\ell'_R)$:
\begin{align} 
\overline{Z_{\text{bulk}}Z'_{\text{bulk}}} &= 
\oint \frac{\dd \lambda_1 \dd \lambda_2}{(2\pi i)^2 \lambda_1 \lambda_2} 
\int \prod_{i=1}^{4} \dd \ell_{i} e^{\ell_{i}}  
\Phi_4(\{ \ell_i \})
O_{\lambda_1}(\ell_L,\,\ell_R)
O_{\lambda_1}(\ell_L',\,\ell_R')
\\ 
&=
\int \prod_{i=1}^{4} \dd \ell_{i} e^{\ell_{i}}  
\Phi_4(\{ \ell_i \})
\mathcal{O}(\ell_L,\,\ell_R)
\mathcal{O}(\ell_L',\,\ell_R')
.
\end{align}
Now all we need to do is to act on the above result with the differential operator $\mathcal{D}_{\beta_L,\beta_R,\beta_L^{\prime},\beta_R^{\prime}}$. This operator acts only on the $\Phi_4(\{ \ell_i \})$ function under the integral, leading to the result 
\begin{align}
\overline{d(\beta_L, \beta_R)} = 
\int \prod_{i=1}^{4} \dd \ell_{i} e^{\ell_{i}}  
\mathcal{D}\left(\Phi_4(\{ \ell_i \})\right)
\mathcal{O}(\ell_L,\,\ell_R)
\mathcal{O}(\ell_L',\,\ell_R'),
\end{align}
where 
\begin{align}
   \mathcal{D}\left(\Phi_4(\{ \ell_i \})\right) 
    &\equiv \mathcal{D}_{\beta_L,\beta_R,\beta_L^{\prime},\beta_R^{\prime}} \Phi_4(\{ \ell_i \} |_{\beta'_{L/R} = \beta_{L/R}}
    \\
    &=\int \prod_{i=1}^4 \dd s_i \varphi_{s_i}(\ell_i) e^{-\frac{\beta}{4}s_i^2} \overline{\rho(s_L)\rho(s_R)\rho({s'_L})\rho({s'_R})} s_L^2(s_R^2-{s'_R}^2) e^{-\frac{\beta_L}{2}(s_L^2+{s'_L}^2)} e^{-\frac{\beta_R}{2}(s_R^2+{s'_R}^2)} \,.
\end{align}
Assembling everything together and noticing that $\mathcal{O}(\ell_1,\ell_2)\mathcal{O}(\ell_3,\ell_4)$ actually factorises to four identical pieces at leading order in $K$, one can conclude that the differential equation evaluates to zero because the integral is fully symmetric in exchanging any $s_i$'s except for the antisymmetric piece $s_L^2(s_R^2-{s'_R}^2)$. This leads us to conclude that at leading order in $K$ but to all orders in $e^{-S_0}$\,,
\be 
\overline{d(\beta_L, \beta_R)}= 0\,,
\ee
with all $1/K$ corrections vanishing in the large $K$ limit.

\paragraph{Proving step 3:} With this, we are now ready to establish the Hilbert space factorisation in each member of the ensemble. We do this by showing
\be 
\label{eq:d-squared-definition}
\overline{d(\beta_L, \beta_R)^2} = \mathcal{D}_{\beta_L^{(1)},\beta_R^{(1)},\beta_L^{(2)},\beta_R^{(2)}} \mathcal{D}_{\beta_L^{(3)},\beta_R^{(3)},\beta_L^{(4)},\beta_R^{(4)}}
\overline{Z_{\text{bulk}}Z'_{\text{bulk}}Z''_{\text{bulk}}Z'''_{\text{bulk}}}
|_{\beta_{L/R}^{(1\dots 4)}=\beta_{L/R}}= 0 
.
\ee
The computation now is exactly analogous to the proof of \textbf{Step 2}, with the only difference being that we now need to consider the fourth power of $Z_{\text{bulk}}$. The leading order in $K$ again allows us to use the cutout pinwheel construction, which now expresses the final answer in terms of an eight boundary partition function $\Phi_8 (\{ \ell_i \})$
\begin{align} 
\overline{Z_{\text{bulk}}Z'_{\text{bulk}}Z''_{\text{bulk}}Z'''_{\text{bulk}}} =
\int \prod_{i=1}^{8} \dd \ell_{i} e^{\ell_{i}}  
\Phi_8(\{ \ell_i \})
\mathcal{O}(\ell_1 ,\ell_2)
\mathcal{O}(\ell_3 ,\ell_4)
\mathcal{O}(\ell_5 ,\ell_6)
\mathcal{O}(\ell_7 ,\ell_8)
,
\end{align} 
where $\Phi_8 (\{ \ell_i \})$ can also be simply expressed in terms of eight-point spectral correlator $\overline{\prod_{i=1}^8 \rho(E_i)}$:
\begin{align}
\Phi_8(\{\ell_i\})
&\equiv 
\overline{\left(    \prod_{i=1}^4 
\varphi_{\beta_L^{(i)}}(\ell_{2i-1}) \varphi_{\beta_R^{(i)}}(\ell_{2i}) \right)
} 
 = \int \left(\prod_{i=1}^8 dE_i \varphi_{E_i}(\ell_i) \right)\, \overline{\prod_{i=1}^8 \rho(E_i)} \times \nonumber
e^{-\sum_{i=1}^4 \left(\beta_L^{(i)} E_{2i-1}+ \beta_R^{(i)} E_{2i}\right)}.
\end{align}
Applying the differential operators in \eqref{eq:d-squared-definition} to the equation above and using the fact that the spectral correlator $\overline{\prod_{i=1}^8 \rho(E_i)}$ is symmetric under the exchange of any two of the energies, we again find a vanishing answer. Thus, at leading order in $K$, the symmetry of the resulting integral leads us to conclude 
\be 
\overline{d(\beta_L, \beta_R)^2} = 0\,,
\ee
where, once again, all $1/K$ corrections vanish in the large $K$ limit. 
A similar argument applies for an arbitrary product of bulk traces in the $K\to \infty$ limit. Explicitly, we find 
\begin{align}
&\hspace{0.5cm}\overline{\Tr_{\mHb}(e^{-\beta_L^{(1)} H_L} e^{-\beta_R^{(1)} H_R})\dots \Tr_{\mHb}(e^{-\beta_L^{(n)} H_L} e^{-\beta_R^{(n)} H_R})} = \nn \\&= 
\overline{\Tr_{L}(e^{-\beta_L^{(1)} H_L}) \Tr_R(e^{-\beta_R^{(1)} H_R})\dots \Tr_{L}(e^{-\beta_L^{(n)} H_L}) \Tr_R( e^{-\beta_R^{(n)} H_R})}\nonumber \\ &= \int dE_{L}^{(1)} \dots dE_{L}^{(n)} dE_{R}^{(1)} \dots dE_{R}^{(n)} \overline{\rho(E_L^{(1)})\dots \rho(E_R^{(n)})} e^{-\beta_L^{(1)} E_L^{(1)} -\dots - \beta_R^{(n)} E_R^{(n)}}\,,
\end{align}
which further confirms that the factorisation of the Hilbert space occurs in each member of the ensemble or before coarse-graining.

At large $K$, factorisation is achieved due to the fact that the spectral correlator is symmetric under the exchange of energies. This should be contrasted to the limit considered in section \ref{sec:fact-leading-order}, where knowing the exact contribution of each geometry was necessary in order to determine the transition point at which the bulk trace factorises. This suggests that, at large $K$, factorisation is generic. For example, even in our two-dimensional gravitational theory one might be concerned about the effect of dynamical matter fields on the wormholes that are not supported by any matter geodesics. Such geometries are known to have divergent contributions as the size (i.e., the length of the closed geodesic) of the wormhole shrinks \cite{Saad:2019lba}. Presumably, UV corrections regulate such divergences, though, at this time, the form of the necessary corrections is unknown. Nevertheless, while such a regularization can correct the spectral correlator, it should not change its symmetry property under energy exchange. Because of this, the factorisation of the Hilbert space should continue to hold.

\subsection{A classification of geometries at subleading orders in $K$}
\label{sec:classif-geometries-subleading-in-K}

To finalize our analysis, we wish to clarify which geometries will survive at different orders in the two limits ($K \to \infty$, $d^2\to \infty$, and $K/d^2 \sim O(1)$, or $K/d^2 \to \infty$) considered thus far. This will clarify why it was correct to consider the wormhole geometries discussed in section \ref{sec:fact-through-non-fact} and \ref{sec:fact-for-each-member-of-the-ensemble} and will help us systematically understand how to further make progress when $K/d^2 \sim O(1)$.  To begin, let us list all the types of geometries that contribute to the exact resolvent. The following classification proves useful: we will classify geometries by whether the contraction of the matter indices is planar or non-planar, whether there are wormhole connections not supported by matter geodesics among the various disconnected components in the resolvent, and whether after cutting along the matter geodesics connecting the different boundaries we obtain a set of surfaces with trivial or non-trivial topology. Thus, we have:
\begin{itemize}
    \item[(a1)] \textbf{Planar trivial topology}. These are precisely the kinds of geometries that we have considered in the resolvent in section \ref{sec:fact-leading-order}. If one cuts the geometries along the matter geodesics, one obtains a set of disconnected surfaces with trivial topology. After the cut, the loops appearing in the summation over the matter indices should yield a planar graph. Examples of such surfaces when focusing on cases with four boundaries include, 
 \be   \includegraphics[valign=c,width=0.25\textwidth]{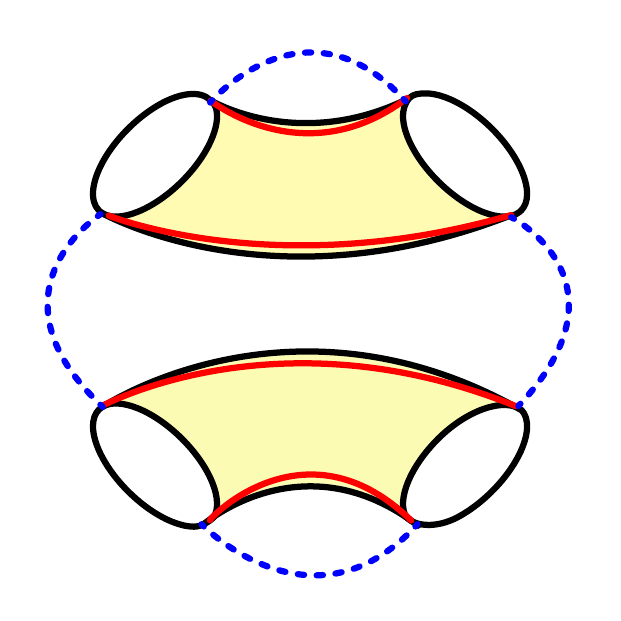},\qquad \includegraphics[valign=c,width=0.25\textwidth]{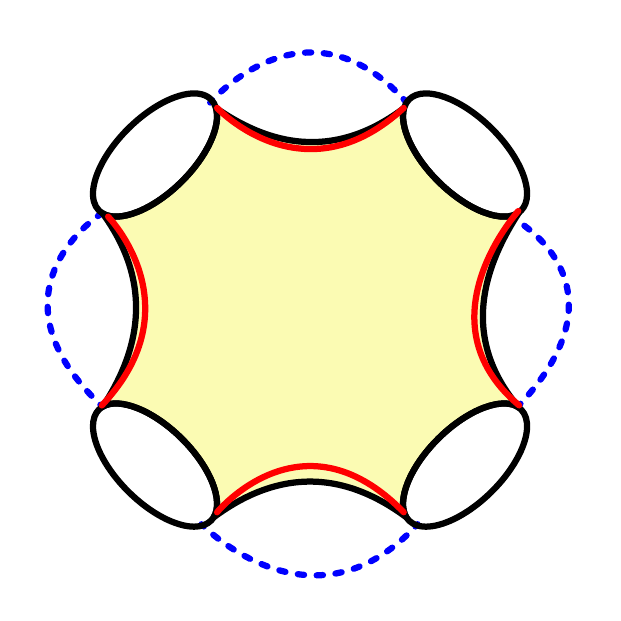}
 \ee
 All such surfaces contribute to leading order in $K$ or $d^2$ in the limit $K \to \infty$, $d^2\to \infty$, and $K/d^2 \sim O(1)$. The fully connected ones contribute to leading order in $K$ and then leading order in $d^2$ when  $K \to \infty$, $d^2\to \infty$, and $K/d^2 \to \infty$.

\item[(a2)] \textbf{Non-planar trivial topology}. If one cuts the geometries along the matter geodesics, one once again obtains a set of disconnected surfaces with trivial topology. However, after the cut, the loops appearing in the summation over the matter indices should yield a non-planar graph. With four boundaries, we should, for example, include 
     \be   \includegraphics[valign=c,width=0.25\textwidth]{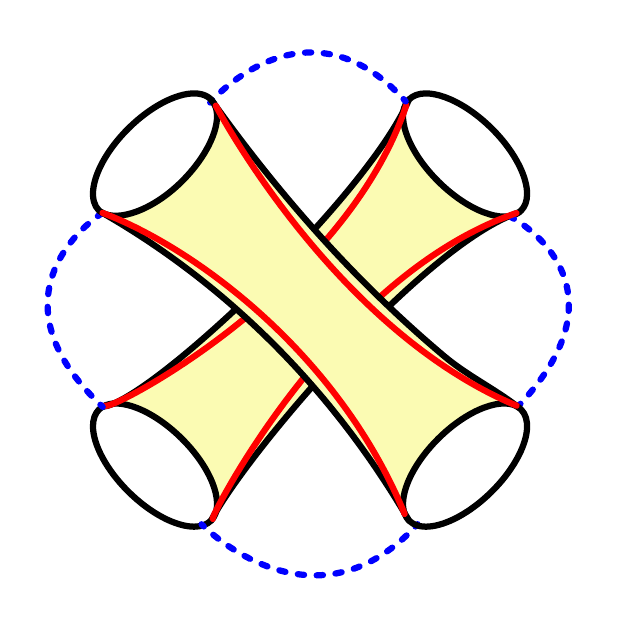},\qquad 
 \ee
Compared to the geometries from (a1) that we included in section \ref{sec:fact-leading-order}, such surfaces contribute to subleading order in $K$ or $d^2$ in the limit $K \to \infty$, $d^2\to \infty$ and $K/d^2 \sim O(1)$, starting at order $1/K$ in the resolvent and the bulk trace, and only contribute to subleading orders $K$, starting at $O(1/K)$ in the resolvent and the bulk trace, when $K/d^2 \to \infty$. For this reason, we have not included such surfaces in our analysis thus far. 
    \item[(b1)] \textbf{Planar disconnected nontrivial topology.} Planar disconnected geometries with non-trivial topology are obtained by taking the geometries in point (a1) and adding handles such that any two surfaces that were disconnected in point (a1) remain disconnected. In this construction, if one cuts the geometries along the matter geodesics, one obtains a set of surfaces where at least some have non-trivial topology, but the loop appearing in their index contraction still forms a planar graph.  Such geometries include, 
         \be   \includegraphics[valign=c,width=0.25\textwidth]{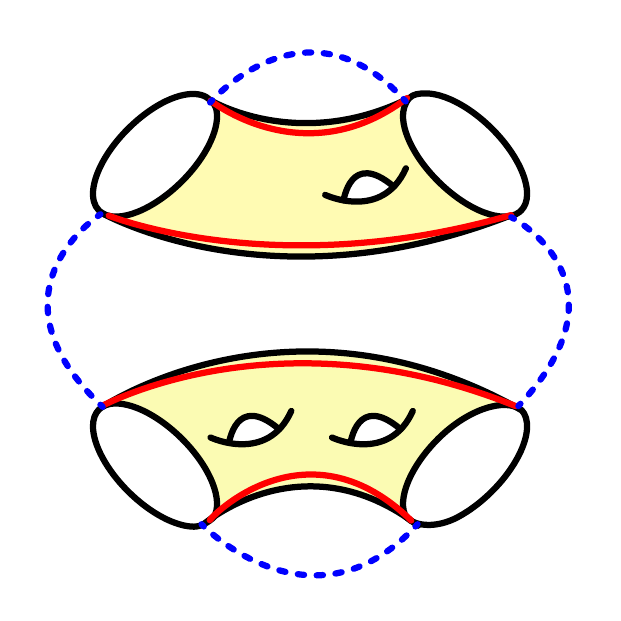},\qquad \includegraphics[valign=c,width=0.25\textwidth]{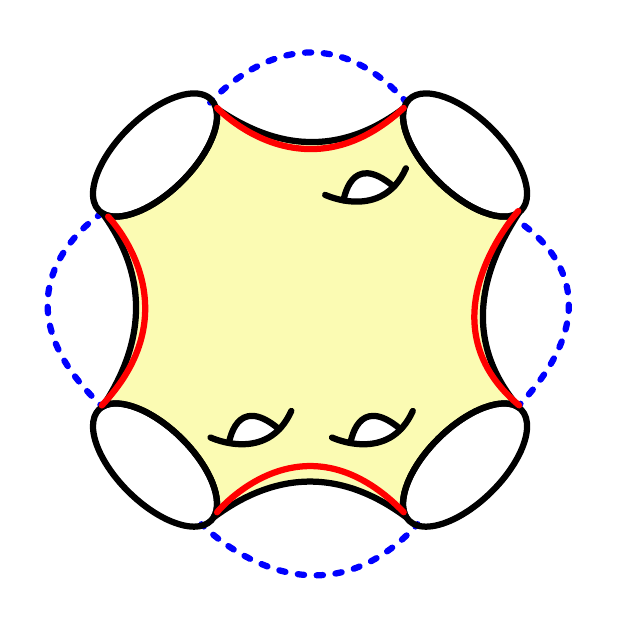}
 \ee
Compared to the geometries from (a1) that we included in section \ref{sec:fact-leading-order},  such surfaces contribute to the subleading orders in $K$ or $d^2$, starting at $O(1/K)$ in both the resolvent and the bulk trace, in the limit $K \to \infty$, $d^2\to \infty$, and $K/d^2 \sim O(1)$. However, in the limit $K/d^2 \to \infty$, when such geometries are fully connected, they do contribute to leading order in $K$, albeit to subleading order in $d^2$. This is the reason why we included these geometries in our analysis in section \ref{sec:fact-through-non-fact} and \ref{sec:fact-for-each-member-of-the-ensemble}. 
    
    \item[(b2)] \textbf{Non-planar disconnected nontrivial topology}. The geometries in (b2) are the same as those in (b1) both before and after cutting along the matter geodesic, but the loop appearing in their index contraction forms a non-planar graph. Examples of such geometries include, 
    \be   \includegraphics[valign=c,width=0.25\textwidth]{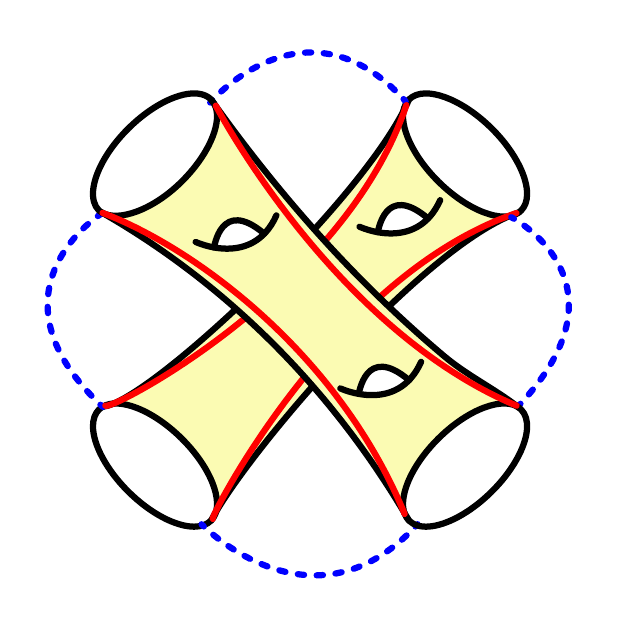},\qquad 
 \ee
 Once again, such surfaces contribute to the first subleading order in $K$ both in the limit $K \to \infty$, $d^2\to \infty$, and $K/d^2 \sim O(1)$, starting at $O(1/K^2)$ in both the resolvent and the trace,  and when $K\to \infty$, starting at $O(1/(d^2 K^2))$. This is why we have neglected these geometries thus far. 
    
    \item[(c1)] \textbf{Planar connected nontrivial topology.} Planar-connected geometries with non-trivial topology are obtained by taking the geometries in point (b1) and adding handles between any two of the disconnected surfaces considered in (b1).  Such geometries include, 
        \be   \includegraphics[valign=c,width=0.25\textwidth]{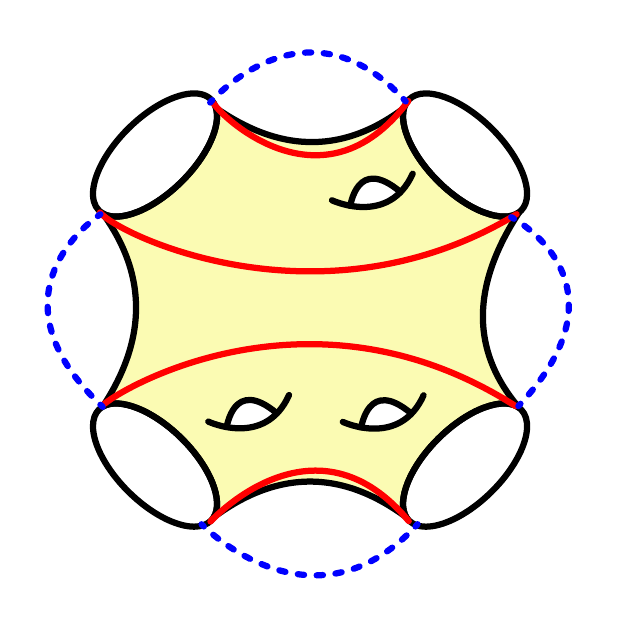},\qquad 
 \ee
 and contribute at subleading order in $K$, starting at $O(1/(K d^2))$ for both the resolvent and the trace, in the limit $K \to \infty$, $d^2\to \infty$, and $K/d^2 \sim O(1)$ or $K/d^2 \to \infty$. Once again, we have not considered any one of these geometries so far. 
    \item[(c2)] \textbf{Non-planar connected nontrivial topology.}  Once again, the geometries in (c2) are the same as those in (c1) both before and after cutting along the matter geodesic, but the loop appearing in their index contraction forms a non-planar graph. These include, 
    \be   \includegraphics[valign=c,width=0.25\textwidth]{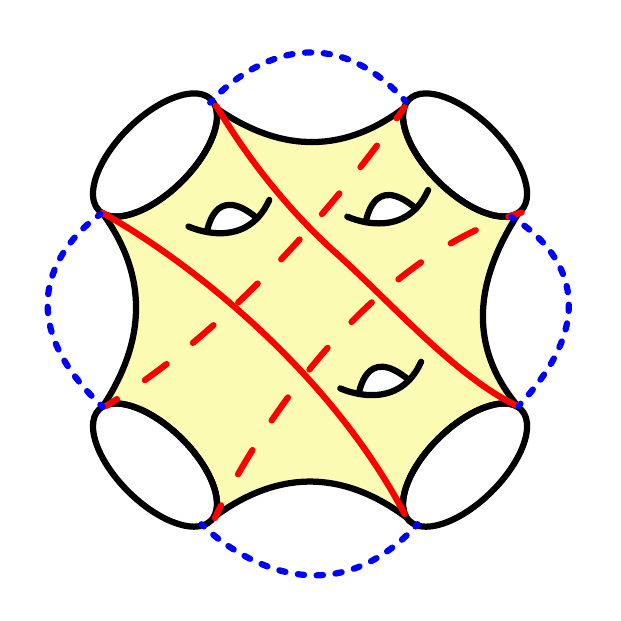}.\qquad 
 \ee
 They contribute to subleading order in $K$ both in the limit $K \to \infty$, $d^2\to \infty$, and $K/d^2 \sim O(1)$, starting at $O(1/K^3)$ in both the resolvent and the trace,  and when $K\to \infty$, starting at $O(1/(d^2 K^2))$.
\end{itemize} 
The classification above confirms that the only geometries that contribute to the bulk trace at leading order in $K$ in the limit $K \to \infty$, $d^2\to \infty$, and $K/d^2 \to \infty$ are of type (a1) and (b1), confirming that the computations in section \ref{sec:fact-through-non-fact} and \ref{sec:fact-for-each-member-of-the-ensemble} are indeed correct. To further understand the effects of the remaining geometries on the resolvent and the trace, we will comment on how they correct the Schwinger-Dyson equation and its relation to the bulk trace. Nevertheless, solving these equations to arbitrary order in $K$ and $d^2$ goes beyond the purpose of this paper. To exemplify the power of the classification above, we can resum the geometries above and find the next order correction to the resolvent in the large $K/d^2$ limit. 
Surprisingly, by resuming the geometries (a1), (a2), (b1) and (c1) that all appear at the next order in $K$ (i.e. order $K^{-1}$), we find that close to $\lambda=0$, the resolvent takes a very similar form:
\be 
\label{eq:R(lambda)-subleading}
\overline{R}(\lambda) = \underbrace{\frac{K - \overline{d^2}}{\lambda}}_{\substack{\equiv R_{(a1)\&(b1)}(\lambda) \\ \text{Resummation of}\\ \text{(a1) \& (b1)}}} - \underbrace{R_0}_{\substack{\text{Corrections from}\\ \text{(a1), (a2), (b1), \& (c1) }}}  + \,\,\quad O\left(\frac{1}{K^2}\right)\,,
\ee 
where $
\overline{d^2} = \int_{\mathcal E} dE_1 dE_2\, \overline{\rho(E_1)\rho(E_2)}
$ and $R_0$ is the analytic piece of the resolvent at $\lambda=0$.  A detailed derivation of this result is provided in appendix $\ref{sec:subleadinginK}$.  The terms in \eqref{eq:R(lambda)-subleading} have a nice physical meaning. The dimension of the Hilbert space is corrected in the topological expansion from $d^2 \to \overline{d^2}$, leading to the change in the first term compared to the result obtained by resumming only the (a1) geometries in section \ref{sec:fact-leading-order}. This change comes from resuming all (a1) and (b1) geometries with an arbitrary genus and an arbitrary index contractions.  Thus, the average number of null states is given by $K - \overline{d^2}$, as expected since the average dimension of the Hilbert space is $\overline{d^2}$. The geometries (a2) and (c1) do not contribute to the residue of the resolvent, and consequently, they do not correct the number of null states. We suspect that this result persists to all perturbative orders in $1/K$ even when including the (b2) and (c2) geometries.

From \eqref{eq:R(lambda)-subleading}, we can also recompute the bulk trace. 
The geometries contributing upto the order $K^{-1}$ are still of the type (a1),(a2),(b1)\& (c1). 
The geometries (a1) \& (b1) are easily resummed, and the relation between the trace and the resolvent remains unchanged compared to the leading order case, meaning 
\begin{equation}
   \overline{\text{Tr}_{\mHb}(k_L k_R)}_{(a1)\& (b1)} =  e^{2S_0} 
\int ds_L ds_R \overline{\rho(s_L) \rho(s_R) }
 \oint_{C_2}
 \frac{\dd \lambda}{2\pi \ii}\,
 \frac{1}{\lambda}
\frac{{R}_{\text{(a1)\&(b1)}}(\lambda) y_k(s_L,s_R)}{1-{R}_{\text{(a1)\&(b1)}} (\lambda) y_q(s_L,s_R)}
\end{equation}
which also equals to $\overline{\text{Tr}_{L}(k_L) \text{Tr}_{R}(k_R)}$ via plugging in ${R}_{\text{(a1)\&(b1)}}(\lambda)$ from \eqref{eq:R(lambda)-subleading}. 

Just like the geometries  (a2),  \& (c1) did not contribute to the residue of the resolvent at $\lambda=0$, we similarly show that they do not affect the bulk trace once $K>\overline{d^2}$. Thus, we find  
\begin{equation}
\begin{aligned}
\overline{\text{Tr}_{\mHb}(k_L k_R)}  & = \overline{\text{Tr}_{\mHb}(k_L k_R)}_{(a1)\& (b1)} + O\left(\frac{1}{K^2}\right)  
\,, 
\end{aligned}
\end{equation} 
This once again confirms the trace factorisation to all non-perturbative orders in $1/G_N$, now to a further subleading order in $1/K$. Importantly, there are no additional corrections at order $K^{-1}$ to $\overline{\text{Tr}_{\mHb}(k_L k_R)} = \overline{\text{Tr}_{L}(k_L) \text{Tr}_{R}(k_R)}$, although we work in the geometries that are supposed to contribute at this order. 
Similar to our expectation for the resolvent, we believe that the trace factorisation relation $\overline{\text{Tr}_{\mHb}(k_L k_R)} = \overline{\text{Tr}_{L}(k_L) \text{Tr}_{R}(k_R)}$ is not affected by the (a2), (b2), (c1), and (c2) geometries even at further perturbatively subleading orders in $1/K$.

The above discussion addresses step 1; for the differential equations in points 2 and 3, our results prompt us to similarly conjecture that only the (a1) and (b1) geometries contribute and $\overline{d(\beta_L,\beta_R)}$ and $\overline{d(\beta_L,\beta_R)^2}$ vanishes without any correction in the sense of $1/K$ perturvative expansion.

\section{Hilbert space factorisation for BPS black holes}
\label{sec:factorisation-BPS-BHs}

We now turn to the case of the ground state  BPS sector in $\mathcal{N}=2$ super JT coupled to matter. Here, as a result of the nonzero energy gap in the spectrum between BPS states and lowest-lying excited states, this setup has the advantage of having a dimension of the BPS Hilbert space that's the same in all members of the ensemble \cite{Iliesiu:2021are, Boruch:2023trc}. This once again means that by choosing the size of the basis $K$ large enough but finite, we can be assured that we're spanning the Hilbert space in all members of the ensemble, in contrast to the finite energy sectors where the number of states within the energy window can fluctuate as we move between members of the ensemble. 

To probe factorisation, we again want to compute the trace over $\mathcal{H}_\text{bulk} (K)$  spanned by states $\ket{q_i}_{i=1,\dots,K}$. The states $\ket{q_i}$ are prepared as before, with the crucial difference that now they are further projected onto the BPS sector by performing infinite Euclidean time evolution on the asymptotic on both sides of the operator insertion. This means that we will not be able to probe factorisation anymore by computing simply $\Tr_{\mH_{\text{bulk}}(K)}(e^{-\beta_L H_L} e^{-\beta_R H_R})$ as the Hamiltonians act trivially on all states. Therefore, in this section, we shall consider two generic matter operators, $k_L$  and $k_R$, acting (geometrically) on the left and right, respectively. The statement of factorisation that we shall thus prove will be that, 
\be
{\Tr_{\mH_{\text{bulk}}}(k_Lk_R)}={\Tr_{\mH_L}(k_L)\Tr_{\mH_R}(k_R)} \,,
\label{eq:factorisation-BPS}
\ee
for any matter operator $k_L$ and $k_R$ in the bulk Hilbert space. The operators $k_L$ and $k_R$ can be arbitrarily complicated and, in particular, to make the analogy between the previous sections and the BPS results clear we can take,
\be 
k_L = e^{-\alpha_L K_L}\,, \qquad k_R = e^{-\alpha_R K_R},
\ee
for two simple operators $K_L$ and $K_R$ projected to the BPS sector. In such a case, our task is thus to prove that the trace factorises into a product of two functions, one of $\alpha_L$ and the other of $\alpha_R$. 
The trace in \eqref{eq:factorisation-BPS} will once again be computed in the $\mathcal H_\text{bulk}(K)$ basis, which becomes the full bulk Hilbert space $\mathcal H_\text{bulk}$ for large enough $K$.
We will show that this indeed happens by expliclty verifying the analogs of steps 1--3 discussed in previous sections. As before, we will restrict ourselves to geometries in which matter geodesics do not intersect each other, a simplification that's expected to be a good approximation to an exact answer for large scaling dimensions of the operators. In previous sections, dropping this assumption is expected to only modify the details for the resolvent in a minor way. In the current computation, however, as we are interested in correlation functions of $k_L$ and $k_R$ operators, we expect there might be an important part of physics captured by the intersecting matter geodesics connecting $k_L$ and $k_R$. As the interplay of this type of geometries with the resolvent is not yet well understood, we will restrict ourselves to the class of operators that cannot be connected together by a matter geodesic. In other words, we're working in the case where 
\be
\label{eq:kL-kR-assumptions}
\langle k_L k_R \rangle_{\text{disk}} 
= \langle k_L O_{i}\rangle_{\text{disk}}  
= \langle k_R O_{i}\rangle_{\text{disk}}
=0. 
\ee
where the operators $O_i$ are the matter excitations used in the construction of $\ket{q_i}$.\footnote{For instance, the two-point functions in \eqref{eq:kL-kR-assumptions} if all $O_i$, $K_L$ and $K_R$  lie in different irreps of the superconfromal group.}
In the following, we will be brief about the details of $\mathcal{N}=2$ super JT and only introduce the aspects relevant for generalizing computations from previous sections. For more detailed explanations of $\mathcal{N}=2$ super JT and its relation to a higher dimensional black hole, we refer the reader to the literature \cite{Boruch:2022tno}.

\subsection{Conventions}
\label{sec:BPSconventions}

Let us briefly describe the main objects required for generalizing previous computations to the BPS sector of $\mathcal{N}=2$ super JT \cite{Stanford:2017thb,Mertens:2017mtv,LongPaper,ShortPaper,Turiaci:2023jfa,Belaey:2024dde}. Working in a sector of a fixed R-charge $j$, quantized in units of $1/\hat{q}$, the canonical quantization of the theory \cite{LongPaper,ShortPaper} leads to a Hamiltonian dependent on two bosonic variables $(\ell_{12}, a_{12})$, which describe the bulk geodesic length variable and the holonomy of the $U(1)$ gauge field between the two boundaries respectively, and two fermionic variables $(\xi_{12},\eta_{12})$ which can be identified as superpartners of the geodesic length with respect to left and right boundary. The Hartle-Hawking wavefunctions are now states with zero energy of the bulk Hamiltonian, which we'll denote as $\Psi^j_{12} \equiv \Psi^j_{12}(\ell_{12},a_{12},\xi_{12},\eta_{12})$ for the ket state and as $\Psi^j_{21} \equiv \Psi^j_{21}(\ell_{12},a_{12},\xi_{12},\eta_{12})$ for the bra state (which are not exactly the same because the explicit form of the wavefunctions distinguishes the boundaries). 
These wavefunctions can be prepared by performing a gravitational path integral on a half-disk, with AdS boundary conditions on the asymptotic boundary of proper length $\beta \rightarrow \infty$, and $(\ell_{12}, a_{12},\xi_{12},\eta_{12})$ boundary conditions on the bulk geodesic boundary. Similarly, the basis states $\ket{q}_i$ can be prepared with the same boundary conditions with an additional insertion of the operator $O_i$, followed by an infinite Euclidean time evolution on both sides
\be
\ket{\Psi}=\includegraphics[valign=c,width=0.2\textwidth]{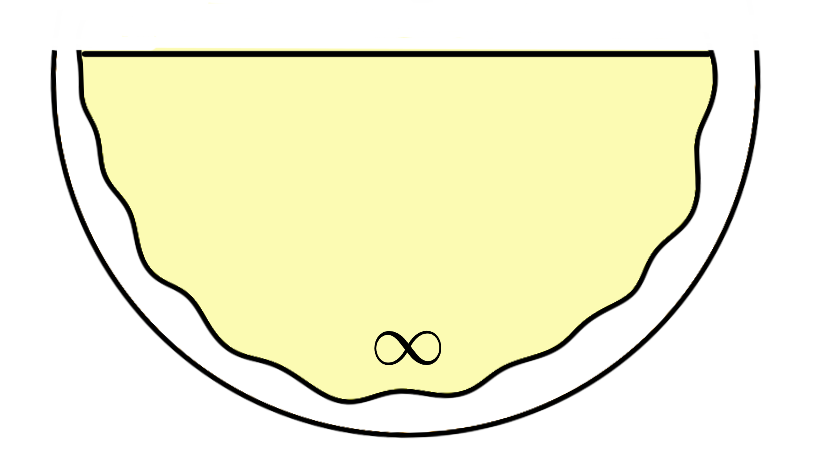}\,,\quad\quad \ket{q_i}
=\includegraphics[valign=c,width=0.2\textwidth]{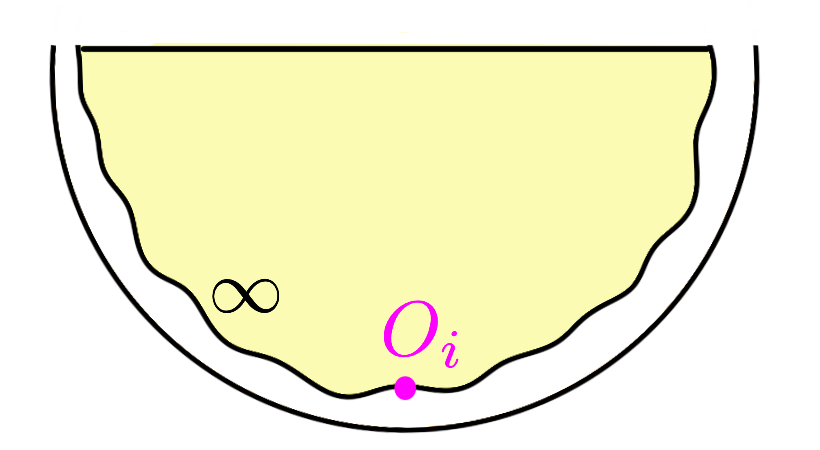}\,.
\label{eq:SUSY-state-qi} 
\ee
Two copies of Hartle-Hawking wavefunctions can now be glued with the measure 
\be  
\int \dd \mu =
\frac{1}{2 \hat{q}} \int_{-\infty}^{\infty} \dd \ell \, \int_{0}^{2\pi \hat{q}} \dd a \, \int \dd \eta \, \int  \dd \xi . 
\ee
Introducing a tilded version of the Hartle-Hawking wavefunctions 
\be 
\Tilde{\Psi}^j_{12} \equiv 
\frac{\Psi^j_{12}}{\cos(\pi j)}  ,
\ee
we have a simple relation 
\be
\int d\mu_{12} \,\Psi^j_{12} \Tilde{\Psi}^{j'}_{21} = \delta_{j\, j'} .
\ee
The two-point functions are again computed by gluing together two wavefunctions with an additional insertion of $e^{-\Delta \ell}$ 
\be
\int d\mu_{12}\,\Psi_{12}^j
\Psi_{21}^j
e^{-\Delta\ell_{12}} =\frac{\cos^2(\pi j)}{2\pi} \frac{\Delta \Gamma(\Delta)^2}{\Gamma(2\Delta)} \Gamma\left(\Delta+\frac{1}{2}\pm j\right) 
.
\ee
With this, by appropriately gluing together Hartle-Hawking states with polygon geometries (for details, see \cite{Boruch:2023trc}), one can easily derive the analog of \eqref{eq:n_boundary_pinwheel} as
\be
Z_{j,n}^{\mathcal{O}} =
\includegraphics[valign=c,width=0.2\textwidth]{3pinwheel.pdf}=e^{2S_0} \cos^2(\pi j) y^n ,
\qquad 
y = e^{-S_0}\frac{\Delta \Gamma(\Delta)^2 \,\Gamma\left(\Delta+\frac{1}{2}\pm j\right)}{2\pi \Gamma(2\Delta)} ,
\ee
and the analogue of the pinwheel with additional $(k_L,k_R)$ operator insertions \eqref{eq:n+1_boundary_pinwheel} 
\begin{align}
Z_{j,n+1}^{\mathcal{O}}(k_L,k_R) &=
\includegraphics[valign=c,width=0.2\textwidth]{4pinwheel.pdf}
=
\includegraphics[valign=c,width=0.2\textwidth]{4pinwheelcut.pdf}
=e^{2S_0} \cos^2(\pi j) \langle k_L\rangle \langle k_R\rangle y^{n+1} .
\end{align}
Importantly, we denoted the final expression for $Z_{j,n+1}^{\mathcal{O}}(k_L,k_R)$ in terms of disk one point functions $\langle k_{L/R} \rangle$ of operators $k_{L/R}$ without any factors of $e^{-S_0}$.

\subsection{Hilbert space factorisation for orthogonal operators}
\label{sec:cutfactorise}
We are now ready to compute the bulk trace through the gravitational path integral via
\be 
\overline{\Tr_{\mHb}(k_L k_R)} 
=\oint \frac{\dd \lambda}{2\pi i \lambda}  \overline{\mathbf{R}_{ij} \langle q_i| k_L k_R | q_j\rangle } .
\ee
The computation closely follows the steps outlined in section \ref{subsec:resolvent-and-bulk-trace} with a few important differences. First of all, as had been explained in \cite{Iliesiu:2021are, Boruch:2023trc, Turiaci:2023jfa}, in the BPS sector, the empty trumpet geometry glued to any other geometry vanishes. This means that there are no empty wormholes in this sector, and we do not need to worry about adding empty handles on top of contributing geometries. We can still, however, have wormholes supported by matter in geometry. In contrast with previous sections where the operators $k_L$ and $k_R$ were constructed by using $H_L$ and $H_R$, now the trace involves the insertion of two matter operators, which can, in principle, lead to wormholes connecting the front and the back of the pinwheel geometry. To deal with this, we will use the cutout pinwheel construction discussed in section \ref{sec:resolvent_as_operator}. Following the steps of the previous sections, the pinwheel with all possible matter handles coming from $k_L$ and $k_R$ operators can be consistently included in a multiplicative factor 
\be 
\overline{Z_{j,n+1}^{\mathcal{O}}}(k_L,k_R) 
=e^{2S_0} \cos^2(\pi j) \overline{\langle k_L\rangle \langle k_R\rangle} y^{n+1} ,
\ee
where $\overline{\langle k_L\rangle \langle k_R\rangle}$ denotes now a gravitational path integral in the BPS sector with two asymptotic boundaries, one containing $k_L$ operator insertion and the other containing $k_R$ operator insertion. This now leads to
\begin{align}
\overline{ \mathbf{R}_{ij} \bra{q_i}k_L k_R \ket{q_j} } = \sum_{n=0}^\infty R^{n+1} \overline{Z_{j,n+1}^{\mathcal{O}}}(k_L,k_R) 
= e^{2S_j} \overline{\langle k_L \rangle \langle k_R \rangle} \frac{Ry}{1-Ry} ,
\end{align}
where $e^{2S_j} \equiv e^{2S_0} \cos^2(\pi j)$.
This, together with the Schwinger-Dyson equations for the resolvent, 
\be 
R(\lambda)=\frac{K}{\lambda} + \frac{1}{\lambda} \sum_{n=1}^\infty Z_n^\mathcal{O} R^{n}=\frac{K}{\lambda} + \frac{e^{2S_j}}{\lambda} \frac{R y}{1-R y} ,
\ee
immediately imply 
\begin{align}
\overline{\Tr_{\mHb(K)}(k_L k_R)} 
&= 
\frac{e^{2S_0}  \overline{\langle k_L \rangle \langle k_R \rangle}}{e^{2S_j}} 
\int_{\lambda_-}^{\lambda_+} d\lambda D(\lambda) 
=e^{2S_0}  \overline{\langle k_L \rangle \langle k_R \rangle} 
 \begin{cases}
 K e^{-2S_j} , &K< e^{2S_j}  
 \\ 
 1 , & K> e^{2S_j} 
 \end{cases} 
 ,
\end{align}
where $D(\lambda)$ is the discontinuity of $R(\lambda)$ along the real axis. We, therefore, explicitly see that once we span the BPS Hilbert space, $K>e^{2S_j}$, the bulk trace is consistent with the factorisation of the Hilbert space, i.e. 
\be 
\overline{\Tr_{\mathcal H_\text{bulk}(K>e^{2S_j})}(k_L k_R)} 
= \overline{\Tr_{\mathcal{H}_L}(k_L) \Tr_{\mathcal{H}_R}(k_R)} .
\ee
for any two operators $k_L$ and $k_R$ that satisfy \eqref{eq:kL-kR-assumptions}.
It turns out, however, that we can do slightly better. As we show below, as long as the operators $k_L$ and $k_R$ cannot be connected by a matter geodesic, there can be no nontrivial handles that would contribute to a connected part of $\overline{\Tr_{\mathcal{H}_L}(k_L) \Tr_{\mathcal{H}_R}(k_R)}$. This means that the right-hand side further factorises, and we can simply write
\be 
\overline{\Tr_{\mH_{\text{bulk}}(K>e^{2S_j})}(k_L k_R)} 
= \overline{\Tr_{\mathcal{H}_L}(k_L)} 
\times \overline{\Tr_{\mathcal{H}_R}(k_R)} .
\label{eq:BPS-to-prove}
\ee
If we manage to prove this, the problem then simply reduces to evaluating the one boundary correlators of the appropriate $k_L$ and $k_R$ operators of our choice.

To prove \eqref{eq:BPS-to-prove} we need to show that any configuration contributing to the connected part of $\overline{\Tr_{\mathcal{H}_L}(k_L) \Tr_{\mathcal{H}_R}(k_R)}$ has an empty handle, a wormhole with a closed geodesic that doesn't intersect any matter propagators. Since all such geometries are known to vanish this would imply that $\overline{\Tr_{\mathcal{H}_L}(k_L) \Tr_{\mathcal{H}_R}(k_R)} = \overline{\Tr_{\mathcal{H}_L}(k_L)} \times \overline{\Tr_{\mathcal{H}_R}(k_R)}$. The complication, however, is that $k_L$ and $k_R$ are not simple operators, and there can be arbitrarily complicated matter propagators starting and ending at the location of the $k_{L,R}$ insertion. For example, consider the expansion of $k_{L, R}$ into the simple operators $K_{L,R}$. It is difficult to tell whether the configuration of handles and matter geodesics between the left and right Hartle-Hawking wavefunctions after cutting out the pinwheels containing the operators $O_i$, 
\be 
X = \includegraphics[valign=c,width=0.3\textwidth]{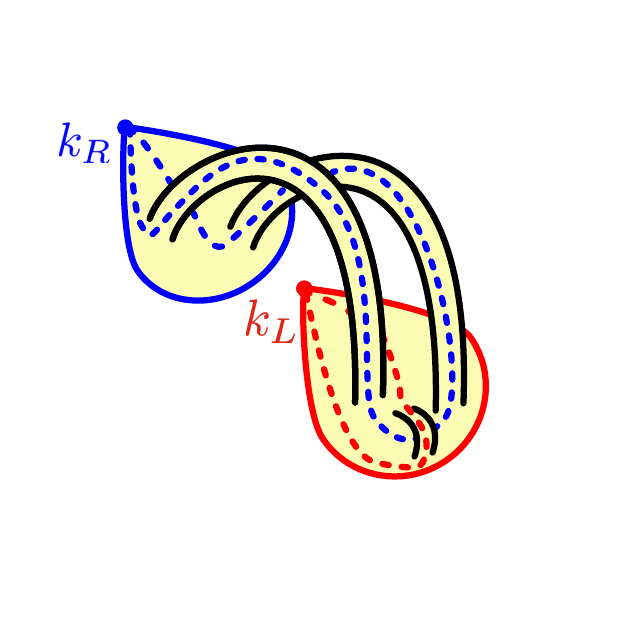}\,,
\ee 
has an empty handle that would cause the geometry to vanish. Above, we have represented the geodesic boundary of each Hartle-Hawking wavefunction by a solid curve and the matter geodesics by dotted curves. 

The proof that such an empty handle always exists proceeds as follows. Suppose we have $g$ handles connecting the left and right Hartle-Hawking wavefunctions after cutting out the pinwheels containing the operators $O_i$: 
\be
X = \includegraphics[valign=c,width=0.3\textwidth]{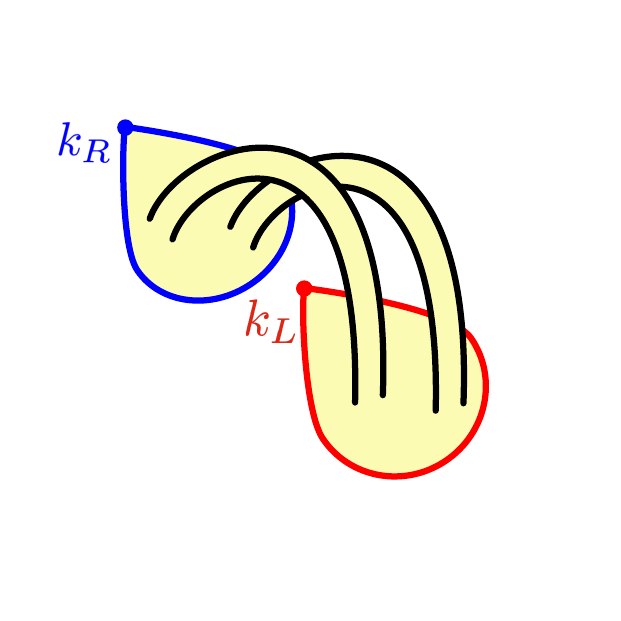}\,,
\label{eq:X-surface}
\ee
where the blue and red are geodesic boundaries starting and ending at the location of the $k_{L,R}$ insertion. Here, we have not drawn the matter geodesics starting and ending at $k_{L,R}$.
$X$ is topologically a genus-$g$ surface with two disks cut out. As in \eqref{eq:X-surface}, we color one of the boundary circles red, and the other boundary circle blue. 
\be
X=\includegraphics[valign=c,width=0.6\textwidth]{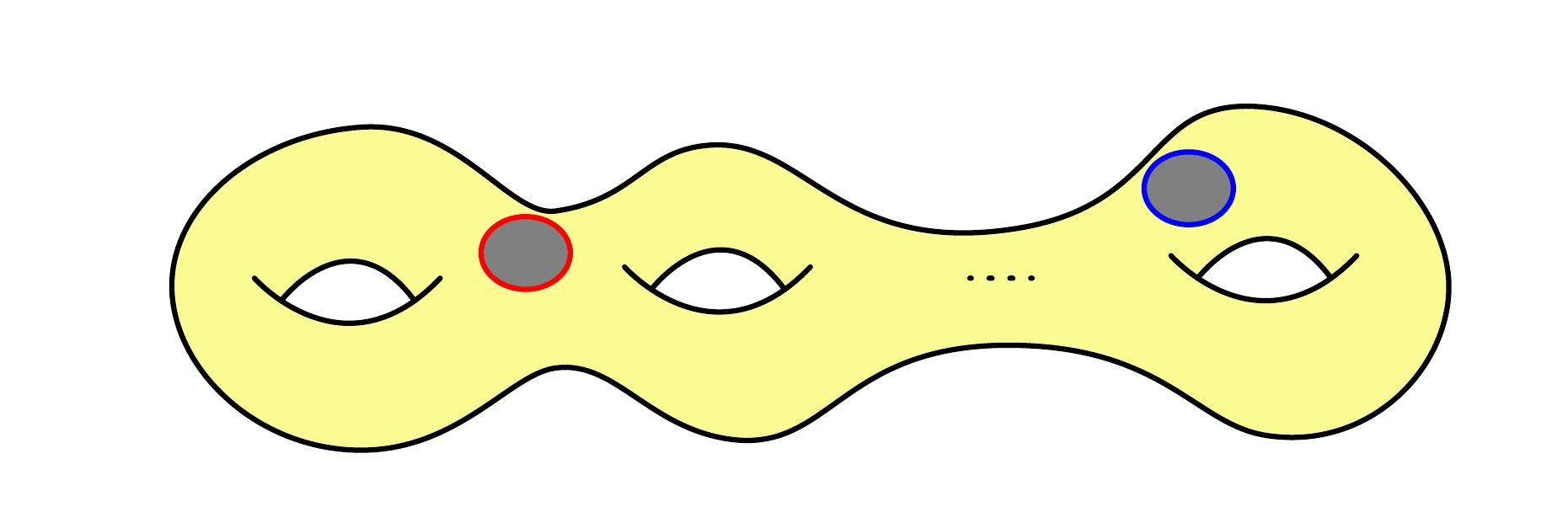}
\ee
We shall cut $X$ along the matter geodesics appearing in the expansion of $k_{L,R}$ into simple operators, 
\be
    \includegraphics[valign=c,width=0.6\textwidth]{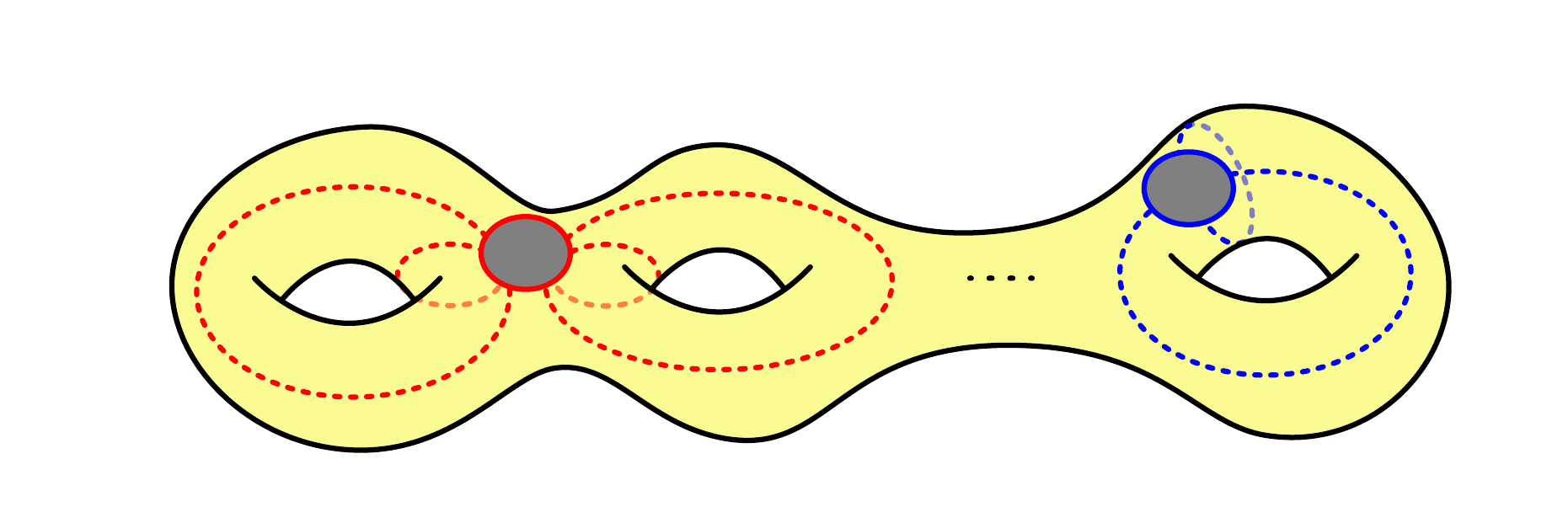}\,.
\ee
These geodesics start and end on the boundaries of $X$, and the result would be non-vanishing if, after cutting along these geodesics, we obtain surfaces that have the topology of disks. As we shall see, this is only possible by having a geodesic connecting the two boundaries. Because we choose the two-point function $k_L$ and $k_R$ to be vanishing on the disk, there can be no such matter geodesics connecting the two boundaries, and therefore \eqref{eq:BPS-to-prove} would follow.

To prove the fact that we are forced to include a geodesic between the two boundaries, let us color one boundary red and the other blue \footnote{We want to thank Ciprian Manolescu for providing this proof idea.}. For the sake of contradiction, suppose that there are only red geodesics starting and ending on the red boundary and blue geodesics starting and ending on the blue boundary.

Suppose that our end result after cutting along all the matter geodesics starting and ending at $k_{L, R}$ is one polygon. Since all the red lines are connected, all the blue lines are connected, but these two groups are disconnected, we end up with a red polygon or a blue polygon. This is a contradiction because we started with both red and blue boundaries. 

Suppose that our end result is multiple polygons. Then we have some red polygons and some blue polygons. But then each red edge can be glued back to some red edge, and similarly for the blue edges. Thus our starting geometry should have been at least two disconnected surfaces, which is again a contradiction since the surface $X$ is connected.

Thus, it follows that 
\be
\overline{\Tr_{\mH_{\text{bulk}}(K>e^{2S_j})}(k_L k_R)} 
= \overline{\Tr_{\mathcal{H}_L}(k_L)} 
\times \overline{\Tr_{\mathcal{H}_R}(k_R)} ,
\ee
and the argument above can be extended for an arbitrary number of traces, 
\begin{align}
&\overline{\Tr_{\mH_{\text{bulk}}(K>e^{2S_j})}(k_L^{(1)} k_R^{(1)}) \dots \Tr_{\mH_{\text{bulk}}(K>e^{2S_j})}(k_L^{(n)} k_R^{(n)})} 
= \nonumber 
\\ &\qquad = \overline{\Tr_{\mathcal{H}_L}(k_L^{(1)})} 
\times \overline{\Tr_{\mathcal{H}_R}(k_R^{(1)})} \dots \overline{\Tr_{\mathcal{H}_L}(k_L^{(n)})} 
\times \overline{\Tr_{\mathcal{H}_R}(k_R^{(n)})} 
.
\end{align}
Choosing $n=2$ or $4$ and $k_{L, R}^{(i)} = e^{-\alpha_{L, R}^{(i)} K_{L, R}^{(i)}}$ and applying the analogous differential as in \textbf{Step 2} and \textbf{Step 3}, \eqref{eq:diff-equations-vanish-1} and \eqref{eq:diff-equations-vanish}, with the inverse temperatures now replaced by the parameters $\alpha^{(i)}_{L,R}$ we find that 
\be
\overline{d(\alpha_L, \alpha_R)} = 0\,, \qquad \overline{\left(d(\alpha_L, \alpha_R)\right)^2} = 0\,,
\ee
once again guaranteeing that factorisation occurs beyond the coarse-grained level.

\section{Discussion}
\label{sec:discussion}

\subsection{Algebraic classification of gravitational operators \&  factorisation }
\label{subsec:algebraic-classification-of-bulk}

Having shown how the bulk two-sided trace factorises due to wormhole corrections, we will now briefly revisit one of our original goals of understanding the algebraic classification of one-sided gravitational observables. First, we should revisit the link between the factorisation of the trace and that of the Hilbert space. In section \ref{sec:fact-leading-order} -- \ref{sec:fact-to-all-orders}, we have shown that taking a trace for an arbitrary function of the left and right ADM Hamiltonians yields a factorising answer. Assuming that there are no algebra elements that commute with these Hamiltonians at a non-perturbative level, which would thus make the eigen energies non-degenerate, the factorisation of the trace then implies the factorisation of the Hilbert space. To make this assumption, we have considered the matter content of the gravitational theory to lack any gauge fields that could give rise to conserved boundary charges (we shall, however, revisit this assumption shortly in section \ref{subsec:bulk-gauge-fields}). There is, however, an additional subtlety that needs to be addressed. At the perturbative level, in gravitational theories, there oftentimes do exist operators that do commute with the two ADM Hamiltonians. One example of such an operator is the Casimir operator of the $SL(2, \mathbb R)$ isometry group that acts on $\mathcal H_\text{matter}$ in JT gravity coupled to matter. While the Casimir operator commutes with the two Hamiltonians at the perturbative level, wormhole corrections clearly indicate that this is no longer the case non-perturbatively. If the two were to commute, this would result in a bulk trace that is infinite within any fixed energy window; instead of finding such a divergent trace, we find a finite factorising answer. Similarly, if the matter fields coupled to gravity exhibit any global symmetry whose associated charges commute with $H_L$ and $H_R$ at the perturbative level, the charge would no longer be conserved non-perturbatively \cite{Harlow:2020bee, Hsin:2020mfa, Chen:2020ojn, Bah:2022uyz}. Thus, in the absence of gauge fields, we can safely conclude that the factorisation of the trace implies the factorisation of the Hilbert space. 

Because the Hilbert space factorises, this implies that we can define pure one-sided states. This can be done by considering a linear combination of our states $\ket{q_i}$ that gives an unentangled state between the two sides and then by tracing out one of the sides. The existence of such pure one-sided states, in turn, implies that the algebra of one-sided gravitational operators forms a type I factor, in contrast to a type II factor that one typically recovers in the perturbative gravitational theory. Nevertheless, our calculation cannot determine which explicit linear combination of the states $\ket{q_i}$ yields such an unentangled state; instead, our computation simply proves the existence of such states. We can also understand the fact that the algebra of one-sided operators is type I by checking that there now exists a well-defined one-sided trace rather than the trace being defined up to proportionality constant, as is the case for type II algebras.\footnote{We thank J.~Sorce for discussion about this point.} This is already clear from the factorisation of the trace in \eqref{eq:kL-kR-factorisation}. The bulk trace on the left-hand side is uniquely defined in terms of two-sided states; setting $k_L = k_R$ results in a uniquely defined one-sided trace on the right-hand side.

Finally, we can analyze our results in the context of the axioms required for a gravitational theory to have a type I algebra for one-sided operators \cite{Colafranceschi:2023urj,Marolf:2024adj}. While our theory does not technically satisfy all the axioms described in \cite{Colafranceschi:2023urj,Marolf:2024adj} since we do not assume the factorization of multi-boundary observables, we expect that the axioms are indeed satisfied for each ensemble member or before coarse-graining. The main difference in our case, however, is that we find that due to the presence of matter, the algebra of one-sided observables is a type I factor rather than just a type I algebra, which can have an arbitrary sum of type I factors. This is important since in AdS/CFT, at finite $N$, one expects the algebra of boundary operators to be a type I factor rather than a sum of type I factors, with the two cases being distinguished by whether the two-sided trace $\Tr_{\mathcal H_\text{bulk}}(k_L k_R)$ factorises for any $k_L$ and $k_R$. For example, we can once again contrast the theory that we studied with pure gravity when going to a single member of the ensemble -- the algebra of one-sided operators is type I in both cases, but in the latter, it is a sum of trivial type I factors with the Hamiltonian being the center element of the algebra.\footnote{Comments along these lines were already discussed in \cite{Colafranceschi:2023urj} for JT gravity  with matter. However, there, the factorisation of the Hilbert space was a priori assumed in each member of the ensemble. }

\subsection{Null states}
\label{subsec:null-states}

Since the Hilbert space factorises, the additional labels that are present when studying states at the perturbative level, that, for instance, labeled the scaling dimension of the matter wavefunction, are not present at the non-perturbative level. This implies that in the presence of matter, there is a tremendous number of null states when going between the perturbative and non-perturbative Hilbert spaces. By understanding such null states one can consequently find the non-isometric code that maps the perturbative (naive bulk) Hilbert space to the non-perturbative (genuine bulk or boundary) Hilbert space \cite{Akers:2022qdl}.  These null states were already explicitly found in \cite{Iliesiu:2024cnh}, where the factorisation of the Hilbert space was assumed in the presence of matter. Having now derived this feature of the theory, it is useful to briefly review what these null states are by summarizing the results of \cite{Iliesiu:2024cnh} for the reader's convenience. Consider an arbitrary state $\ket{\psi}$,  that is a linear combination of states with and without any matter insertions:
\begin{equation}
   |\psi\rangle = \int d\ell \,\psi_0(\ell) |\ell\rangle + \sum_\Delta \sum_m \int d\ell \, \psi_\Delta(\ell,m) |\Delta;\ell,m\rangle.
\end{equation}
In the two-sided non-perturbative Hilbert space whose states we label by $|E_i,E_j\rangle$, this state becomes \cite{Iliesiu:2024cnh} 
\begin{equation}
\label{eq:what-psi-is}
    |\psi\rangle = e^{-S_0/2} \sum_{i,j}\left[\delta_{i,j} \hat{\psi}_0(E_i)  + \sum_\Delta \frac{\mathcal{O}^\Delta_{ij}}{\Gamma^\Delta_{E_i,E_j}} \hat{\psi}_\Delta(E_i,E_j)\right]|E_i,E_j\rangle,
\end{equation}
where $\mathcal{O}^\Delta_{ij}$ is the matrix element of the matter operator used to create the state and where $\hat{\psi}_0(E_i) $ and $\hat{\psi}_\Delta(E_i,E_j)$ are transforms of the functions $\psi_0(\ell)$ and $\psi_\Delta(\ell,m) $ with respect to the wavefunctions  $\phi_E(\ell)$ and $\phi^\Delta_{E_L,E_R}(\ell,m)$ appearing in \eqref{eq:phi-E-l} and \eqref{eq:Delta-EL-ER},
\be 
        \hat{\psi}(E) := \int d\ell \,\phi_E(\ell) \psi_0(\ell)\,, \qquad  \hat{\psi}_\Delta(E_L,E_R) = \sum_m \int d\ell \,\phi^\Delta_{E_L,E_R}(\ell,m) \psi_\Delta(\ell,m).
\ee
The norm of the state $\ket{\psi}$ then becomes
\be 
 \qquad \langle \psi |\psi \rangle =  e^{-S_0}\sum_{i,j}\left|\delta_{i,j} \hat{\psi}_0(E_i)  + \sum_\Delta \frac{\mathcal{O}^\Delta_{ij}}{\Gamma^\Delta_{E_i,E_j}} \hat{\psi}_\Delta(E_i,E_j)\right|^2
\ee
which vanishes if each absolute value vanishes. There are countless of options through which this can happen: we can choose all wavefunctions to vanish on the energies $E_i$ but we can also choose a complicated combination of $\hat{\psi}_\Delta(E_i,E_j)$ which makes the sum of all terms in each absolute value vanish. \eqref{eq:what-psi-is} also tells us how to express an unentangled state between left and right as a linear combination of the states $\ket{\tilde q_i}$ or $\ket{ q_i}$. Since unentangled states should not have an Einstein-Rosen(ER) bridge connecting the two boundaries, the fact that such a linear combination exists confirms, from the bulk perspective, that there is no linear operator that probes the spacetime connectedness of the two boundaries \cite{Jafferis:2017tiu}. An even more surprising feature that can be seen through the results of section \ref{sec:checking-factorisation-is-basis-indep} is that the entire Hilbert space can be spanned by a set of states whose ER bridge has a fixed length.\footnote{For this, one simply sets $\ell_i = \ell$ for the $\ket{\tilde q_i}$ basis.} Thus, an unentangled state with no ER bridge can be expressed as a linear combination of states whose ER bridge has a fixed length. Since, in the former case, there is no length to even discuss, the length operator in the presence of matter only physically makes sense in a code subspace and not in the entire Hilbert space. With this new understanding in hand, it would be good to revisit the discussion of superselection sectors in AdS/CFT presented in \cite{Marolf:2012xe}.

\subsection{Bulk gauge fields and charged fields}
\label{subsec:bulk-gauge-fields}

As mentioned above, in our proof of factorisation, we assumed that the gravitational theory has no gauge fields in its low-energy description. If gauge fields are present, the story is slightly changed because there are additional gauge constraints in the Hartle-Hawking state that we start with: not only is $H_L=H_R$, but when gauge fields are present, we also have $Q_L = Q_R$ for any of the charges associated to the gauge field. Therefore, to prove factorisation we should check that ${\rm Tr}_{\cH_\text{bulk}}(e^{-\beta_L H_L - \alpha_L Q_L} e^{-\beta_R H_R- \alpha_R Q_R}) = {\rm Tr}_{\mathcal H_L} (e^{-\beta_L H_L-\alpha_L Q_L} ) {\rm Tr}_{\mathcal H_R}(e^{-\beta_R H_R-\alpha_R Q_R})$. This can only happen if we have states in $\mHb$ that break the gauge constraint $Q_L = Q_R$, and charged operators in every irreducible representation of the gauge field need to be used in the construction of $\ket{q_i}$ for factorisation to even stand a chance \cite{Guica:2015zpf, Harlow:2015lma}. In the case of JT gravity coupled to a gauge field and charged matter fields in all representations of the gauge group, the same resolvent technique that we have used throughout this paper is seemingly sufficient for ${\rm Tr}_{\cH_\text{bulk}}(e^{-\beta_L H_L - \alpha_L Q_L} e^{-\beta_R H_R- \alpha_R Q_R})$ to factorise without any additional constraints on the matter fields aside from the existence of charges in all representations. It would be interesting to see what wormhole corrections yield (as a function of $\beta_{L,R}$ and $\mu_{L,R}$) when more complicated operators are also inserted in the trace. For example, consider  ${\rm Tr}_{\cH_\text{bulk}}\left[ \chi_R\left(\mathcal P e^{i \int_C A}\right) e^{-\beta_L H_L - \alpha_L Q_L} e^{-\beta_R H_R- \alpha_R Q_R}\right]$ for a Wilson line going along a contour $\mathcal C$ between the two sides of the black hole. This would more closely parallel the setups of \cite{Guica:2015zpf, Harlow:2015lma} discussed in past literature in the context of factorisation.

\subsection{Generalization to higher dimensions}
\label{subsec:generalization-to-higher-dimensions}

Lastly, our results should be generalizable to higher dimensional setups. Whilst there, the sum over bulk geometries is not as well understood as in the context of JT gravity, one could work in saddle point approximation to leading order in $1/G_N$. In particular, the generalization of the overcomplete basis of states $\ket{q_i}$ was studied in \cite{Balasubramanian:2022gmo,Balasubramanian:2022lnw,Climent:2024trz}, and takes the form of the bag of gold states formed using spherical dust shells. With this, one can compute the trace over bulk Hilbert space and investigate steps 1 through 3 using wormhole saddles similar to the pinwheel geometries discussed in this paper. This would suffice to prove factorisation to leading order in $\frac{1}{G_N}$ as we did in section \ref{sec:fact-leading-order}. From this perspective, the results in section \ref{sec:fact-leading-order} can be viewed as an $s$-wave realization of the above models.  Furthermore, the fact that our main arguments presented in section \ref{sec:fact-to-all-orders} for the vanishing of the differential equation in steps 2 and 3  relied on very general symmetry properties of the energy integral leads us to believe that one should be able to determine the factorization of the Hilbert space from the gravitational point of view to all orders in $1/G_N$. It would nevertheless be interesting to concretely study the higher dimensional geometries that contribute to the bulk trace and verify these claims in higher dimensional theories with a known CFT dual.

\section*{Acknowledgments}

We are thankful to David Kolchmeyer for initial discussions and valuable comments on the draft. We also thank Xi Dong, Roberto Emparan, Don Marolf, Daniel Harlow, Jonathan Sorce, Douglas Stanford, Geoff Penington, Masamichi Miyaji, Arvin Shahbazi-Moghaddam, Adam Levine, Henry Maxfield, Mykhaylo Usatyuk, and Jinzhao Wang, for their useful comments. CY wants to give special thanks to Eleny Ionel and Ciprian Manolescu for discussions. LVI was supported by the Simons Collaboration on Ultra-Quantum Matter, a Simons Foundation Grant with No.~651440, during part of this work. CY was supported in part by the Heising-Simons Foundation, the Simons Foundation, and grant no. PHY-2309135 to the Kavli Institute for Theoretical Physics (KITP). This research was also conducted at a workshop also supported by grant NSF PHY-2309135 to the Kavli Institute for Theoretical Physics (KITP).

\appendix

\section{The resolvent at leading order in $K$}
\label{ap:analytic}
In this appendix, we give the details about the analytic structure of the resolvent, based on  the Schwinger-Dyson equation,
\begin{equation}\label{eq:DSeqap}
    R(\lambda)=\frac{K}{\lambda}+ \frac{e^{2S_0}}{\lambda} \int_{\mathcal{E}} \dd E_L \dd E_R \, \rho(E_L) \rho(E_R) \frac{R(\lambda) y(E_L,E_R)}{1- R(\lambda) y(E_L,E_R) }\,,
\end{equation}
where, throughout this section, for brevity we will denote $y\equiv y_q$ 
\begin{equation}
    y(E_L,E_R) = e^{-S_0} e^{-\beta E_L/2} e^{-\beta E_R/2} \frac{\Gamma(\Delta\pm i \sqrt{E_L}\pm i \sqrt{E_R})}{2^{2\Delta-1}\Gamma(2\Delta)} 
\end{equation}
and in our notation we shall always suppress the energy window $\mathcal{E}$ and $\dd E_L \dd E_R$ for simplicity.

\paragraph{Solving the Schwinger-Dyson equation near $\lambda=0$}
\ \\
\indent $\boldsymbol{K>d^2}$: We make an ansatz $R(\lambda)\sim \frac{\#}{\lambda}$ and plug it in to \eqref{eq:DSeqap}.  In the limit of $\lambda\rightarrow 0$, $\frac{Ry}{1-Ry}$, the integrand of the energy integral, becomes $-1$, and we are left with $\frac{K-\int \rho(E_L)\rho(E_R)}{\lambda}$ on the RHS which matches the LHS of the SD equation from our ansatz.
This confirms that the ansatz gives a consistent solution $R(\lambda)\sim \frac{K-d^2}{\lambda}$.

$\boldsymbol{K<d^2}$: We would like to argue that $R(\lambda=0) = R_0$ which is a finite number in this case. Since this analytic behavior is drastically different from the above $K\geq d^2$ case, it should be helpful to address the distinction. 
Let us first explain why a (negative) finite $R_0$ is an acceptable solution: the reason is that $\frac{Ry}{1-R y}$, the integrand of the energy integral, has an absolute value smaller than 1. This implies the inequality for the integral,
\be 
\Big|\int \rho(E_L) \rho(E_R) \frac{Ry}{1-R y}\Big| < d^2
.
\ee 
The value of $R_0$ in such a case can be adjusted such that we get exactly $ - K$, making the pole at $\lambda=0$ canceled. 
We emphasise that such a cancellation can only happen if $K<d^2$ which implies that when $K>d^2$ the solution discussed above is unique.

\paragraph{Approximation of $R_0$ when $K<d^2$}
\ \\
\indent One can even further approximate $R_0$ by $\frac{-K}{\int \rho(E_L)\rho(E_R)y(E_L,E_R)}$ when $K \ll d^2$, which can be justified as follows
\begin{equation}
\begin{aligned}
 & \lim_{\lambda\rightarrow 0} \lambda R(\lambda) \approx K -  \frac{K}{ \int  \rho(E_L) \rho(E_R) y(E_L,E_R) } \int \rho(E_L) \rho(E_R) y(E_L,E_R) \left(1+ \cdots \right)=0 \,.
\end{aligned}
\end{equation}
And $R_0$ gets a larger absolute value when $K$ approaches $d^2$. 
The $K \rightarrow d^2$ limit of $R_0$ will be divergent as $(K-d^2)^{-1}$ as will be proved as follows. 
We take the ansatz 
\be
R_0=\frac{r_0}{K-d^2}
\ee
where $r_0$ is independent of $K$. To solve for $r_0$ we plug the ansatz into equation (\ref{lambdaSD}) setting $\lambda=0$
\begin{align}
    K &= -e^{2 S_0} \int \rho(E_L) \rho(E_R) \frac{R_0 y(E_L,E_R)}{1-R_0 y(E_L,E_R) }= -e^{2 S_0} \int \rho(E_L) \rho(E_R) \frac{\frac{r_0}{K-d^2} y(E_L,E_R)}{1-\frac{r_0}{K-d^2} y(E_L,E_R)}\\
      &\approx e^{2 S_0}\int \rho(E_L) \rho(E_R)\left(1+\frac{K-d^2}{r_0 y(E_L,E_R)}\right)
      =d^2+\frac{K-d^2}{r_0} e^{2 S_0} \int  \frac{\rho(E_L) \rho(E_R)}{y(E_L,E_R)}
\end{align}
which then gives
\be
r_0 =  e^{2 S_0} \int d s_L d s_R\, \frac{\rho(s_L) \rho(s_R)}{y_q(s_L,s_R)}\,.
\ee
This is indeed independent of $K$, so we have a consistency check of the ansatz.

\paragraph{No other pole or branch cut}
\ \\
\indent We would like to prove that there are no other poles and branch cuts in the resolvent except for the possible pole at $\lambda=0$, the branch cut in $(\lambda_{-},\lambda_{+})$, and the pole at $\lambda=\infty$.
The reason why we care about the absence of additional non-analytical features is that we need to perform a contour deformation in the $\text{Tr}_{\mHb}$ calculation, which could pick up additional contributions in the presence of other poles or branch cuts. In particular, we want to show that the factor $\left(R(\lambda) y_q(s_L,s_R)-1\right)^{-1}$ is never divergent and will therefore not lead to any additional poles or branch-cuts. 
To show that, we multiply both sides by of the SD equation \eqref{eq:DSleadingorder} by $\lambda$ to get 
\be
    \lambda R(\lambda)=K +  \int \rho(E_L) \rho(E_R) \frac{R(\lambda) y (E_L,E_R)}{1-R(\lambda) y(E_L,E_R)}\label{lambdaSD}\,.
\ee
Suppose for the sake of contradiction,  $R(\lambda)y(E_L,E_R)=1$ for some values of $\lambda=\lambda_*$, $E_{L,R}$, and so $R(\lambda_*)=\frac{1}{y(E_L,E_R)}$. Then the LHS of (\ref{lambdaSD}) is finite at $\lambda_*$, but the RHS is infinite which is a contradiction. 

The conclusion that $1-R(\lambda) y(E_L,E_R)$ is never zero will also be useful in the following branch cut discussion below. 

\paragraph{The range of branch cut}
\ \\
\indent Now, we give the approximation of the range of branch cut. Such a range depends on the monotonically decreasing property of the $y(E_L,E_R)$ function when increasing energy. 
In particular, we define $y_{\rm max}=y(E_{\rm min},E_{\rm min})$ and $y_{\rm min}=y(E_{\rm max},E_{\rm max})$. In the $K>d^2$ case, we would like to argue that the branch cut goes between
\be 
  (\lambda_{-},\lambda_{+}) \approx \left((K-d^2)y_{\rm min}, (K-d^2)y_{\rm max}\right)\,.
\ee 
To probe the branch points, we still try to manipulate the SD equation as 
\begin{equation}
    R(\lambda) + \lambda \frac{d R}{d\lambda}=\frac{d}{d\lambda} (\lambda R(\lambda))= \frac{dR}{d\lambda}  \int \rho(E_L)\rho(E_R) \frac{y(E_L,E_R)}{\left(1- R(\lambda) y(E_L,E_R)\right)^2}\,.
\end{equation}
For the branch point, we need $\frac{d R}{d\lambda}=\infty$ while $R(\lambda)$ is finite. 
Therefore, $\lambda_{\rm bp}$ (bp for branch-point) has to satisfy
\begin{equation}\label{eq:lambdabp}
    \lambda_{\rm bp}= \int \rho(E_L)\rho(E_R) \frac{y(E_L,E_R)}{\left(1- R(\lambda_{\rm bp}) y(E_L,E_R)\right)^2}\,.
\end{equation}
This equation, together with the Schwinger-Dyson equation, allows us to analyze the locations of the two branch points. 

We now find an approximation for $\lambda_{\rm bp}$. For that, we look at $R(\lambda_{\rm bp})$.
Our claim is that the $R(\lambda_{\rm bp})$ should be sufficiently close to $y_\text{min}^{-1}$ or $y_\text{max}^{-1}$ to satisfy the Schwinger-Dyson equation. 
To verify this, one can rewrite the Schwinger-Dyson equation at the branch-point as 
\begin{equation}\label{eq:bpDSeq}
    K-d^2+ \int \rho(E_L)\rho(E_R) \frac{2}{1-R(\lambda_{\rm bp}) y(E_L,E_R)} = \int  \rho(E_L)\rho(E_R) \frac{1}{(1-R(\lambda_{\rm bp}) y(E_L,E_R) )^2} \,.
\end{equation}
and note that the integrand on the RHS has the chance to be much larger than the LHS when $R(\lambda_{\rm bp})$ is close to $y_{\rm min}$ or $y_{\rm max}$. 
If $R(\lambda_{\rm bp})$ was nowhere close to the extremal value of $y$, then the two integrals would be roughly $O(d^2)$, and could be much smaller than $K$. 
This consequently means that the SD equation would not be satisfied. 
On the other hand, if we simply set $R(\lambda_{bp})$ to be exactly the extremal value of $y$, then the RHS of \eqref{eq:bpDSeq} would be divergent. 
For example, let us consider analyzing $\lambda_{+}$ and the corresponding $R(\lambda_{+})$. 
If $R(\lambda_{+})$ was $y^{-1}_{\rm max}$, then the integral near $E_L=E_R=E_{\rm min}$ can be approximated by 
\begin{equation}
    \int \dd \Delta E_L \dd \Delta E_R \frac{\rho(E_{\rm min}+\Delta E_L) \rho(E_{\rm min}+\Delta E_R) }{\left(\partial_{E_{L}}{y(E_L,E_R)} \Delta E_L+\partial_{E_{R}}{y(E_L,E_R)}\Delta E_R \right)^2}
\end{equation}
where both $\Delta E_L$ and $\Delta E_R$ are small of order $\varepsilon$. 
Then the integral is roughly $\propto \int (\epsilon \dd\epsilon) {\epsilon^{-2}}$, which is divergent. 
Meanwhile, the LHS integral still stays $O(d^2)$, which cannot compete with the RHS integral. 
In short, the branch-point Schwinger-Dyson equation \eqref{eq:bpDSeq} can be satisfied by setting $R(\lambda_{\rm bp})$ very close to $y_{\rm min}$ or $y_{\rm max}$ can make the RHS integral nearly divergent to match the possibly large $K-d^2$ term on the LHS. 
This analysis also tells us that 
\begin{equation}
    K-d^2 \approx \int  \rho(E_L)\rho(E_R) \frac{1}{(1-R(\lambda_{\rm bp}) y(E_L,E_R) )^2} \,.
\end{equation}
Comparing this equation with \eqref{eq:lambdabp}, and keeping in mind that the most important part of the contribution to the integral comes from the corner $E_L\approx E_R\approx E_{\rm min,max}$, we thus see 
\begin{equation}
    (\lambda_{-},\lambda_{+}) \approx \left((K-d^2)y_{\rm min}, (K-d^2)y_{\rm max}\right)\,.
\end{equation}

The discussion for $K<d^2$ is basically identical, so we omit any further details.

\section{Basis prepared using operators with different dimensions}
\label{sec:basis-prepared-with-different-dim}

In this appendix, we extend the computation of the bulk trace to a more general case, with basis $\{q_i\}_{i=1,\dots,K}$ prepared by operators $O_i$ with different scaling dimensions $\Delta_i$ and Euclidean time evolutions $ \beta_L^{(i)}/4$ and $ \beta_R^{(i)}/4$ to the left and right of the operator insertion. 
More precisely, we define the set of states 
\begin{equation}
    |q_i\rangle = O_{\Delta_i} |HH\rangle_{ \beta_L^{(i)},  \beta_R^{(i)}} = \ket{\Delta_i; \frac{\beta_L^{(i)}}4,  \frac{\beta_R^{(i)}}4 }  \,, \text{ with } i= 1,2,\ldots, K \,,
\end{equation}
and work in the same limit as in Section~\ref{sec:fact-leading-order}. 

In the spirit of \eqref{eq:DSpicture}, we write down the Schwinger-Dyson equation 
\begin{equation}
    R_{ij}(\lambda) = \frac{\delta_{ij}}{\lambda} + \frac{1}{\lambda} \sum_n \sum_{k} W_{n,ik}^{O} R_{kj} \,, 
\end{equation}
where $W_{n,ik}^{O}$ is defined as the irreducible part of the resolvent, given by
\begin{equation}
    W_{n,ik}^O = \delta_{ik} \sum_{i_1,i_2,\cdots,i_{n-1}} Z_{n,k i_1 \cdots i_{n-1} } R_{i_1 i_1}  R_{i_2 i_2} \cdots  R_{i_{n-1} i_{n-1}} \,,
\end{equation}
in which $Z_{n,k i_1 \cdots i_{n-1} }$ is a pinwheel like \eqref{eq:n_boundary_pinwheel} but the $n$ geodesics are connecting operators with scaling dimensions $\Delta_k,\Delta_{i_1},\ldots, \Delta_{i_k}$. 
This will modify the $y$ definition, since it is $\Delta$ dependent, in the pinwheel definition. 
Moreover, the $n$ asymptotic boundaries can have different sizes, because they have different Euclidean time for state preparation. 
Putting these two modifications together, we define 
\begin{equation}
\label{eq:yqDelta_i}
    y_{q_i}(s_L,s_R) = e^{-S_0} e^{-\beta_L^{(i)} s_L^2/4} e^{-\beta_R^{(i)} s_R^2/4}  \gamma_{\Delta_i} (s_L, s_R) \,, \ \   y_{q_i}(E_L,E_R) \equiv  y_{q,\Delta_i}(s_L(E_L),s_R(E_R)) \,, 
\end{equation}
so that 
\begin{equation}
    Z_{n,k i_1 \cdots i_{n-1} } = \int \rho(E_L)\rho(E_R) y_{q_k}  y_{q_{i_1}}  \ldots y_{q_{i_{n-1}}} \,. 
\end{equation}
Putting everything together, we get, utilizing a trick replacing a sum over product with a product of sums, 
\begin{equation}\label{eq:Rijdifferenty}
\begin{aligned}
    W_{n,ik}^O  & = \delta_{ik} \int  \rho(E_L)\rho(E_R) y_{q_k}(E_L,E_R) \left(:{y_q R}: \right)^{n-1}\\
    R_{ij}(\lambda) & = \frac{\delta_{ij}}{\lambda} + \frac{e^{2 S_0}}{\lambda} \sum_n \int \rho(E_L)\rho(E_R) \left(:{y_q R}:  \right)^{n-1} y_{q_i}(E_L,E_R) R_{ij} \\
    & = \frac{\delta_{ij}}{\lambda} + \frac{e^{2 S_0}}{\lambda} \int  \rho(s_L)\rho(s_R) 
    \frac{y_{q_i}(E_L,E_R) R_{ij}}{1-:{y_q R}:  } 
\end{aligned}
\end{equation}
where we have defined\footnote{We use the $:{\dots}:$ notation to distinguish it from the $\overline{R}$ notation, with the latter being an average over operator indices and the latter being the gravitational path integral result. }
\be 
:{y_{q} R}: \equiv \sum_{i=1}^{K} y_{q_i}(E_L,E_R) R_{ii}. 
\ee
A simple observation from \eqref{eq:Rijdifferenty} is that $R_{ij}=0$ if $j\neq i$. 

Now we analyze the analytic structure, in particular the behavior near $\lambda=0$, of the resolvent $ R_{ij}(\lambda)$ using the Schwinger-Dyson equation \eqref{eq:Rijdifferenty}. 
Note that in the special case where all the $y_{q_i}$'s are the same, there is only one function, $R(\lambda)$, to solve for.
However, in the case of distinct $y_{q_i}$, all the equations for $R_{ii}$ have to be used together, because the variables are $R_{ii}, i\in 1,\ldots, K$.

\paragraph{$R_{ii}$ for $K\geq d^2$:} 
We would like to make the ansatz that $R_{ii}(\lambda) \sim  {c_i}\lambda^{-1}$ for $\lambda\rightarrow 0$.
Substituting this ansatz in, we see that 
\begin{equation}\label{eq:RiiKgeqdsq}
\begin{aligned}
     R_{ii}(\lambda) \sim \frac{c_i}{\lambda} 
  & =  \frac{1}{\lambda} + \frac{e^{2 S_0}}{\lambda}\int \rho(E_L) \rho(E_R) 
    \frac{y_{q_i}(E_L,E_R) c_i}{\lambda - \sum_j c_j y_{q_j} (E_L,E_R) } \\
    &\approx  \frac{1}{\lambda} - \frac{e^{2 S_0}}{\lambda}\int \rho(E_L) \rho(E_R) 
    \frac{y_{q_i} c_i }{ \sum_j c_j y_{q_j} (E_L,E_R) } \,,
\end{aligned}
\end{equation}
resulting in 
\begin{equation}\label{eq:cieqns}
    c_i \left(1 + e^{2 S_0} \int \rho(E_L) \rho(E_R) 
    \frac{y_{q_i}(E_L,E_R) }{ \sum_j c_j y_{q_j} (E_L,E_R) }  \right) = 1 \,, \quad i = 1,2,\ldots,K \,.
\end{equation}
There are $K$ equations and $K$ variables in this rather complicated set of equations, but one can in principle find the solution to these residues.
A reasonable expectation, which is consistent with the equations (recall that $y_{q_j}>0$), is that $c_i \in (0,1)$, which represents the probability of the $M_{ii}$ element corresponds to a zero eigenvalue. 
In the case of identical $\Delta_i$, which we have discussed in the main text, $c_i=\frac{K-d^2}{K}$, consistent with the fact that the probability of any one of the equivalent $K$ states being one of the $K-d^2$ null states is evenly distributed to be $\frac{K-d^2}{K}$. 

One constraint, from \eqref{eq:cieqns}, is that 
\begin{equation}
    \sum_i  c_i \left(1 + e^{2 S_0} \int  \rho(E_L) \rho(E_R) 
    \frac{y_{q_i}(E_L,E_R) }{ \sum_j c_j y_{q_j} (E_L,E_R) }  \right)  = K\,, 
\end{equation}
which tells us 
\begin{equation}
    \sum_i c_i = K- d^2  \ \Rightarrow \ \text{Res}_{\lambda=0} R(\lambda)= K-d^2\,,
\end{equation}
indicating that the total number of null states is $K-d^2$. Since $d^2$ is still the dimension of the two-sided Hilbert space, even with this change of basis, this is exactly the expected result. As we shall see shortly, while \eqref{eq:cieqns} can, in principle, be solved we will not need to determine the solution in order to compute $ {\rm Tr}_{\mHb}(k_L k_R)$ when $K>d^2$. 

\paragraph{$R_{ii}$ for $K <  d^2$:}
We wish to argue that $R_{ii}(\lambda=0)$ is finite in the $K<d^2$ case. 
Let us show that $R_{ii}(\lambda=0)$ being finite is indeed an acceptable solution. 
Rewrite the Schwinger-Dyson equation as 
\begin{equation}\label{eq:RiiKlessdsqDS}
\begin{aligned}
     & \lambda R_{ii}(\lambda) = 1 + R_{ii}(\lambda) e^{2 S_0} \int \rho(E_L) \rho(E_R) \frac{y_{q_i}(E_L,E_R)}{ 1 - :y_{q}R:} \\
     & \Rightarrow \ 1 = R_{ii}(\lambda=0) e^{2 S_0}\int \rho(E_L) \rho(E_R) \frac{y_{q_i}(E_L,E_R)}{ 1 -:{y_q R}(\lambda=0): }  \,. 
\end{aligned}
\end{equation}
An acceptable set of solution is to have $R_{ii}(\lambda=0)$ to be negative, and one can argue that 
\begin{equation}\label{eq:Riilambda0}
    R_{ii}(\lambda=0) < - \left( e^{2 S_0}\int \rho(E_L) \rho(E_R) y_{q_i}(E_L,E_R) \right)^{-1}\,.
\end{equation}
If $K$ is not so close to $d^2$, $R_{ii}(\lambda=0)$ will not be so away from the value in \eqref{eq:Riilambda0} because the fraction in the integrand would be positive but not so different from 1, which can be checked by summing over $i$ on both sides of \eqref{eq:RiiKlessdsqDS} to confirm that $:y_{q}R:$ is  $O(K/d^2)$.

\paragraph{From the resolvent to the bulk trace:}  

We can now show that factorisation is still achieved when $K\geq d^2$. 
Using the same trick leading us to \eqref{eq:Rijdifferenty}, one can write the bulk trace as,
\begin{equation}
    {\rm Tr}_{\mHb}(k_L k_R) = \int_{C_2}\frac{\dd \lambda}{2\pi i } \frac{1}{\lambda} e^{2 S_0} \int \rho(E_L)\rho(E_R) \frac{e^{-\beta_L E_L} e^{-\beta_R E_R} :{y_q R}: }{1-:y_{q}R:}\,.
    \label{eq:TrHbulk-diff-basis}
\end{equation}
We see that this expression is essentially the same as what we got in Section~\ref{sec:fact-leading-order}. The complication of having different $y_{q_i}$ and different $R_{ii}(\lambda)$ is hidden by the sum over $i$ in $:y_{q}R:=\sum_i y_{q_i} R_{ii} $. 
If $K>d^2$, $:y_{q}R:=\sum_i y_{q_i} R_{ii} $ behaves as $(K-d^2)\lambda^{-1}$ as $\lambda\to 0 $. Thus,  \be 
\int_{C_2}\frac{\dd \lambda}{2\pi i } \frac{1}{\lambda}  \frac{e^{-\beta_L E_L} e^{-\beta_R E_R} :{y_q R}: }{1-:y_{q}R:} =e^{-\beta_L E_L} e^{-\beta_R E_R} 
\ee
giving factorisation in \eqref{eq:TrHbulk-diff-basis}. 
On the other hand, if $K<d^2$, $:y_{q}R:$ is finite, so that the correlation between $E_{L,R}$ from $y(E_L,E_R)$ won't drop out and we get no factorisation. 

\paragraph{Another change of basis -- defining states on geodesic slices:}  Finally, we can discuss a final change of basis defining our states on geodesic slice, instead of asymptotic boundaries. Thus, we can define:
\begin{equation}
\label{eq:tilde-q-basis}
    |\tilde q_i\rangle = O_{\Delta_i} |HH\rangle_{ \beta_L^{(i)},  \beta_R^{(i)}} = \ket{\Delta_i; \ell_i, m_i }  \,, \text{ with } i= 1,2,\ldots, K \,.
\end{equation}
The only difference this makes in the calculations above is that the function $y_{q_i}(s_L, s_R)$ defined in \eqref{eq:yqDelta_i} is replaced by:
\begin{equation}
    y_{\tilde q_i}(s_L,s_R) = e^{-S_0} \phi_{\frac{s_L^2}2, \frac{s_R^2}2}^\Delta(\ell_i, m_i) \gamma_{\Delta_i} (s_L, s_R) \,, \ \   y_{\tilde q_i}(E_L,E_R) \equiv  y_{\tilde q_i}(s_L(E_L),s_R(E_R)) \,, 
\end{equation}
where $\phi_{E_L, E_R}^\Delta(\ell_i, m_i)$ is the change of basis matrix  defined by \eqref{eq:Delta-EL-ER}.  Since the derivation of factorisation when $K>d^2$ is independent of the exact form of $y_{\tilde q_i}(s_L,s_R)$, we can apply the same arguments as above when working with the basis \eqref{eq:tilde-q-basis}.

\section{The resolvent at higher orders in $K$}\label{sec:subleadinginK}

In this appendix, we give the calculation of the resolvent and trace when including higher orders in $K$. To compute the correction to the next order in $1/K$, we need to inspect the (a2) and (c1) geometries.  This is because the other two geometry classes, (b2) and (c2), only contribute to even higher orders in $K$. 
Based on the classification of geometries, the salient feature of these subleading geometries is to connect disconnected pinwheels through wormholes. 
Specifically, for (a2), the wormholes are matter supported and for (c1) the wormholes are empty.\footnote{
Note that the second kind of wormhole does not exist in the BPS sector.}
We will treat the contributions from (a2) and (c1) separately. 

The idea for performing the computation is to start from the planar pinwheel geometries, from which the contribution to the resolvent 
\begin{equation}
    R_{(a1)}(\lambda)=\frac{K}{\lambda} + \frac{1}{\lambda}\sum_{n=0}^{\infty} Z_{n} R^n(\lambda)
\end{equation}
and put a bar between $Z$ and $R$ or between $R$ and $R$. This is geometrically equivalent to building a wormhole connecting two disconnected pieces and then replacing $ZR$ or $RR$  with the ``connected correlator" between two pinwheels $\overline{Z R}\text{ or } \overline{RR}$.
Before jumping into details, it should also be helpful to summarize the results in the $\lambda\rightarrow 0$ limit, which captures the null states that we are particularly interested in. 
\begin{equation}
\begin{aligned}
    R_{(c1)}(\lambda) & \sim 
    O(\lambda^0)\,, \quad      R_{(a2)}(\lambda) & \sim  O(\lambda^0) \,.
\end{aligned}
\end{equation}
Putting the two results together, we will find that the subleading order correction to the resolvent $\overline{R(\lambda)}$ is finite, so that 
\begin{equation}\label{eq:R-correction-from-(c1)-and-(a2)}
    \text{Res}_{\lambda=0}\overline{R(\lambda)}=\text{Res}_{\lambda=0}{R_{(a1)\&(b1)\&(c1)\&(a2)}(\lambda)}=\text{Res}_{\lambda=0} R_{(a1)\&(b1)}(\lambda) = K - \overline{d^2} \,, 
\end{equation}
meaning the number of null states is not affected by the $(c1)$ and $(a2)$ type geometries in the sense of $1/K$ series expansion.

This has a nice physical explanation. For a given $K$, the average number of null states can be computed as 
\begin{equation}
     \overline{\text{$\#$ of null states}} =\int_0^K  \dd \mu (d^2)  (K-d^2) 
\end{equation}
where $\mu(d^2)$ is the probability distribution of the dimension of the two-sided Hilbert space. 
We also know that
\begin{equation}
   \int_0^\infty \dd\mu(d^2) = 1\,, \quad \int_0^\infty \dd\mu(d^2) d^2 = \overline{d^2} \,.
\end{equation}
It is straightforward to see that, the number of null states, which equals to $\text{Res}_{\lambda=0}\overline{R(\lambda)}$, deviates $K-\overline{d^2}$ by only a tiny amount, since 
\begin{equation}
- (K-\overline{d^2})
    +\int_0^K  \dd \mu (d^2)  (K-d^2)  = \int_{K}^{\infty} \dd \mu(d^2) (d^2-K)\,.
\end{equation}
The RHS is expected to be at least exponentially suppressed in $K$, because the distribution of the dimension $d^2$ should be approximately a Gaussian. 
Under this assumption, we plug in $\dd \mu(d^2)= \text{exp}\left(-\frac{(d^2-\overline{d^2})^2}{\sigma^2}\right) \dd (d^2) $, from which
\begin{equation}
    \int_{K}^{\infty} \dd \mu(d^2) (d^2-K) \sim \frac{\sigma^4}{4K^2} \text{exp}\left(-\frac{(K-\overline{d^2})^2}{\sigma^2}\right)\,.
\end{equation}
This expression has no perturbative expansion in $1/K$. This explains the absence of the term like $\lambda^{-1}K^{-1}$ in the resolvent characterizing the change of number of null states. 

Now, we lay out the computational details for the (c1) and (a2) geometries. 
\paragraph{The (c1) geometry contribution to $R(\lambda)$:}

In (c1) geometries, the wormholes appearing in $\overline{ZR}$ are empty wormholes, and one can first write down explicitly the $\overline{Z_m Z_n}$ (which is the building block of  the $\overline{ZR}$ or $\overline{RR}$ correlators)
\begin{equation}
    \overline{Z_m Z_n}_{(c1)} =  \int \left(\delta \rho^2\right)^2 y(E_L,E_R)^{m} y(E_L^{\prime},E_R^{\prime})^{n} \,,
\end{equation}
where
\begin{equation}
   \left(\delta \rho^2 \right)^2 \equiv  \overline{\rho(E_L) \rho(E_R) \rho(E_L^{\prime}) \rho(E_R^{\prime})} - \overline{\rho(E_L) \rho(E_R) } \times \overline{\rho(E_L^{\prime}) \rho(E_R^{\prime})} \,.
\end{equation}

With such building blocks, we can do some combinatoric work to get $\overline{ZR}$ or $\overline{RR}$. Note that connections between more than two connected components are further suppressed in $K$: for example, if there are three boundaries, the leading in $K$ geometry (connected three boundary pinwheel) is giving $K^3 e^{-S_0}$ while the (c1) type geometry with a pair of pants connecting three disks is $1/K^2$ suppressed.
Let us start from the $\overline{ZR}$ calculation
\begin{equation}
    \lambda \overline{Z_m R}_{(c1)}=\left(\overline{Z_m \sum_{n=1}^{\infty} Z_n R^n }\right)_{(c1)}= \sum_{n}\overline{Z_m Z_n }_{(c_1)} R^n + \sum_{n} n \overline{Z_m R }_{(c1)} Z_n R^{n-1}
\end{equation}
so that 
\begin{equation}
    \overline{Z_m R}_{(c1)} = \frac{\sum_n \overline{Z_m Z_n}_{(c1)} R^n }{\lambda -X}
    \,.
\end{equation}
where for the simplicity of notations, we can define $X= \frac{d}{dR} \sum_{m}Z_m R^m$ and also $Y=\sum_{m,n} \overline{Z_m Z_n}_{(c1)} R^m R^n$ useful later. 

Furthermore, one can study the $\overline{RR}$ term 
\begin{equation}
\begin{aligned}
    \lambda^2 \overline{RR}_{(c1)} &= \overline{\sum_{m=1}^{\infty} Z_{m} R^m \sum_{n=1}^{\infty} Z_n R^n} \\
    & = \sum_{m,n} \overline{Z_m Z_n}_{(c1)} R^m R^n + 2 \sum_{m,n}m Z_m R^{m-1} \overline{R Z_n}_{(c1)} R^n +  \sum_{m,n} m n  Z_m Z_n R^{m-1} R^{n-1} \overline{R R }_{(c1)}
\end{aligned}
\end{equation}
so that 
\begin{equation}\label{eq:RRa2}
\begin{aligned}
    \overline{RR}_{(c1)} & = \frac{\sum_{m,n} \overline{Z_m Z_n}_{(c1)} R^m R^n + 2 X \sum_{n}  \overline{R Z_n}_{(c1)} R^n}{\lambda^2-X^2 }\\
    & = \frac{\sum_{m,n} \overline{Z_m Z_n}_{(c1)} R^m R^n }{\left(\lambda-X\right)^2}
    \,.
\end{aligned}
\end{equation}
With these two ingredients, we can write down the contribution to $R$ from the (c1) geometries\footnote{Note that in the last line, the $R$ derivative is a formal one, in the sense that it does not act on $\lambda$. The reason to introduce it is simply to give a more compact expression.}
\begin{equation}\label{eq:c1final}
\begin{aligned}
    \text{(c1) contribution} & = \frac{1}{\lambda}\sum_{m} m \overline{Z_m R} R^{m-1} + \frac{1}{\lambda}\sum_{m} \frac{m(m-1)}{2} Z_m R^{m-2} \overline{R R } \\
    & = \frac{1}{2 \lambda}\frac{ \partial_R Y}{ \lambda  - X} + \frac{1}{2\lambda} \partial_R X \frac{Y}{(\lambda - X)^2} \bigg|_{R=R(\lambda)}
    \,. 
\end{aligned}
\end{equation}
where 
\begin{equation}
\begin{aligned}
      X & = \int \overline{\rho(E_L) \rho(E_R) } \frac{ y(E_L,E_R)}{(1-R y(E_L,E_R))^2}  \rightarrow \frac{\lambda^2}{(K-\overline{d^2})^2} \int \overline{\rho(E_L)\rho(E_R)}y(E_L,E_R)^{-1}\\
      Y & = \int \left(\delta \rho^2\right)^2  \frac{R y(E_L,E_R)}{1-R y(E_L,E_R)} \frac{R y(E_L^{\prime},E_R^{\prime})}{1-R y(E_L^{\prime},E_R^{\prime})} \rightarrow  \int \left(\delta \rho^2\right)^2   \\
      \partial_R X & = \int \overline{\rho(E_L) \rho(E_R) } \frac{ 2 y^2(E_L,E_R)}{(1-R y(E_L,E_R))^3}  \rightarrow \frac{-2\lambda^3}{(K-\overline{d^2})^3 } \int \overline{\rho(E_L)\rho(E_R)}y(E_L,E_R)^{-1} \\
      \partial_R Y & = \int \left(\delta \rho^2\right)^2  \frac{R y(E_L,E_R)}{1-R y(E_L,E_R)} \frac{ y(E_L^{\prime},E_R^{\prime})}{(1-R y(E_L^{\prime},E_R^{\prime}))^2} \rightarrow  \frac{\lambda^2}{(K-\overline{d^2})^2} \int \left(\delta \rho^2\right)^2 y(E_L^{\prime},E_R^{\prime})^{-1}
\end{aligned}
\end{equation}
in which the limit are taken to be $\lambda\rightarrow 0$ under the condition $K>d^2$. 
Thus, we can see that the 
\begin{equation}
  \text{(c1) contribution}  \sim \frac{1}{K^2}  O(\lambda^0)\,.
\end{equation}

\paragraph{The (a2) geometry contribution to $R(\lambda)$:}

The derivation for the (a2) geometry contribution is similar, but one need to be careful about avoiding overcounting geometries which are actually of (a1) type. 

The starting point is again $\overline{Z_m Z_n}$, and the geometry now looks like $Z_{m+n}$ because the wormhole is matter supported
\begin{equation}
    \overline{Z_m Z_n}_{(a2)} = Z_{m+n} = \int \overline{\rho(E_L) \rho(E_R) } y(E_L,E_R)^{m+n}\,.
\end{equation}

When considering $\overline{Z_m R}$, we need to exclude the possible geometries connecting adjacent connected components, because such a connection will lead to (a1) geometries. 
Then for $\overline{Z_m R}$, we should distinguish between $\overline{Z_m R}_{\rm adjacent}$ and $\overline{Z_m R}_{\rm non-adjacent}$.
The reason of making such a distinction will be transparent in the following derivation. 
Let us start from a discussion on $\overline{Z_m R}$ where the resolvent $R$ is right next to a pinwheel $Z_m$, and we denote this as the adjacent contribution. One get 
\begin{equation}\label{eq:zmRa2adj}
\begin{aligned}
   \overline{ Z_m R}_{(a2),{\rm adjacent}} & = \overline{ Z_m \times\text{irreducible part of $R$} \times R  } \\
   & = \overline{ Z_m R}_{(a2),{\rm adjacent}} \times \text{irreducible part of $R$} + \overline{ Z_m \text{irreducible part of $R$}}_{(a2)}\times R
\end{aligned}
\end{equation}
where the irreducible part of $R$ is taking the following form 
\begin{equation}
\begin{aligned}
    \text{irreducible part of $R$} & = \frac{1}{\lambda}\sum_{n=1}^{\infty} Z_n R^{n-1} \\
    & =  \includegraphics[valign=c,width=0.22\textwidth]{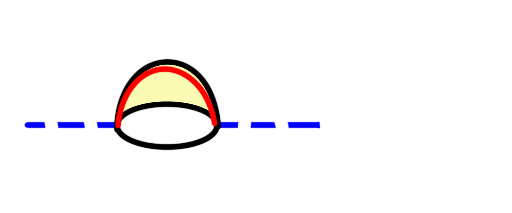}\hskip -30pt +\includegraphics[valign=c,width=0.28\textwidth]{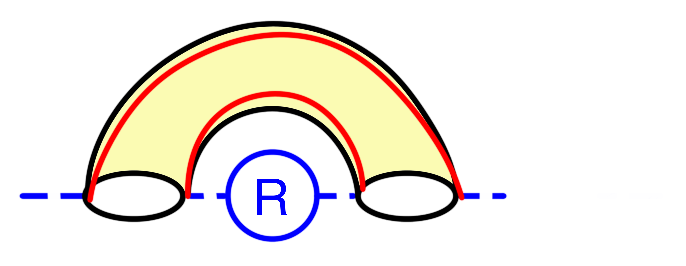} \hskip -30pt 
 +\cdots\\
    &= \frac{1}{\lambda R} \int \overline{\rho(R_L)\rho(E_R)} \frac{R  y(R_L,E_R)}{1- R y(E_L,E_R)}  
\end{aligned}
\end{equation}
This means 
\begin{equation}\label{eq:zmRirr}
\begin{aligned}
    \lambda \overline{ Z_m \text{irreducible part of $R$}}_{(a2)} &  = \underbrace{\textcolor{gray}{\sum_{n=1}^{\infty} \overline{Z_m Z_n}_{(a2)} R^{n-1}} }_{\text{absent}} + \sum_{n=1}^{\infty} (n-1) \overline{Z_m R}_{(a2),\rm non-adjacent} Z_n R^{n-2} \\
    & = \underbrace{\textcolor{gray}{\sum_{n=1}^{\infty} \overline{Z_m Z_n}_{(a2)} R^{n-1}} }_{\text{absent}}  +  \frac{\overline{Z_m R}_{(a2),\rm non-adjacent}}{R}  \left(\sum_{n}n Z_n R^{n-1} -\frac{1}{R} \sum_{n}Z_n R^{n}\right)\,,
\end{aligned}
\end{equation}
in which we highlight, using the gray color, the terms that are naively included but actually absent terms because they are of the (a1) terms.

To derive the non-adjacent one, one just notice that there is a one-to-one map between the $\overline{Z R}_{(a2),{\rm non-adjacent}}$ and  $\overline{Z R}_{(c1)}$ by replacing the empty wormhole in the later with a matter-geodesic-supported one. Then 
\begin{equation}
    \overline{Z_m R}_{(a2),{\rm non-adjacent}} = \frac{\sum_n \overline{Z_m Z_n}_{(a2)} R^n }{\lambda - X}
    \,.
\end{equation}
And we can plug this back to  \eqref{eq:zmRirr} and furthermore \eqref{eq:zmRa2adj} to get $\overline{ Z_m R}_{(a2),{\rm adjacent}}$, which in the end takes the form 
\begin{equation}
\begin{aligned}
    \overline{ Z_m R}_{(a2),{\rm adjacent}} & = \frac{\lambda R}{K}  \, \overline{ Z_m \text{irreducible part of $R$}}_{(a2)} \times R \\
    & = \frac{ R}{K} \sum_{n=1}^{\infty} \overline{Z_m Z_n}_{(a2)} R^{n} \times \left(-1 + 1+ \frac{1}{\lambda -\sum_n n Z_n R^{n-1}}\left( X-\frac{1}{R}\sum_n Z_n R^{n}\right)\right)\\
    & =   \frac{ \sum_{n=1}^{\infty} \overline{Z_m Z_n}_{(a2)} R^{n} }{\lambda - X } - \frac{R}{K} \sum_{n=1}^{\infty} \overline{Z_m Z_n}_{(a2)} R^{n} \,.
\end{aligned}
\end{equation}

For $\overline{RR}_{(c1)}$, one can argue that it is simply replacing $\overline{Z_m Z_n}_{(a2)}$ with $\overline{Z_m Z_n}_{(c1)}$ in \eqref{eq:RRa2} because only $ \overline{ Z_m R}_{(a2),{\rm non-adjacent}} $ appears
\begin{equation}\label{eq:RRc1}
\begin{aligned}
    \overline{RR}_{(a2)} & = \frac{\sum_{m,n} \overline{Z_m Z_n}_{(a2)} R^m R^n + 2 \left(\sum_m m Z_m R^{m-1}\right) \sum_{n}  \overline{R Z_n }_{(a2)} R^n}{\lambda^2-\left(\sum_{n}n Z_n R^{n-1} \right)^2 }\\
    & = \frac{\sum_{m,n} \overline{Z_m Z_n}_{(a2)} R^m R^n }{\left(\lambda-\sum_m m Z_m R^{m-1}\right)^2}
    \,.
\end{aligned}
\end{equation}

In the end, summarizing the above result, we again have 
\begin{equation}
    \text{(a2) contribution}= \frac{1}{\lambda}\sum_{m} m \overline{Z_m R}_{(a2), {\rm adjacent}} R^{m-1} + \frac{1}{\lambda}\sum_{m} \frac{m(m-1)}{2} Z_m R^{m-2} \overline{R R }_{(a2)} \,. 
\end{equation}
and the approximation in the $\lambda\rightarrow 0$ limit gives 
\begin{equation}
    \text{(a2) contribution} \sim  - \frac{1}{(K-\overline{d^2})^2} \frac{(\overline{d^2})^2}{K(K-\overline{d^2})} \int \overline{\rho(E_L)\rho(E_R)} \frac{1}{y(E_L,E_R)} \sim  \frac{1}{K^4}O(\lambda^0) \,.
\end{equation}
Thus, putting the results about the geometries \textbf{(c1)} and \textbf{(a2)} together, we find that the correction to the resolvent close to $\lambda =0$ is finite so that \eqref{eq:R-correction-from-(c1)-and-(a2)} is true.

\paragraph{The resulting $\text{Tr}_{\mathcal{H}_{bulk}}$ computation:}
Following our the resolvent computation, one might ask whether $\text{Tr}_{\mathcal{H}_{bulk}}$ is actually modified. 
We will see that the answer is \textbf{no}.
\begin{equation}\label{eq:a2final}
\begin{aligned}
\overline{\text{Tr}_{\mHb}(k_L k_R)}  &=
e^{2S_0} 
\int \overline{\rho(E_L) \rho(E_R) }
 \oint_{C_2}
 \frac{\dd \lambda}{2\pi \ii}\,
 \frac{1}{\lambda}
\frac{{R}_{\text{(a1)\&(b1)}}(\lambda) y_k(s_L,s_R)}{1-{R}_{\text{(a1)\&(b1)}} (\lambda) y_q(s_L,s_R)}  \\ 
& \quad + \oint_{C_2} \frac{\dd \lambda}{2\pi \ii}\,
 \frac{1}{\lambda} \left(\text{higher resolvent moments from (a2) and (c1) geometries}\right)\\
& = e^{2S_0} \int \overline{\rho(E_L) \rho(E_R) } \frac{y_k(E_L,E_R)}{y_q(E_L,E_R)} =  e^{2S_0} \int \overline{\rho(E_L) \rho(E_R) } e^{-\beta_L E_L} e^{-\beta_R E_R} \,.
\end{aligned}
\end{equation}

The resummation of (c1) and (a2) geometries is similar to the above resolvent derivation, and we simply  present a bit detail for (c1) as
\begin{equation}\label{eq:c1final2}
\begin{aligned}
     \text{Tr}_{\mathcal{H}_{bulk}}(\text{(c1) contribution}) & = \oint_{C_2} \frac{\dd \lambda}{2\pi \ii}\,
 \frac{1}{\lambda}  \left\{ \sum_{m} m \overline{Z_m(K_L,k_R) R} R^{m-1} + \sum_{m} \frac{m(m-1)}{2} Z_m(K_L,k_R) R^{m-2} \overline{R R } \right\} \\
    & = \oint_{C_2} \frac{\dd \lambda}{2\pi \ii} \left\{\frac{1}{\lambda}\frac{ \widetilde{\partial_R Y} }{ \lambda  - X} + \frac{1}{2\lambda} \partial_R X \frac{\tilde{Y} }{(\lambda - X)^2} \bigg|_{R=R(\lambda)} \right\}
    \,,
\end{aligned}
\end{equation}
where 
\begin{equation}
\begin{aligned}
      \tilde{Y} & = \int \left(\delta \rho^2\right)^2  \frac{R y_{k}(E_L,E_R)}{1-R y_q(E_L,E_R)} \frac{R y_q(E_L^{\prime},E_R^{\prime})}{1-R y_q(E_L^{\prime},E_R^{\prime})} \rightarrow  \int \left(\delta \rho^2\right)^2 e^{-\beta_L E_L} e^{-\beta_R E_R}  \\
      \widetilde{\partial_R Y} & = \int \left(\delta \rho^2\right)^2  \frac{R y_k(E_L,E_R)}{1-R y_k(E_L,E_R)} \frac{ y_q(E_L^{\prime},E_R^{\prime})}{(1-R y_k(E_L^{\prime},E_R^{\prime}))^2} \rightarrow  \frac{\lambda^2}{(K-\overline{d^2})^2} \int \left(\delta \rho^2\right)^2 \frac{e^{-\beta_L E_L} e^{-\beta_R E_R} }{y_q(E_L^{\prime},E_R^{\prime})}\,.
\end{aligned}
\end{equation}
And this leads to 
\begin{equation}
   \text{Tr}_{\mathcal{H}_{bulk}}(\text{(c1) contribution}) = \oint_{C_2} \frac{\dd \lambda}{2\pi \ii} \left\{\frac{1}{\lambda}\frac{ \widetilde{\partial_R Y} }{ \lambda  - X} + \frac{1}{2\lambda} \partial_R X \frac{\tilde{Y} }{(\lambda - X)^2} \bigg|_{R=R(\lambda)} \right\} = 0\,.
\end{equation}
because the integrand in the bracket is regular at $\lambda=0$. 
The (a2) geometries gives no correction either.

We have one final comment on the these corrections. Recall that when performing the analytic continuation, we need to smoothly deform the contour from the brach cut to the $\lambda=0$ pole. 
However, from the expression \eqref{eq:c1final2}, we see a possible pole at $\lambda= X = \int \overline{\rho(E_L)\rho(E_R)}\frac{y(E_L,E_R)}{(1-R(\lambda) y(E_L,E_R)}$. 
Fortunately, from our discussion previously on the branch cut in Appendix~\ref{ap:analytic}, in particular \eqref{eq:lambdabp}. 
The condition $\lambda = X $ is actually the condition for branch points.
Therefore, the contour surrounding the branch cut also includes these additional poles, and there is no additional singularity structure from introducing the (c1) and (a2) geometries.

\pagebreak

\bibliographystyle{jhep}
\bibliography{biblio.bib}

\providecommand{\href}[2]{#2}\begingroup\raggedright\begin{thebibliography}{10}

\bibitem{Maldacena:1997re}
J.M.~Maldacena, \emph{{The Large N limit of superconformal field theories and supergravity}}, \href{https://doi.org/10.1023/A:1026654312961, 10.4310/ATMP.1998.v2.n2.a1}{\emph{Int. J. Theor. Phys.} {\bfseries 38} (1999) 1113} [\href{https://arxiv.org/abs/hep-th/9711200}{{\ttfamily hep-th/9711200}}].

\bibitem{Gubser:1998bc}
S.S.~Gubser, I.R.~Klebanov and A.M.~Polyakov, \emph{{Gauge theory correlators from noncritical string theory}}, \href{https://doi.org/10.1016/S0370-2693(98)00377-3}{\emph{Phys. Lett.} {\bfseries B428} (1998) 105} [\href{https://arxiv.org/abs/hep-th/9802109}{{\ttfamily hep-th/9802109}}].

\bibitem{Witten:1998qj}
E.~Witten, \emph{{Anti-de Sitter space and holography}}, \href{https://doi.org/10.4310/ATMP.1998.v2.n2.a2}{\emph{Adv. Theor. Math. Phys.} {\bfseries 2} (1998) 253} [\href{https://arxiv.org/abs/hep-th/9802150}{{\ttfamily hep-th/9802150}}].

\bibitem{Maldacena:2001kr}
J.M.~Maldacena, \emph{{Eternal black holes in anti-de Sitter}}, \href{https://doi.org/10.1088/1126-6708/2003/04/021}{\emph{JHEP} {\bfseries 04} (2003) 021} [\href{https://arxiv.org/abs/hep-th/0106112}{{\ttfamily hep-th/0106112}}].

\bibitem{Maldacena:2013xja}
J.~Maldacena and L.~Susskind, \emph{{Cool horizons for entangled black holes}}, \href{https://doi.org/10.1002/prop.201300020}{\emph{Fortsch. Phys.} {\bfseries 61} (2013) 781} [\href{https://arxiv.org/abs/1306.0533}{{\ttfamily 1306.0533}}].

\bibitem{Guica:2015zpf}
M.~Guica and D.L.~Jafferis, \emph{{On the construction of charged operators inside an eternal black hole}}, \href{https://doi.org/10.21468/SciPostPhys.3.2.016}{\emph{SciPost Phys.} {\bfseries 3} (2017) 016} [\href{https://arxiv.org/abs/1511.05627}{{\ttfamily 1511.05627}}].

\bibitem{Harlow:2015lma}
D.~Harlow, \emph{{Wormholes, Emergent Gauge Fields, and the Weak Gravity Conjecture}}, \href{https://doi.org/10.1007/JHEP01(2016)122}{\emph{JHEP} {\bfseries 01} (2016) 122} [\href{https://arxiv.org/abs/1510.07911}{{\ttfamily 1510.07911}}].

\bibitem{Harlow:2018tqv}
D.~Harlow and D.~Jafferis, \emph{{The Factorization Problem in Jackiw-Teitelboim Gravity}}, \href{https://doi.org/10.1007/JHEP02(2020)177}{\emph{JHEP} {\bfseries 02} (2020) 177} [\href{https://arxiv.org/abs/1804.01081}{{\ttfamily 1804.01081}}].

\bibitem{Penington:2023dql}
G.~Penington and E.~Witten, \emph{{Algebras and States in JT Gravity}},  \href{https://arxiv.org/abs/2301.07257}{{\ttfamily 2301.07257}}.

\bibitem{Leutheusser:2021qhd}
S.~Leutheusser and H.~Liu, \emph{{Causal connectability between quantum systems and the black hole interior in holographic duality}},  \href{https://arxiv.org/abs/2110.05497}{{\ttfamily 2110.05497}}.

\bibitem{Witten:2021unn}
E.~Witten, \emph{{Gravity and the Crossed Product}},  \href{https://arxiv.org/abs/2112.12828}{{\ttfamily 2112.12828}}.

\bibitem{Chandrasekaran:2022eqq}
V.~Chandrasekaran, G.~Penington and E.~Witten, \emph{{Large N algebras and generalized entropy}}, \href{https://doi.org/10.1007/JHEP04(2023)009}{\emph{JHEP} {\bfseries 04} (2023) 009} [\href{https://arxiv.org/abs/2209.10454}{{\ttfamily 2209.10454}}].

\bibitem{Bahiru:2022mwh}
E.~Bahiru, \emph{{Algebra of operators in an AdS-Rindler wedge}}, \href{https://doi.org/10.1007/JHEP06(2023)197}{\emph{JHEP} {\bfseries 06} (2023) 197} [\href{https://arxiv.org/abs/2208.04258}{{\ttfamily 2208.04258}}].

\bibitem{Jensen:2023yxy}
K.~Jensen, J.~Sorce and A.J.~Speranza, \emph{{Generalized entropy for general subregions in quantum gravity}}, \href{https://doi.org/10.1007/JHEP12(2023)020}{\emph{JHEP} {\bfseries 12} (2023) 020} [\href{https://arxiv.org/abs/2306.01837}{{\ttfamily 2306.01837}}].

\bibitem{Witten:2023qsv}
E.~Witten, \emph{{Algebras, regions, and observers}}, {\emph{Proc. Symp. Pure Math.} {\bfseries 107} (2024) 247} [\href{https://arxiv.org/abs/2303.02837}{{\ttfamily 2303.02837}}].

\bibitem{Witten:2023xze}
E.~Witten, \emph{{A background-independent algebra in quantum gravity}}, \href{https://doi.org/10.1007/JHEP03(2024)077}{\emph{JHEP} {\bfseries 03} (2024) 077} [\href{https://arxiv.org/abs/2308.03663}{{\ttfamily 2308.03663}}].

\bibitem{Kolchmeyer:2023gwa}
D.K.~Kolchmeyer, \emph{{von Neumann algebras in JT gravity}}, \href{https://doi.org/10.1007/JHEP06(2023)067}{\emph{JHEP} {\bfseries 06} (2023) 067} [\href{https://arxiv.org/abs/2303.04701}{{\ttfamily 2303.04701}}].

\bibitem{Kudler-Flam:2023qfl}
J.~Kudler-Flam, S.~Leutheusser and G.~Satishchandran, \emph{{Generalized Black Hole Entropy is von Neumann Entropy}},  \href{https://arxiv.org/abs/2309.15897}{{\ttfamily 2309.15897}}.

\bibitem{Engelhardt:2023xer}
N.~Engelhardt and H.~Liu, \emph{{Algebraic ER=EPR and Complexity Transfer}},  \href{https://arxiv.org/abs/2311.04281}{{\ttfamily 2311.04281}}.

\bibitem{Soni:2023fke}
R.M.~Soni, \emph{{A type I approximation of the crossed product}}, \href{https://doi.org/10.1007/JHEP01(2024)123}{\emph{JHEP} {\bfseries 01} (2024) 123} [\href{https://arxiv.org/abs/2307.12481}{{\ttfamily 2307.12481}}].

\bibitem{AliAhmad:2023etg}
S.~Ali~Ahmad and R.~Jefferson, \emph{{Crossed product algebras and generalized entropy for subregions}}, \href{https://doi.org/10.21468/SciPostPhysCore.7.2.020}{\emph{SciPost Phys. Core} {\bfseries 7} (2024) 020} [\href{https://arxiv.org/abs/2306.07323}{{\ttfamily 2306.07323}}].

\bibitem{Akers:2024bel}
C.~Akers and J.~Sorce, \emph{{Relative state-counting for semiclassical black holes}},  \href{https://arxiv.org/abs/2404.16098}{{\ttfamily 2404.16098}}.

\bibitem{Cirafici:2024jdw}
M.~Cirafici, \emph{{On the Nonequilibrium Dynamics of Gravitational Algebras}},  \href{https://arxiv.org/abs/2402.03939}{{\ttfamily 2402.03939}}.

\bibitem{Iliesiu:2024cnh}
L.V.~Iliesiu, A.~Levine, H.W.~Lin, H.~Maxfield and M.~Mezei, \emph{{On the non-perturbative bulk Hilbert space of JT gravity}},  \href{https://arxiv.org/abs/2403.08696}{{\ttfamily 2403.08696}}.

\bibitem{Almheiri:2019psf}
A.~Almheiri, N.~Engelhardt, D.~Marolf and H.~Maxfield, \emph{{The entropy of bulk quantum fields and the entanglement wedge of an evaporating black hole}}, \href{https://doi.org/10.1007/JHEP12(2019)063}{\emph{JHEP} {\bfseries 12} (2019) 063} [\href{https://arxiv.org/abs/1905.08762}{{\ttfamily 1905.08762}}].

\bibitem{Penington:2019npb}
G.~Penington, \emph{{Entanglement Wedge Reconstruction and the Information Paradox}}, \href{https://doi.org/10.1007/JHEP09(2020)002}{\emph{JHEP} {\bfseries 09} (2020) 002} [\href{https://arxiv.org/abs/1905.08255}{{\ttfamily 1905.08255}}].

\bibitem{Penington:2019kki}
G.~Penington, S.H.~Shenker, D.~Stanford and Z.~Yang, \emph{{Replica wormholes and the black hole interior}}, \href{https://doi.org/10.1007/JHEP03(2022)205}{\emph{JHEP} {\bfseries 03} (2022) 205} [\href{https://arxiv.org/abs/1911.11977}{{\ttfamily 1911.11977}}].

\bibitem{Almheiri:2020cfm}
A.~Almheiri, T.~Hartman, J.~Maldacena, E.~Shaghoulian and A.~Tajdini, \emph{{The entropy of Hawking radiation}}, \href{https://doi.org/10.1103/RevModPhys.93.035002}{\emph{Rev. Mod. Phys.} {\bfseries 93} (2021) 035002} [\href{https://arxiv.org/abs/2006.06872}{{\ttfamily 2006.06872}}].

\bibitem{Hsin:2020mfa}
P.-S.~Hsin, L.V.~Iliesiu and Z.~Yang, \emph{{A violation of global symmetries from replica wormholes and the fate of black hole remnants}}, \href{https://doi.org/10.1088/1361-6382/ac2134}{\emph{Class. Quant. Grav.} {\bfseries 38} (2021) 194004} [\href{https://arxiv.org/abs/2011.09444}{{\ttfamily 2011.09444}}].

\bibitem{Boruch:2023trc}
J.~Boruch, L.V.~Iliesiu and C.~Yan, \emph{{Constructing all BPS black hole microstates from the gravitational path integral}},  \href{https://arxiv.org/abs/2307.13051}{{\ttfamily 2307.13051}}.

\bibitem{Teitelboim}
C.~Teitelboim, \emph{{Gravitation and Hamiltonian Structure in Two Space-Time Dimensions}}, \href{https://doi.org/10.1016/0370-2693(83)90012-6}{\emph{Phys. Lett.} {\bfseries B126} (1983) 41}.

\bibitem{Jackiw}
R.~Jackiw, \emph{{Lower Dimensional Gravity}}, \href{https://doi.org/10.1016/0550-3213(85)90448-1}{\emph{Nucl. Phys.} {\bfseries B252} (1985) 343}.

\bibitem{Henneaux:1985nw}
M.~Henneaux, \emph{{QUANTUM GRAVITY IN TWO-DIMENSIONS: EXACT SOLUTION OF THE JACKIW MODEL}}, \href{https://doi.org/10.1103/PhysRevLett.54.959}{\emph{Phys. Rev. Lett.} {\bfseries 54} (1985) 959}.

\bibitem{Louis-Martinez:1993bge}
D.~Louis-Martinez, J.~Gegenberg and G.~Kunstatter, \emph{{Exact Dirac quantization of all 2-D dilaton gravity theories}}, \href{https://doi.org/10.1016/0370-2693(94)90463-4}{\emph{Phys. Lett. B} {\bfseries 321} (1994) 193} [\href{https://arxiv.org/abs/gr-qc/9309018}{{\ttfamily gr-qc/9309018}}].

\bibitem{AlmheiriPolchinski}
A.~Almheiri and J.~Polchinski, \emph{{Models of AdS$_{2}$ backreaction and holography}}, \href{https://doi.org/10.1007/JHEP11(2015)014}{\emph{JHEP} {\bfseries 11} (2015) 014} [\href{https://arxiv.org/abs/1402.6334}{{\ttfamily 1402.6334}}].

\bibitem{Yang:2018gdb}
Z.~Yang, \emph{{The Quantum Gravity Dynamics of Near Extremal Black Holes}}, \href{https://doi.org/10.1007/JHEP05(2019)205}{\emph{JHEP} {\bfseries 05} (2019) 205} [\href{https://arxiv.org/abs/1809.08647}{{\ttfamily 1809.08647}}].

\bibitem{Maldacena:2016upp}
J.~Maldacena, D.~Stanford and Z.~Yang, \emph{{Conformal symmetry and its breaking in two dimensional Nearly Anti-de-Sitter space}}, \href{https://doi.org/10.1093/ptep/ptw124}{\emph{PTEP} {\bfseries 2016} (2016) 12C104} [\href{https://arxiv.org/abs/1606.01857}{{\ttfamily 1606.01857}}].

\bibitem{Kitaev:2017awl}
A.~Kitaev and S.J.~Suh, \emph{{The soft mode in the Sachdev-Ye-Kitaev model and its gravity dual}}, \href{https://doi.org/10.1007/JHEP05(2018)183}{\emph{JHEP} {\bfseries 05} (2018) 183} [\href{https://arxiv.org/abs/1711.08467}{{\ttfamily 1711.08467}}].

\bibitem{Kitaev:2018wpr}
A.~Kitaev and S.J.~Suh, \emph{{Statistical mechanics of a two-dimensional black hole}}, \href{https://doi.org/10.1007/JHEP05(2019)198}{\emph{JHEP} {\bfseries 05} (2019) 198} [\href{https://arxiv.org/abs/1808.07032}{{\ttfamily 1808.07032}}].

\bibitem{Saad:2019pqd}
P.~Saad, \emph{{Late Time Correlation Functions, Baby Universes, and ETH in JT Gravity}},  \href{https://arxiv.org/abs/1910.10311}{{\ttfamily 1910.10311}}.

\bibitem{Mertens:2022irh}
T.G.~Mertens and G.J.~Turiaci, \emph{{Solvable Models of Quantum Black Holes: A Review on Jackiw-Teitelboim Gravity}},  \href{https://arxiv.org/abs/2210.10846}{{\ttfamily 2210.10846}}.

\bibitem{Jafferis:2022wez}
D.L.~Jafferis, D.K.~Kolchmeyer, B.~Mukhametzhanov and J.~Sonner, \emph{{Jackiw-Teitelboim gravity with matter, generalized eigenstate thermalization hypothesis, and random matrices}}, \href{https://doi.org/10.1103/PhysRevD.108.066015}{\emph{Phys. Rev. D} {\bfseries 108} (2023) 066015} [\href{https://arxiv.org/abs/2209.02131}{{\ttfamily 2209.02131}}].

\bibitem{Kitaev:2017hnr}
A.~Kitaev, \emph{{Notes on $\widetilde{\mathrm{SL}}(2,\mathbb{R})$ representations}},  \href{https://arxiv.org/abs/1711.08169}{{\ttfamily 1711.08169}}.

\bibitem{Lin:2019qwu}
H.W.~Lin, J.~Maldacena and Y.~Zhao, \emph{{Symmetries Near the Horizon}}, \href{https://doi.org/10.1007/JHEP08(2019)049}{\emph{JHEP} {\bfseries 08} (2019) 049} [\href{https://arxiv.org/abs/1904.12820}{{\ttfamily 1904.12820}}].

\bibitem{Stanford:2020wkf}
D.~Stanford, \emph{{More quantum noise from wormholes}},  \href{https://arxiv.org/abs/2008.08570}{{\ttfamily 2008.08570}}.

\bibitem{Balasubramanian:2022gmo}
V.~Balasubramanian, A.~Lawrence, J.M.~Magan and M.~Sasieta, \emph{{Microscopic Origin of the Entropy of Black Holes in General Relativity}}, \href{https://doi.org/10.1103/PhysRevX.14.011024}{\emph{Phys. Rev. X} {\bfseries 14} (2024) 011024} [\href{https://arxiv.org/abs/2212.02447}{{\ttfamily 2212.02447}}].

\bibitem{Balasubramanian:2022lnw}
V.~Balasubramanian, A.~Lawrence, J.M.~Magan and M.~Sasieta, \emph{{Microscopic Origin of the Entropy of Astrophysical Black Holes}}, \href{https://doi.org/10.1103/PhysRevLett.132.141501}{\emph{Phys. Rev. Lett.} {\bfseries 132} (2024) 141501} [\href{https://arxiv.org/abs/2212.08623}{{\ttfamily 2212.08623}}].

\bibitem{Bousso:2023efc}
R.~Bousso and M.~Miyaji, \emph{{Fluctuations in the Entropy of Hawking Radiation}},  \href{https://arxiv.org/abs/2307.13920}{{\ttfamily 2307.13920}}.

\bibitem{Saad:2019lba}
P.~Saad, S.H.~Shenker and D.~Stanford, \emph{{JT gravity as a matrix integral}},  \href{https://arxiv.org/abs/1903.11115}{{\ttfamily 1903.11115}}.

\bibitem{Iliesiu:2021are}
L.V.~Iliesiu, M.~Kologlu and G.J.~Turiaci, \emph{{Supersymmetric indices factorize}},  \href{https://arxiv.org/abs/2107.09062}{{\ttfamily 2107.09062}}.

\bibitem{Boruch:2022tno}
J.~Boruch, M.T.~Heydeman, L.V.~Iliesiu and G.J.~Turiaci, \emph{{BPS and near-BPS black holes in $AdS_5$ and their spectrum in $\mathcal{N}=4$ SYM}},  \href{https://arxiv.org/abs/2203.01331}{{\ttfamily 2203.01331}}.

\bibitem{Stanford:2017thb}
D.~Stanford and E.~Witten, \emph{{Fermionic Localization of the Schwarzian Theory}}, \href{https://doi.org/10.1007/JHEP10(2017)008}{\emph{JHEP} {\bfseries 10} (2017) 008} [\href{https://arxiv.org/abs/1703.04612}{{\ttfamily 1703.04612}}].

\bibitem{Mertens:2017mtv}
T.G.~Mertens, G.J.~Turiaci and H.L.~Verlinde, \emph{{Solving the Schwarzian via the Conformal Bootstrap}}, \href{https://doi.org/10.1007/JHEP08(2017)136}{\emph{JHEP} {\bfseries 08} (2017) 136} [\href{https://arxiv.org/abs/1705.08408}{{\ttfamily 1705.08408}}].

\bibitem{LongPaper}
H.W.~Lin, J.~Maldacena, L.~Rozenberg and J.~Shan, \emph{{Looking at supersymmetric black holes for a very long time}},  \href{https://arxiv.org/abs/2207.00408}{{\ttfamily 2207.00408}}.

\bibitem{ShortPaper}
H.W.~Lin, J.~Maldacena, L.~Rozenberg and J.~Shan, \emph{{Holography for people with no time}},  \href{https://arxiv.org/abs/2207.00407}{{\ttfamily 2207.00407}}.

\bibitem{Turiaci:2023jfa}
G.J.~Turiaci and E.~Witten, \emph{{$\mathcal{N}=2$ JT Supergravity and Matrix Models}},  \href{https://arxiv.org/abs/2305.19438}{{\ttfamily 2305.19438}}.

\bibitem{Belaey:2024dde}
A.~Belaey, F.~Mariani and T.G.~Mertens, \emph{{Gravitational wavefunctions in JT supergravity}},  \href{https://arxiv.org/abs/2405.09289}{{\ttfamily 2405.09289}}.

\bibitem{Harlow:2020bee}
D.~Harlow and E.~Shaghoulian, \emph{{Global symmetry, Euclidean gravity, and the black hole information problem}}, \href{https://doi.org/10.1007/JHEP04(2021)175}{\emph{JHEP} {\bfseries 04} (2021) 175} [\href{https://arxiv.org/abs/2010.10539}{{\ttfamily 2010.10539}}].

\bibitem{Chen:2020ojn}
Y.~Chen and H.W.~Lin, \emph{{Signatures of global symmetry violation in relative entropies and replica wormholes}}, \href{https://doi.org/10.1007/JHEP03(2021)040}{\emph{JHEP} {\bfseries 03} (2021) 040} [\href{https://arxiv.org/abs/2011.06005}{{\ttfamily 2011.06005}}].

\bibitem{Bah:2022uyz}
I.~Bah, Y.~Chen and J.~Maldacena, \emph{{Estimating global charge violating amplitudes from wormholes}}, \href{https://doi.org/10.1007/JHEP04(2023)061}{\emph{JHEP} {\bfseries 04} (2023) 061} [\href{https://arxiv.org/abs/2212.08668}{{\ttfamily 2212.08668}}].

\bibitem{Colafranceschi:2023urj}
E.~Colafranceschi, X.~Dong, D.~Marolf and Z.~Wang, \emph{{Algebras and Hilbert spaces from gravitational path integrals: Understanding Ryu-Takayanagi/HRT as entropy without invoking holography}},  \href{https://arxiv.org/abs/2310.02189}{{\ttfamily 2310.02189}}.

\bibitem{Marolf:2024adj}
D.~Marolf and D.~Zhang, \emph{{When left and right disagree: Entropy and von Neumann algebras in quantum gravity with general AlAdS boundary conditions}},  \href{https://arxiv.org/abs/2402.09691}{{\ttfamily 2402.09691}}.

\bibitem{Akers:2022qdl}
C.~Akers, N.~Engelhardt, D.~Harlow, G.~Penington and S.~Vardhan, \emph{{The black hole interior from non-isometric codes and complexity}},  \href{https://arxiv.org/abs/2207.06536}{{\ttfamily 2207.06536}}.

\bibitem{Jafferis:2017tiu}
D.L.~Jafferis, \emph{{Bulk reconstruction and the Hartle-Hawking wavefunction}},  \href{https://arxiv.org/abs/1703.01519}{{\ttfamily 1703.01519}}.

\bibitem{Marolf:2012xe}
D.~Marolf and A.C.~Wall, \emph{{Eternal Black Holes and Superselection in AdS/CFT}}, \href{https://doi.org/10.1088/0264-9381/30/2/025001}{\emph{Class. Quant. Grav.} {\bfseries 30} (2013) 025001} [\href{https://arxiv.org/abs/1210.3590}{{\ttfamily 1210.3590}}].

\bibitem{Climent:2024trz}
A.~Climent, R.~Emparan, J.M.~Magan, M.~Sasieta and A.~Vilar~L\'opez, \emph{{Universal construction of black hole microstates}}, \href{https://doi.org/10.1103/PhysRevD.109.086024}{\emph{Phys. Rev. D} {\bfseries 109} (2024) 086024} [\href{https://arxiv.org/abs/2401.08775}{{\ttfamily 2401.08775}}].

\end{thebibliography}\endgroup

\end{document}